\documentclass[a4paper,11pt]{article}
\pdfoutput=1
\usepackage{jheppub}
\usepackage[ascii]{inputenc}
\usepackage{bbm,braket,microtype,mathtools}
\usepackage[T1]{fontenc}
\usepackage{lmodern}
\usepackage[USenglish]{babel}
\DeclareMathOperator{\tr}{tr}
\DeclareMathOperator{\Prob}{Prob}
\DeclareMathOperator{\rank}{rank}
\DeclareMathOperator{\Sym}{Sym}
\newcommand{\gav}[1]{\overline {#1}^{\text{good}}}
\newcommand{\nav}[1]{\overline {#1}^{\neq 0}}
\allowdisplaybreaks[4]

\begin{document}

\title{Holographic duality from random tensor networks}

\author{Patrick Hayden,}
\author{Sepehr Nezami,}
\author{Xiao-Liang Qi,}
\author{Nathaniel Thomas,}
\author{Michael Walter}
\author{and Zhao Yang}

\affiliation{Stanford Institute for Theoretical Physics, \\
Physics Department, Stanford University, CA 94304-4060, USA}

\abstract{Tensor networks provide a natural framework for exploring
  holographic duality because they obey entanglement area laws. They
  have been used to construct explicit toy models realizing many of
  the interesting structural features of the AdS/CFT correspondence,
  including the non-uniqueness of bulk operator reconstruction in the
  boundary theory. In this article, we explore the holographic
  properties of networks of random tensors. We find that our models naturally incorporate many features that are analogous to those of the AdS/CFT correspondence. When the bond dimension of the tensors is large, we show that the entanglement entropy of all boundary regions, whether connected or not, obey the Ryu-Takayanagi entropy formula, a fact closely related to known properties of the multipartite entanglement of assistance.
  We also discuss the behavior of R\'enyi entropies in our models and contrast it with AdS/CFT.
  Moreover, we find that each boundary region faithfully encodes the physics of the entire bulk entanglement wedge, {\it i.e.,} the bulk region enclosed by the boundary region and the minimal surface. Our method is to interpret the average over random tensors
  as the partition function of a classical ferromagnetic Ising model,
  so that the minimal surfaces of Ryu-Takayanagi appear as domain
  walls.  Upon including the analog of a bulk field, we find that our
  model reproduces the expected corrections to the Ryu-Takayanagi
  formula: the bulk minimal surface is displaced and the entropy is
  augmented by the entanglement of the bulk field. Increasing the
  entanglement of the bulk field ultimately changes the minimal surface behavior topologically, in a way similar to the effect of creating a black hole. Extrapolating bulk correlation functions to the boundary
  permits the calculation of the scaling dimensions of boundary
  operators, which exhibit a large gap between a small number of
  low-dimension operators and the rest.
  While we are primarily motivated by the AdS/CFT duality, the
  main results of the article define a more general form of
  bulk-boundary correspondence which could be useful for extending
  holography to other spacetimes.}

\maketitle

\section{Introduction}

Tensor networks have been proposed~\cite{swingle2012} as a helpful tool for understanding holographic duality~\cite{maldacena1998,witten1998,gubser1998} due to the intuition that the entropy of a tensor network is bounded by an area law that agrees with the Ryu-Takayanagi (RT) formula~\cite{ryu2006}.
In general, the area law only gives an upper bound to the entropy~\cite{swingle2012}, which for particular tensor networks and choices of regions has been shown to be saturated~\cite{pastawski2015}.
Tensor networks can also be used to build holographic mappings or holographic codes~\cite{qi2013,pastawski2015,yang2015}, which are isometries between the Hilbert space of the bulk and that of the boundary.
In particular, some of us have recently proposed bidirectional holographic codes built from tensors with particular properties, so-called pluperfect tensors~\cite{yang2015}.
These codes simultaneously satisfies several desired properties, including the RT formula for a subset of boundary states, error correction properties of bulk local operators~\cite{almheiri2014}, a kind of bulk gauge invariance, and the possibility of sub-AdS locality.

The perfect and pluperfect tensors defined in Refs.~\cite{pastawski2015} and \cite{yang2015}, respectively, have entanglement properties that are idealized version of large-dimensional random tensors, which is part of the motivation why it is natural to study these tensor networks.
In this work, we will show that by directly studying networks of large dimensional random tensors, instead of their ``idealized'' counterpart, their properties can be computed more systematically.
Specifically, we will assume that each tensor in the network is chosen independently at random.
We find that the computation of typical R\'{e}nyi entropies and other quantities of interest in the corresponding tensor network states can be mapped to the evaluation of partition functions of classical statistical models, namely generalized Ising models with boundary pinning fields.
When each leg of each tensor in the network has dimension $D$, these statistical models have inverse temperature $\beta\propto \log D$.
For large enough $D$, they are in the long-range ordered phase, and we find that the entropies of a boundary region is directly related to the energy of a domain wall between different domains of the order parameter.
The minimal energy condition for this domain wall naturally leads to the RT formula.%
\footnote{In our models, the RT formula holds for all R\'enyi entropies, which is an important difference from AdS/CFT~\cite{dong2016gravity}. We will discuss this point in more detail further below.}
Besides yielding the RT formula for general boundary subsystems, the technique of random state averaging allows us to study many further properties of a random tensor network:

\begin{enumerate}
\item {\bf Effects of bulk entanglement.}
  Using the random tensor
  network as a holographic mapping rather than a state on the
  boundary, we derive a formula for the entropy of a boundary region
  in the presence of an entangled state in the bulk.  As a special
  example of the effect of bulk entanglement, we show how the behavior
  of minimal surfaces (which are minimal energy domain walls in the
  statistical model) is changed qualitatively by introducing a highly
  entangled state in the bulk.  When the state is sufficiently highly
  entangled, no minimal surface penetrates into this region,
  so that the topology of the space has effectively changed.  This
  phenomenon is analogous to the change of spatial geometry in the Hawking-Page transition~\cite{hawking1983,witten1998b}, where the bulk geometry changes from perturbed AdS to a black hole upon increasing temperature.
\item {\bf Bidirectional holographic code and code subspace.}
By calculating the entanglement entropy between a bulk region and the boundary in a given tensor network, we can verify that the random tensor network defines a bidirectional holographic code (BHC).
When the bulk Hilbert space has a higher dimension than the boundary, we obtain an approximate isometry from the boundary to the bulk.
When we restrict the bulk degrees of freedom to a smaller subspace (``code subspace'', or ``low energy subspace'') which has dimension lower than the boundary Hilbert space dimension, we also obtain an approximate isometry from this bulk subspace to the boundary.
This bulk-to-boundary isometry satisfies the error correction properties defined in Ref.~\cite{almheiri2014}.
To be more precise, all bulk local operators in the entanglement wedge of a boundary region can be recovered from that boundary region.%
\footnote{In this work, the entanglement wedge of a boundary region refers to the spatial region enclosed by the boundary region and the minimal surface homologous to it, rather than to a space-time region.}
\item {\bf Correlation spectrum.}
In addition to entanglement entropies, we can also study properties of boundary multi-point functions.
In particular, we show that the boundary two-point functions are determined by the bulk two-point functions and the properties of the statistical model.
When the bulk geometry is a pure hyperbolic space, the boundary two-point correlations have power-law decay, which defines the scaling dimension spectrum.
We show that in large-dimensional random tensor networks there are two kinds of scaling dimensions, those from the bulk ``low energy'' theory which do not grow with the bond dimension $D$, and those from the tensor network itself which grow $\propto \log D$.
This confirms that the holographic mapping defined by a random tensor network maps a weakly-interacting bulk state to a boundary state with a scaling dimension gap, consistent with the expectations of AdS/CFT.
\end{enumerate}

The use of random matrix techniques has a long and rich history in quantum information theory (see, e.g., the recent review~\cite{collins2016random} and references therein).
Previous work on random tensor network states has originated from a diverse set of motivations, including the construction of novel random ensembles that satisfy a generalized area law~\cite{collins2010random,collins2013area}, the relationship between entropy and the decay of correlations~\cite{hastings2015random}, and the maximum entropy principle~\cite{collins2013matrix}.
 The relation between the Schmidt ranks of tensor network states and minimal cuts through the network has been investigated in~\cite{cui2015quantum}.
While the primary motivation for this work is to better understand holographic duality, its methods and even the nature of many of its conclusions place it squarely in this earlier tradition. In the holographic context, it was in fact previously shown that using a class of pseudo-random tensors known as quantum expanders in a MERA tensor network would reproduce the qualitative scaling of the Ryu-Takayanagi formula~\cite{swingle2012b}.

\medskip

The remainder of the paper is organized as follows.  In
Section~\ref{sec:setup} we define the random tensor networks. We show
how the calculation of the second R\'{e}nyi entropy is mapped to the
partition function of a classical Ising model.  In
Section~\ref{sec:RTformula} we investigate the RT formula in the large
dimension limit of the random tensors, and discuss the effect of bulk
entanglement.  As an explicit example we study the minimal surfaces
for a highly entangled (volume-law) bulk state and discuss the
transition of the effective bulk geometry as
a function of bulk entropy density.  In Section~\ref{sec:bhc} we study
the properties of the holographic mapping defined by random tensor
networks, including boundary-to-bulk isometries and bulk-to-boundary
isometries for the code subspace, and we discuss the recovery of
bulk operators from boundary regions.  In Section~\ref{sec:nthRenyi}
we generalize the calculation of the second R\'{e}nyi entropy to
higher R\'{e}nyi entropies.  We show that the $n$-th R\'{e}nyi entropy
calculation is mapped to the partition function of a statistical model
with a $\Sym_n$ permutation group element at each vertex.  The same
technique also enables us to compute other averaged quantities
involving higher powers of the density operator.  In
Section~\ref{sec:correlationspec} we use this technique to study the
boundary two-point correlation functions.  We show that the boundary
correlation functions are determined by the bulk correlations and the
tensor network, and that a gap in the scaling dimensions opens at
large $D$ in the case of AdS geometry.  In Section~\ref{sec:finiteD}
we bound the fluctuations around the typical values calculated
previously and discuss the effect of finite bond dimensions.
Section~\ref{sec:split-transfer} explains the close relationship
between the random tensors networks of this paper and optimal
multipartite entanglement distillation protocols previously studied in
the quantum information theory literature.  In Section~\ref{sec:stabs}
we consider other ensembles of random
states. 
We find that the RT formula can be exactly satisfied in tensor
networks built from random stabilizer states, which allows for the
construction of exact holographic codes.  Finally,
Section~\ref{sec:conclusion} is devoted to conclusion and discussion.


\section{General setup}\label{sec:setup}

\subsection{Definition of random tensor networks}

We start by defining the most general tensor network states in a language that is suitable for our later discussion.
A rank-$n$ tensor has components $T_{\mu_1\mu_2\dots\mu_n}$ with $\mu_k=1,2,\dots,D_k$.
We can define a Hilbert space $\mathbb{H}_k$ with dimension $D_k$ for each leg of the tensor, and consider the index $\mu_k$ as labeling a complete basis of states $\ket{\mu_k}$ in this Hilbert space.
In this language, $T_{\mu_1\mu_2\dots\mu_n}$ (with proper normalization) corresponds to the wavefunction of a quantum state $\ket{T}=\sum_{\left\{\mu_k\right\}} T_{\mu_1\mu_2\dots\mu_n} \ket{\mu_1} \otimes \ket{\mu_2} \otimes \dots \otimes \ket{\mu_n}$ defined in the product Hilbert space $\bigotimes_{k=1}^{n} \mathbb{H}_k$.

\begin{figure}
\begin{center}
\includegraphics[width=0.55\textwidth]{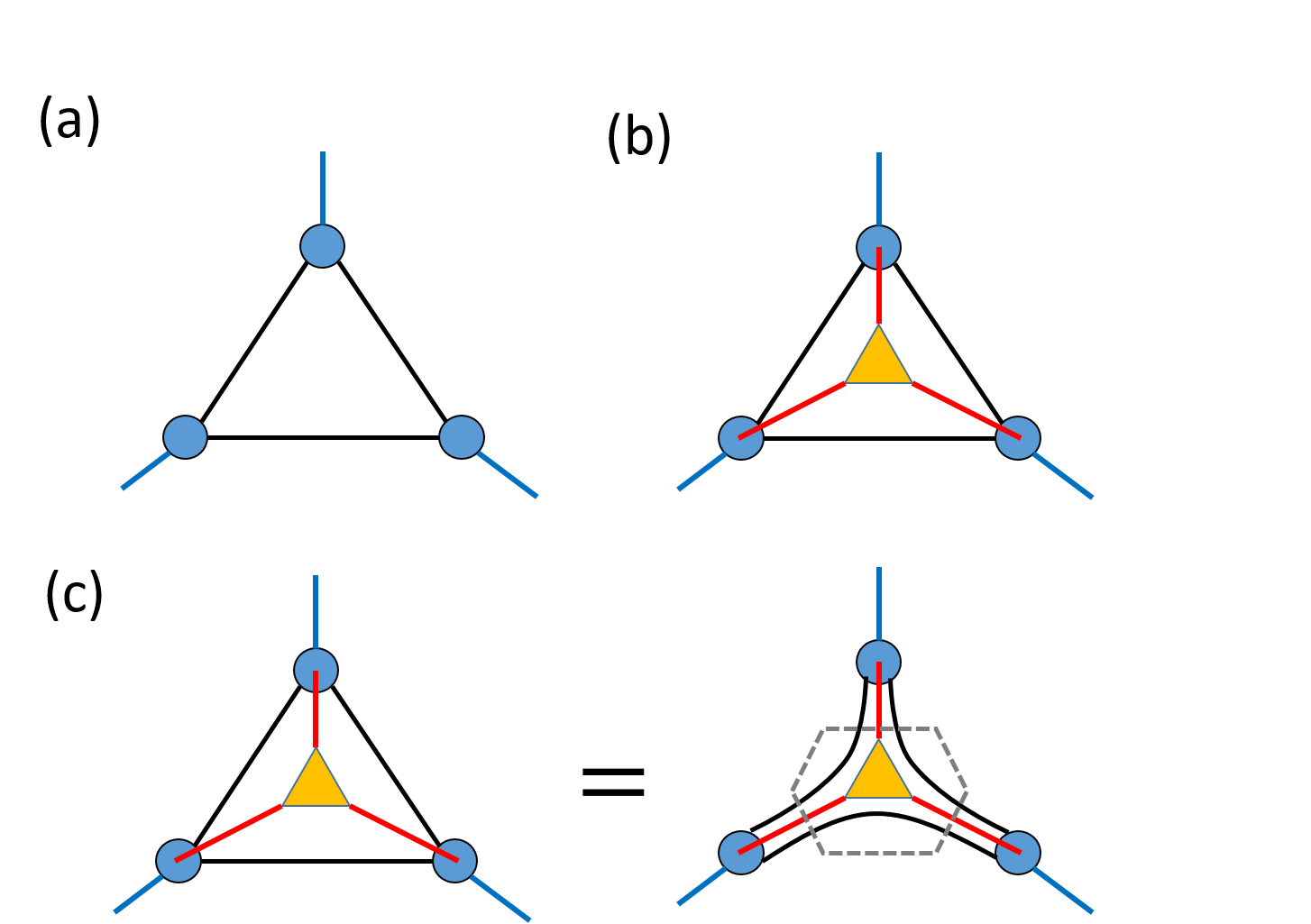}
\caption{(a) A tensor network that defines a state in the Hilbert space of the dangling indices. (b) A tensor network that defines a mapping from bulk legs (red) to boundary legs (blue). An arbitrary bulk state (orange triangle) is mapped to a boundary state. (For simplicity, we have drawn a pure state in the bulk. For a mixed state the map needs to be applied to both indices of the bulk density operator.) (c) The internal lines of the tensor network can always be combined with the bulk state and viewed as a state in an enlarged Hilbert space (enclosed by the dashed hexegon). In this view, each tensor acts independently on this generalized bulk state and maps it to the boundary state.} \label{fig:tensor}
\end{center}
\end{figure}

A tensor network is obtained by connecting tensors, {\it i.e.}, by contracting a common index.
For purposes of illustration, a small tensor network is shown in Fig.~\ref{fig:tensor}\,(a).
Before connecting the tensors, each tensor corresponds to a quantum state, so that the collection of all tensors can be considered as a tensor product state $\bigotimes_x \ket{V_x}$.
Here, $x$ denotes all vertices in the network, and $\left|V_x\right\rangle$ is the state corresponding to the tensor at vertex $x$.
Each leg of a tensor corresponds to a Hilbert space.
We will denote the Hilbert space corresponding to a leg from $x$ to another vertex $y$ by $\mathbb{H}_{xy}$, and its dimension by $D_{xy}$.
If a leg is dangling, {\it i.e.}, not connected to any other vertex, we will denote the corresponding Hilbert space by $\mathbb{H}_{x\partial}$ and its dimension by $D_{x\partial}$.
(Without loss of generality we can assume there is at most one dangling leg at each vertex.)
Connecting two tensors at $x,y$ by an internal line then corresponds to a projection in the Hilbert space $\mathbb{H}_{xy}\otimes \mathbb{H}_{yx}$ onto a maximally entangled state $\ket{xy}=\frac1{\sqrt{D_{xy}}}\sum_{\mu=1}^{D_{xy}}\ket{\mu_{xy}} \otimes \ket{\mu_{yx}}$.
Here $\ket{\mu_{xy}}$ denotes a state in the Hilbert space $\mathbb{H}_{xy}$ and similarly for $\ket{\mu_{yx}}$.
By connecting the tensors according to the internal lines of the tensor network, we thus obtain the state
\begin{equation}
    \ket{\Psi}=\left(\bigotimes_{\langle xy\rangle} \bra{xy}\right)\left(\bigotimes_x \ket{V_x}\right)\label{peps}
\end{equation}
in the Hilbert space corresponding to the dangling legs, $\bigotimes_{x\in \partial} \mathbb{H}_{x\partial}$.
We note that $\ket{\Psi}$ is in general not normalized.
Tensor network states defined in this way are often referred to as projected entangled pair states (PEPS)~\cite{verstraete2004}.

As has been discussed in previous works~\cite{qi2013,pastawski2015,yang2015}, tensor networks can be used to define not only quantum states but also holographic mappings, or holographic codes, which map between the Hilbert space of the bulk and that of the boundary.
Fig.~\ref{fig:tensor}\,(b) shows a very simple ``holographic mapping'' which maps the bulk indices (red lines) to boundary indices (blue lines), with internal lines (black lines) contracted.
A bulk state (orange triangle in the figure) is mapped to a boundary state by this mapping.
Such a boundary state can also be written in a form similar to Eq.~(\ref{peps}).
Instead of viewing the tensor network as defining a mapping, we can equivalently consider it as a quantum state in the Hilbert space $\mathbb{H}_b\otimes \mathbb{H}_\partial$, which is a direct product of the bulk Hilbert space $\mathbb{H}_b$ and the boundary Hilbert space $\mathbb{H}_\partial$.
Denoting the bulk state as $\ket{\Phi_b}$, the corresponding boundary state is
\begin{eqnarray} \label{eqn:boundary-state}
\ket{\Psi}=\left(\bra{\Phi_b} \otimes \bigotimes_{\langle xy\rangle} \bra{xy}\right)\left(\bigotimes_x \ket{V_x}\right)\label{peps2}.
\end{eqnarray}
From this expression one can see that the internal lines of the tensor network can actually be viewed as part of the bulk state. As is illustrated in Fig.~\ref{fig:tensor}\,(c), one can view the maximally entangled states on internal lines together with the bulk state $\ket{\Phi_b}$ as a state in the enlarged ``bulk Hilbert space''. This point of view will be helpful for our discussion. More generally, one can also have a mixed bulk state with density operator $\rho_b$, instead of the pure state $\ket{\Phi_b}$. The most generic form of the boundary state is given by the density operator
\begin{align}
\rho &= \tr_P\left(\rho_P \, \prod_x \ket{V_x}\!\!\bra{V_x}\right),\label{peps3}\\
\rho_P &=\rho_b \otimes \bigotimes_{\langle xy\rangle} \ket{xy}\!\!\bra{xy}.\label{rhoP}
\end{align}
Here the partial trace $\tr_P$ is carried over the bulk and internal legs of all tensors ({\it i.e.}, over all but the dangling legs). In this compact form, one can see that the state $\rho$ is a linear function of the independent pure states of each tensor $\ket{V_x}\!\!\bra{V_x}$.

In this work, we study tensor network states of the form \eqref{peps3}, where the tensors $\ket{V_x}$ are unit vectors chosen independently at random from their respective Hilbert spaces.
We will mostly use the ``uniform'' probability measure that is invariant under arbitrary unitary transformations.
Equivalently, we can take an arbitrary reference state $\ket{0_x}$ and define $\ket{V_x}=U\ket{0_x}$ with $U$ a unitary operator.
The random average of an arbitrary function $f\left(\ket{V_x}\right)$ of the state $\ket{V_x}$ is then equivalent to an integration over $U$ according to the Haar probability measure $\int dU f\left(U\ket{0_x}\right)$, with normalization $\int dU=1$.

All nontrivial entanglement properties of such a tensor network state are induced by the projection, {\it i.e.}, the partial trace with $\rho_P$.
However, the average over random tensors can be carried out before taking the partial trace, since the latter is a linear operation.
This is the key insight that enables the computation of entanglement properties of random tensor networks.

\subsection{Calculation of the second R\'{e}nyi entropy}\label{sec:secondRenyi}

We will now apply this technique to study the second R\'{e}nyi entropies of the random tensor network state $\rho$ defined in Eq.~(\ref{peps3}).
For a boundary region $A$ with reduced density matrix $\rho_A$, the second R\'{e}nyi entropy $S_2(A)$ is given by $e^{-S_2(A)}=\tr \rho_A^2 / (\tr \rho)^2$.\footnote{In the quantum information theory literature, the R\'{e}nyi entropy is usually defined with logarithm in base $2$, $S_n(A)=1/(1-n) \log_2\frac{\tr \rho_A^n}{(\tr \rho_A)^n}$. Here we use base  $e$ to keep the notation consistent with the condensed matter and high energy literature.} It is helpful to write this expression in a different form by using the ``swap trick'',
\begin{eqnarray}
e^{-S_2(A)}=\frac{\tr \left[\left(\rho\otimes \rho\right) \mathcal F_A \right]}{\tr \left[ \rho \otimes \rho \right]}.\label{secondRenyi}
\end{eqnarray}
Here we have defined a direct product $\rho \otimes \rho$ of two copies of the original system, and the operator $\mathcal F_A$ is defined on this two-copy system and swaps the states of the two copies in the region $A$.
To be more precise, its action on a basis state 
of the two-copy Hilbert space is given by $\mathcal F_A (\ket{n_A}_1\otimes \ket{m_{\bar{A}}}_1\otimes \ket{n_A'}_2\otimes \ket{m'_{\bar{A}}}_2) =\ket{n_A'}_1\otimes \ket{m_{\bar{A}}}_1\otimes \ket{n_A}_2\otimes \ket{m'_{\bar{A}}}_2$, where $\bar{A}$ denotes the complement of $A$ on the boundary.

\begin{figure}
\begin{center}
\includegraphics[width=0.75\textwidth]{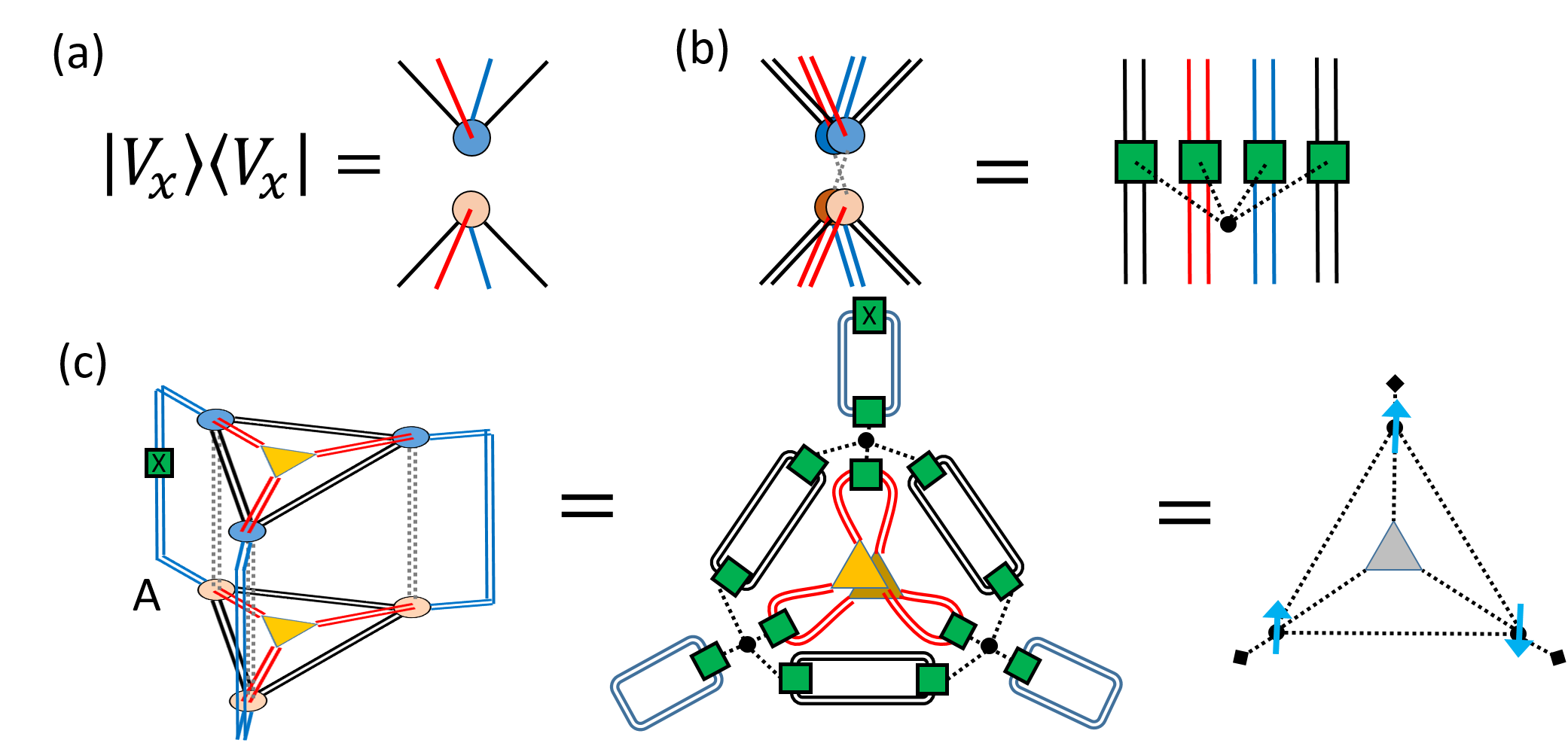}
\caption{(a) Graphic representation of the single site density operator $\ket{V_x}\!\!\bra{V_x}$ for a vertex in the tensor network shown in Fig.~\ref{fig:tensor}. (b) The average over the state $\ket{V_x}\!\!\bra{V_x}\otimes \ket{V_x}\!\!\bra{V_x}$ in the Hilbert space (see Eq.~(\ref{Haaraverage})). On the right side of the equality, the dashed line connected to the black dot stands for a sum over an Ising variable $s_x=\pm 1$. When $s_x=1$ ($s_x=-1$), each green rectangle represents an operator $I_x$ ($\mathcal F_x$), respectively. (c) The state average of $Z_1$ in Eq.~(\ref{Z1average}) for the simple tensor network shown in Fig.~\ref{fig:tensor}. We consider a region $A$ consisting of a single site, and the green rectangle with X represents the swap operator $\mathcal F_A$. After contracting the doubled line loops one obtains the partition function of an Ising model, with the blue arrows representing the Ising variables. The dashed lines in the right of last equality represent three different terms in the Ising model contributed by the links, the bulk state (middle triangle) and the choice of boundary region $A$.} \label{fig:average}
\end{center}
\end{figure}

We are now interested in the typical values of the entropy.
Denote the numerator and denominator resp.\ of Eq.~(\ref{secondRenyi}) by
\begin{align}
  Z_1 &= \tr \left[\left(\rho\otimes \rho\right) \mathcal F_A \right], \label{eq:Z1 second Renyi}\\
  Z_0 &= \tr \left[\rho \otimes \rho\right]. \label{eq:Z0 second Renyi}
\end{align}
These are both functions of the random states $\ket{V_x}$ at each vertex.
We would like to average over all states in the single-vertex Hilbert space.
The variables $Z_1$ and $Z_0$ are easier to average than the entropy, since they are quadratic functions of the single-site density matrix $\ket{V_x}\!\!\bra{V_x}$. The entropy average can be expanded in powers of the fluctuations $\delta Z_1=Z_1-\overline{Z_1}$ and $\delta Z_0=Z_0-\overline{Z_0}$:
\begin{eqnarray}
\overline{S_2(A)}=-\overline{\log\frac{\overline{Z_1}+\delta Z_1}{\overline{Z_0}+\delta Z_0}}
=-\log \frac{\overline{Z_1}}{\overline{Z_0}}+\sum_{n=1}^{\infty}\frac{(-1)^{n-1}}n\left(\frac{\overline{\delta Z_0^n}}{\overline{Z_0}^n}-\frac{\overline{\delta Z_1^n}}{\overline{Z_1}^n}\right)\label{S2approximation}.
\end{eqnarray}
We will later show in Section~\ref{sec:finiteD} that for large enough bond dimensions $D_{xy}$ the fluctuations 
are suppressed. Thus we can approximate the entropy with high probability by the separate averages of $Z_1$ and $Z_0$:
\begin{equation}
\label{eq:secondRenyi approx}
    S_2(A) \simeq -\log \frac {\overline{Z_1}} {\overline{Z_0}}.
\end{equation}
Throughout this article we use $\simeq$ for asymptotic equality as the bond dimensions go to infinity.
In the following we will compute $\overline{Z_1}$ and $\overline{Z_0}$ separately and use \eqref{eq:secondRenyi approx} to determine the typical entropy.
To compute $\overline{Z_1}$, we insert Eq.~\eqref{peps3} into Eq.~\eqref{eq:Z1 second Renyi} and obtain
\begin{eqnarray}
\overline{Z_1}=\tr \left[\left(\rho_P \otimes \rho_P\right) \mathcal F_A \prod_x\overline{\ket{V_x}\!\!\bra{V_x}\otimes \ket{V_x}\!\!\bra{V_x}}\right]\label{Z1average}.
\end{eqnarray}
In this expression we have combined the partial trace over bulk indices in the definition of the boundary state $\rho$ and the trace over the boundary indices in Eq.~\eqref{eq:Z1 second Renyi} into a single trace over all indices.
In the expression it is now transparent that the average over states, one at each vertex, can be carried out independently before couplings between different sites are introduced by the projection.
The average over states can be done by taking an arbitrary reference state $\ket{0_x}$ and setting $\ket{V_x}=U_x\ket{0_x}$.
Then the average is equivalent to an integration over $U_x\in SU\left(D_x\right)$ with respect to the Haar measure.
The result of this integration can be obtained using Schur's lemma (see, e.g., Ref.~\cite{harrow2013}):
\begin{equation}
    \overline{\ket{V_x}\!\!\bra{V_x}\otimes \ket{V_x}\!\!\bra{V_x}}
  = \int dU_x \big( U_x \otimes U_x \big) \left( \ket{0_x}\!\!\bra{0_x} \otimes \ket{0_x}\!\!\bra{0_x} \right) \big( U_x^\dagger \otimes U_x^\dagger \big)
  = \frac {I_x + \mathcal F_x} {D_x^2 + D_x} \label{Haaraverage}.
\end{equation}
Here, $I_x$ denotes the identity operator and $\mathcal F_x$ the swap operator defined in the same way as $\mathcal F_A$ described above, swapping the two copies of Hilbert space of the vertex $x$ (which means all legs connecting to $x$).
The Hilbert space dimension is $D_x=\prod_{y \text{~n.n.~} x}D_{xy}$, the product of the dimensions corresponding to all legs adjacent to $x$, including the boundary dangling legs.
It is helpful to represent Eq.~\eqref{Haaraverage} graphically as in Fig.~\ref{fig:average}\,(a) and (b).

Carrying out the average over states at each vertex $x$, $\overline{Z_1}$ then consists of $2^N$ terms if there are $N$ vertices, with an identity operator or swap operator at each vertex. We can then introduce an Ising spin variable $s_x=\pm 1$, and use $s_x=1$ ($s_x=-1$) to denote the choice of $I_x$ and $\mathcal F_x$, respectively. In this representation, $\overline{Z_1}$ becomes a partition function of the spins $\left\{s_x\right\}$:
\begin{equation*}
    \overline{Z_1} = \sum_{\left\{s_x\right\}}e^{-\mathcal{A}\left[\left\{s_x\right\}\right]},
\end{equation*}
where
\begin{equation*}
    e^{-\mathcal{A}\left[\left\{s_x\right\}\right]} \equiv \frac1{\prod_{x}\left(D_x^2+D_x\right)}\tr \left[\left(\rho_P\otimes \rho_P\right) \mathcal F_A \prod_{x\text{~with~}s_x=-1} \mathcal F_x\right].
\end{equation*}
For each value of the Ising variables $\left\{s_x\right\}$, the operator being traced is now completely factorized into a product of terms, since $\mathcal F_x$ acts on each leg of the tensor independently. This fact is illustrated in Fig.~\ref{fig:average}\,(c). The trace of the swap operators with $\rho_P\otimes \rho_P$ is simply $\exp\left[-S_2\left(\left\{s_x=-1\right\};\rho_P\right)\right]$ with $S_2\left(\left\{s_x=-1\right\};\rho_P\right)$ the second R\'{e}nyi entropy of $\rho_P$ in the Ising spin-down domain defined by $s_x=-1$. The trace on boundary dangling legs gives a factor that is either $D_{x\partial}^2$ or $D_{x\partial}$, depending on the Ising variables $s_x$ and whether $x$ is in $A$. To be more precise, we can define a boundary field
\begin{eqnarray}
h_x=\left\{\begin{array}{cc}+1,&x\in \bar{A}\\-1,&x\in A\end{array}\right.
\end{eqnarray}
Then the trace at a boundary leg $x\partial$ gives $D_{x\partial}^{\frac12\left(3+h_xs_x\right)}$.
Taking a product of these two kinds of terms in the trace, we obtain the Ising action
\begin{equation*}
\mathcal{A}\left[\left\{s_x\right\}\right] = S_2\left(\left\{s_x=-1\right\};\rho_P\right)-\sum_{x\in\partial}\frac12\log D_{x\partial}\left(3+h_xs_x\right)+\sum_x\log\left(D_x^2+D_x\right).
\end{equation*}
The form of the action can be further simplified by recalling that $\rho_P$ has the direct product form in Eq.~(\ref{rhoP}). Therefore the second R\'{e}nyi entropy factorizes into that of the bulk state $\rho_b$ and that of the maximally entangled states at each internal line $xy$. The latter is a standard Ising interaction term, since the entropy of either site is $\log D_{xy}$ while the entropy of the two sites together vanishes. Therefore
\begin{equation}
\label{Isingaction2}
\begin{split}
\mathcal{A}\left[\left\{s_x\right\}\right] = &-\sum_{\left\langle xy\right\rangle}\frac12\log D_{xy}\left(s_xs_y-1\right)-\sum_{x\in\partial}\frac12\log D_{x\partial}\left(h_xs_x-1\right) \\
&+S_2\left(\left\{s_x=-1\right\};\rho_b\right)+\text{const.}
\end{split}
\end{equation}
Here we have omitted the details of the constant term since it plays no role in later discussions.
Eq.~(\ref{Isingaction2}) is the foundation of our later discussion.
The same derivation applies to the average of the denominator $Z_0 = \tr \left[ \rho \otimes \rho \right]$ in Eq.~(\ref{secondRenyi}), which leads to the same Ising partition function with a different boundary condition $h_x=1$ for all boundary sites, since there is no swap operator $\mathcal F_A$ applied.
One can define $F_1=-\log\overline{Z_1}, ~F_0=-\log \overline{Z_0}$, such that $F_1$ and $F_0$ are the free energy of the Ising model with different boundary conditions.%
\footnote{The standard definition of free energy should be $-\beta^{-1}\log \overline{Z_1}$ but it is more convenient for us to define it without the temperature prefactor.}
Then Eq.~\eqref{eq:secondRenyi approx} reads
\[ S_2(A) \simeq F_1-F_0. \]
That is, the typical second R\'{e}nyi entropy is given by the difference of the two free energies, {\it i.e.}, the ``energy cost'' induced by flipping the boundary pinning field to down ($-1$) in region $A$, while keeping the remainder of the system with a pinning field up ($+1$).

In summary, what we have achieved is that the second R\'{e}nyi entropy is related to the partition function of a classical Ising model defined on the same graph as the tensor network. Besides the standard two-spin interaction term, the Ising model also has an additional term in its energy contributed by the second R\'{e}nyi entropy of the bulk state $\rho_b$, and the Ising spins at the boundary vertices are coupled to a boundary ``pinning field'' $h_x$ determined by the boundary region $A$. If the bulk contribution from $\rho_b$ is small (which means major part of quantum entanglement of the boundary states is contributed by the tensor network itself), one can see that the parameters $\log D_{xy}$ and $\log D_{x\partial}$ determine the effective temperature of the Ising model. For simplicity, in the following we assume $D_{xy}=D_{x\partial}=D$ for all internal legs and boundary dangling legs. In this case we can take $\beta=\frac 12\log D$ as the inverse temperature of the classical Ising model.

\section{Ryu-Takayanagi formula}\label{sec:RTformula}

Once the mapping to the classical Ising model is established, it is easy to see how the Ryu-Takayanagi formula emerges.
In the large $D$ limit, the Ising model is in the low-temperature long-range ordered phase (as long as the bulk has spatial dimension $\geq 2$), so that the Ising action can be estimated by the lowest energy configuration.
The boundary pinning field $h_x$ leads to the existence of an Ising domain wall bounding the boundary region $A$, and in the absence of a bulk contribution the minimal energy condition of the domain wall is exactly the RT formula.
In this section we will analyze this emergence of the Ryu-Takayanagi formula and corrections due to bulk entanglement in more detail. 

\subsection{Ryu-Takayanagi formula for a bulk direct-product state}

\begin{figure}[!ht]
\centering
\includegraphics[width=0.95\textwidth]{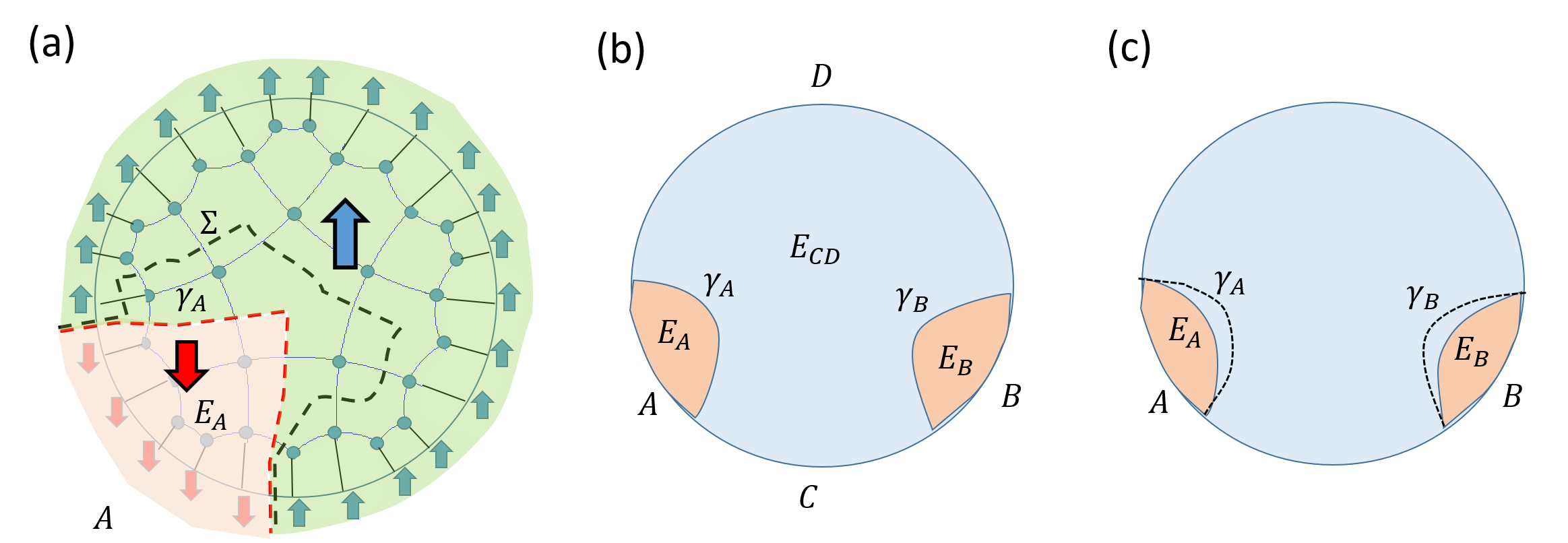}
\caption{(a) An example of Ising spin configuration with boundary fields down ($h_x=-1$) in $A$ region and up ($h_x=+1$) elsewhere. $\gamma_A$ is the boundary of minimal energy spin-down domain configuration. $\Sigma$ (black dashed line) is an example of other domain wall configurations with higher energy. The spin-down domain $E_A$ is called the entanglement wedge of $A$. (b) The minimal surfaces bounding two far-away regions $A$ and $B$, which are also the boundary of the entanglement wedge of the completement region $CD$. (c) The effect of bulk entanglement in the same configuration as panel (b). The entanglement wedges are deformed.} \label{fig:Isingdomain}
\end{figure}

We first consider the simplest situation with the bulk state a pure direct-product state $\rho_b=\bigotimes_x\ket{\phi_x}\!\!\bra{\phi_x}$. In this case one can contract the bulk state at each site with the tensor of that site, which leads to a new tensor with one fewer legs. Since each tensor is a random tensor, the new tensor obtained from contraction with the bulk state is also a random tensor. Therefore the holographic mapping with a pure direct-product state in the bulk is equivalent to a purely in-plane random tensor network, similar to a MERA, or a ``holographic state'' defined in Ref.~\cite{pastawski2015}. The second R\'{e}nyi entropy of such a tensor network state is given by the partition function of Ising model in Eq.~(\ref{Isingaction2}) without the $\rho_b$ term. Omitting the constant terms that appears in both $Z_0$ and $Z_1$, the Ising action can be written as
\begin{eqnarray}
\mathcal{A}\left[\left\{s_x\right\}\right]=-\frac12\log D\left[\sum_{\langle xy\rangle}\left(s_xs_y-1\right)+
\sum_{x\in\partial}\left(h_xs_x-1\right)\right].\label{Isingaction3}
\end{eqnarray}
In the large $D$ limit, the Ising model is in the low temperature limit, and the partition function is dominated by the lowest energy configuration. As illustrated in Fig.~\ref{fig:Isingdomain}\,(a), the ``energy'' of an Ising configuration is determined by the number of links crossed by the domain wall between spin-up and spin-down domains, with the boundary condition of the domain wall set by the boundary field $h_x$. For the calculation of denominator $Z_0$, $h_x=+1$ everywhere, so that the lowest energy configuration is obviously $s_x=+1$ for all $x$, with energy $F_0=0$. For $F_1$, the nontrivial boundary field $h_x=-1$ for $x\in A$ requires the existence of a spin-down domain. Each link $\langle xy\rangle$ with spins anti-parallel leads to an energy cost of $\log D$. Therefore the R\'{e}nyi entropy in large $D$ limit is
\begin{eqnarray}
S_2(A) = F_1 - F_0 \simeq \log D \min_{\Sigma\text{~bound~}A}\left|\Sigma\right|\equiv \log D\left|\gamma_A\right|.
\end{eqnarray}
The minimization is over surfaces $\Sigma$ such that $\Sigma\cup A$ form the boundary of a spin-down domain, and $|\Sigma|$ denotes the area of $\Sigma$, {\it i.e.}, the number of edges that cross the surface. Therefore the minimal area surface, denoted by $\gamma_A$, is the geodesic surface bounding $A$ region. Here we have assumed that the geodesic surface is unique. More generally, if there are $k$ degenerate minimal surfaces (as will be the case for a regular lattice in flat space), $F_1$ is modified by $-\log k$.

With this discussion, we have proved that Ryu-Takayanagi formula
applies to the second R\'{e}nyi entropy of a large dimensional random
tensor network, with the area of geodesic surface given by the graph
metric of the network. As will be discussed later in
Section~\ref{sec:nthRenyi}, the higher R\'{e}nyi entropies take the
same value in the large $D$ limit, and it can also be extended to the
von Neumann entropy, at least if the minimal geodesics are unique (see
Section~\ref{sec:finiteD}). However, the triumph that the second
  R\'{e}nyi entropy is equal to the area of the minimal surface in the
  graph metric is in fact a signature that the random tensor
  construction deviates from the holographic theory. The holographic
  calculation of the second R\'{e}nyi entropy amounts to evaluating 
  the Euclidean action of the two-fold replica geometry, which
  satisfies the Einstein equation everywhere in the bulk. Thus, in
  general, the second R\'{e}nyi entropy does not exactly correspond to the
  area of the minimal surface in the original geometry. Due to the
  back-reaction of the gravity theory, the $n$-fold replica geometry
  is in general different from the geometry constructed by simply
  gluing $n$ copies of the original geometry around the minimal
  surfaces, the discrepancy between which can be seen manifestly from
  the $n$-dependence of the holographic R\'enyi-$n$ entropy.
  We will see in Section~\ref{sec:nthRenyi} that our random tensor model can reproduce the correct R\'enyi entropies for a
  single boundary region if we replace the bond states $\ket{xy}$ by appropriate short-range entangled states with non-trivial entanglement
  spectrum. However, this does not resolve the problem for multiple
  boundary regions, for which we will have a more detailed discussion in
  Section~\ref{sec:nthRenyi}.

To compare with the RT formula defined on a continuous manifold, one can consider a triangulation of a given spatial manifold and define a random tensor network on the graph of the triangulation. (See~\cite[Appendix A]{bao2015cone} for further discussion of the construction of the triangulation graph.) Denoting by $l_g$ the length scale of the triangulation (the average distance between neighboring triangles), the area $\left|\gamma_A\right|$ in our formula is dimensionless and the area $\left|\gamma_A^c\right|$ defined on the continuous Riemann manifold is given by $\left|\gamma_A^c\right|=l_g^{d-1}\left|\gamma_A\right|$ (when the spatial dimension of bulk is $d$). Therefore $S(A)=l_g^{1-d}\log D\left|\gamma_A^c\right|$, and we see that $l_g^{1-d}\log D$ corresponds to the gravitational coupling constant $\frac1{4G_N}$.

Compared to previous results about the RT formula in tensor networks~\cite{pastawski2015,yang2015}, our proof of RT formula has the following advantages:
Firstly, our result does not require the boundary region $A$ to be a single connected region on the boundary.
Since the entropy in the large $D$ limit is always given by the Ising spin configuration with minimal energy, the result applies to multiple boundary regions.
Secondly, our result does not rely on any property of the graph structure, except for the uniqueness of the geodesic surface (if this is not satisfied then the entropy formula acquires corrections as discussed above; cf.~Section~\ref{sec:stabs}).
If we obtain a graph by triangulation of a manifold, our formula applies to manifolds with zero or positive curvature, even when the standard AdS/CFT correspondence does not apply.
In addition to these two points, we will also see in later discussions that our approach allows us to study corrections to the RT formula systematically. Notice that we are not limited to two-dimensional manifolds. One can consider a higher dimensional manifold and construct a graph approximating its geometry. It follows from our results that the entropy of a subregion of the boundary state is given by the size of the minimum cut on the graph, i.e., the area of the minimal surface in the bulk homologous to the boundary region.

\subsection{Ryu-Takayanagi formula with bulk state correction}\label{sec:RTwithbulk}

If we do not assume the bulk state to be a pure direct-product state, the bulk entropy term in Eq.~(\ref{Isingaction2}) is nonzero. If we still take the $D\rightarrow \infty$ limit, the Ising model free energy is still determined by the minimal energy spin configuration, which is now determined by a balance between the area law energy $\log D\left|\Sigma\right|$ for a domain wall $\Sigma$, and the energy cost from bulk entropy.
We can define the spin-down region in such a minimal energy configuration as $E_A$, which bounds the boundary region $A$, and corresponds to the region known as the entanglement wedge in the literature~\cite{hubeny2007,heemskerk2012}. The second R\'{e}nyi entropy is then given by
\begin{equation}
\label{eq:second renyi}
S_2(A) \simeq \log D\left|\gamma_A\right|+S_2\left(E_A;\rho_b\right).
\end{equation}
The bulk contribution has two effects. First it modifies the position of the minimal energy domain wall $|\gamma_A|$, and thus modified the area law (RT formula) term of the entropy. Second it gives an additional contribution to the entanglement entropy of the boundary region. This is similar to how bulk quantum fields contribute corrections to the RT formula in AdS/CFT~\cite{faulkner2013quantum}.

To understand the consequence of this bulk correction, we consider an example shown in Fig.~\ref{fig:Isingdomain}\,(b) and (c), where $A$ and $B$ are two distant disjoint regions on the boundary. If the bulk entanglement entropy vanishes, the RT formula applies and the entanglement wedges $E_A$ and $E_B$ are disjoint. Therefore we find that $S_2(A) + S_2(B) = S_2(AB)$ and so the ``mutual information'' between the two intervals $I_2(A:B) = S_2(A) + S_2(B) - S_2(AB)$ vanishes in the large $D$ limit.%
\footnote{The mutual information for R\'{e}nyi entropy is generally not an interesting quantity, but it is meaningful in our case since it approaches the von Neumann mutual information for large $D$.}
When the bulk state is entangled, if we assume the entanglement is not too strong, so that the entanglement wedges remain disjoint, the minimal energy domain walls $\gamma_A$ and $\gamma_B$ may change position, but remain disconnected. Therefore:
\begin{align*}
S_2(AB) &\simeq \log D \left(\left|\gamma_A\right|+\left|\gamma_B\right|\right)+S_2(E_A\cup E_B;\rho_b), \\
I_2(A:B) &\simeq S_2(E_A;\rho_b)+S_2(E_B;\rho_b)-S_2(E_A\cup E_B;\rho_b)= I_2(E_A:E_B;\rho_b).
\end{align*}
From this equation, we see that even if a small bulk entanglement entropy may only lead to a minor correction to the minimal surface location, it is the only source of mutual information between two far-away regions in the large $D$ limit. (If we consider a large but finite $D$, and include spin fluctuations of the Ising model, we obtain another source of mutual information between far-away regions, which vanishes exponentially with $\log D$.) The suppression of mutual information between two far-away regions implies that the correlation functions between boundary regions $A$ and $B$ are suppressed, even if each region has a large entanglement entropy in the large $D$ limit. In the particular case when the bulk geometry is a hyperbolic space, the suppression of two-point correlations discussed here translates into the scaling dimension gap of boundary operators, which is known to be a required property for CFTs with gravity duals~\cite{heemskerk2009,elshowk2012,benjamin2015}. A more quantitative analysis of the behavior of two-point correlation functions and scaling dimension gap will be postponed to Section~\ref{sec:correlationspec}.

\subsection{Phase transition of the effective bulk geometry induced by bulk entanglement}

We have shown that a bulk state with nonzero entanglement entropy
gives rise to corrections to the Ryu-Takayanagi formula.  In the
discussion in Section~\ref{sec:RTwithbulk}, we assumed that the bulk
entanglement was small enough that the topology of the minimal
surfaces remained the same as those in the absence of bulk
entanglement.  Alternatively, one can also consider the opposite
situation when the bulk entanglement entropy is not a small
perturbation compared to the area law term
$\log D \, \lvert \gamma_A \rvert$, in which case the behavior of the
minimal surfaces may change qualitatively.  In this subsection, we
will study a simple example of this phenomenon, with the bulk state
being a random pure state in the Hilbert space of a subregion in the
bulk.  As is well-known, a random pure state is nearly maximally
entangled~\cite{P93}, which we will use as a toy model of a thermal
state ({\it i.e.}, of a pure state that satisfies the eigenstate
thermalization hypothesis~\cite{deutsch1991,srednicki1994}).  The
amount of bulk entanglement can be controlled by the dimension of the
Hilbert space $D_b$ of each site.  We will show that the topologies of
minimal surfaces experience phase transitions upon increasing $D_b$
which qualitatively reproduces the transition of the bulk geometry
in the Hawking-Page phase transition~\cite{hawking1983,witten1998b}. 
  To be more precise, the entropy of the boundary region receives two contributions: the area of the minimal surfaces in the
  AdS background and the bulk matter field correction. However, above
  a critical value of $D_b$, the minimal
  surface tends to avoid the highly entangled region in the bulk, such
  that there is a region which no minimal surface ever penetrates
  into, and the minimal surface jumps discontinuously from one side of
  the region to the other side as the boundary region size increases
  to half of the system.
  This is qualitatively similar to how a black
hole horizon emerges from bulk entanglement. (A black hole cannot be
identified conclusively in the absence of causal structure, however,
so our conclusions in this section are necessarily tentative.)

\begin{figure}[!ht]
\centering
\includegraphics[width=0.95\textwidth]{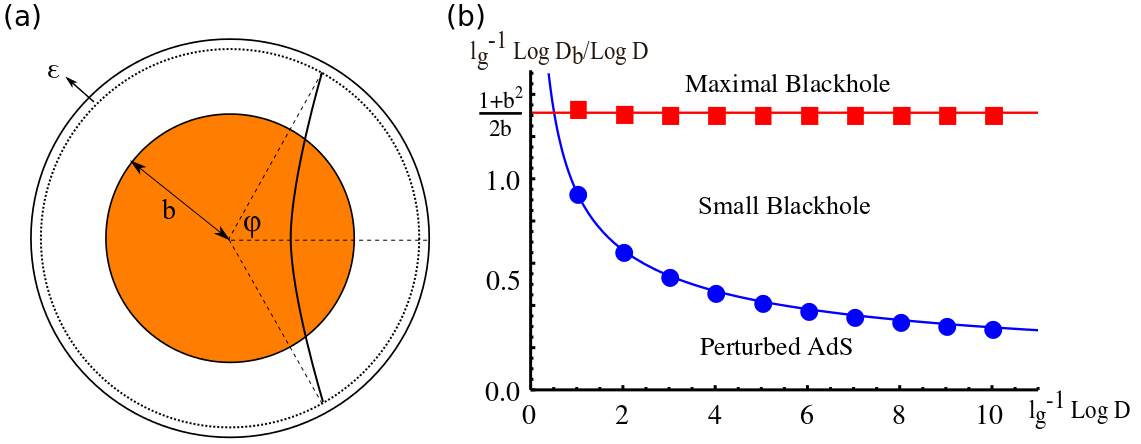}
\caption{(a) Illustration of the setup.
The orange disk-shaped bulk region of radius $b$ is in a random pure state.
We study the second R\'{e}nyi entropy of a boundary region $\theta\in[-\varphi,\varphi]$ at radius $r=1-\epsilon$.
(b) The phase diagram of the boundary state as parametrized by the bond dimensions $D$ and $D_b$, corresponding to in-plane and bulk degrees of freedom, respectively.
The blue line, obtained numerically, describes the phase boundary that separates the perturbed AdS phase and the small black hole phase.
The red line distinguishes the small black hole phase and the maximal black hole phase.
The three phases are discussed in more detail in the main text.}
\label{SBHPro}
\end{figure}

We consider a tensor network which is defined on a uniform triangulation of a hyperbolic disk.
Each vertex is connected to a bulk leg with dimension $D_b$ in addition to internal legs between different vertices.
Then we take a disk-shaped region, as shown in Fig.~\ref{SBHPro}\,(a).
We define the bulk state to be a random state in the disk region, and a direct-product state outside:
\begin{equation*}
  \ket{\Psi}_{\text{bulk}} = \left( \bigotimes_{\lvert \vec{x} \rvert>b} \ket{\psi_{\vec{x}}} \right) \otimes \ket{\psi_{\lvert \vec{x} \rvert < b}}.
\end{equation*}
The second R\'{e}nyi entropy of a boundary region is determined by the Ising model partition function with the action (\ref{Isingaction2}).\footnote{{For readers more comfortable with the graph theoretic description, here is a sketch in that language of the entropy calculation in the presence of a bulk random state. Because any vertex corresponds to projection to a random state, the insertion of a random bulk state amounts to connecting the bulk dangling legs to a single new vertex. Therefore, the study of entropies will be equivalent to the study of minimum cuts in the modified graph.} }

The bulk contribution $S_2\left(\left\{ s_x=-1\right\};\rho_b\right)$ for a random state with large dimension only depends on the volume of the spin-down domain in the disk region, since all sites a play symmetric role. After an average over random states, the entropy of a bulk region with $N$ sites is given by~\cite{LL93}
\begin{equation*}
S_2\left(N\right)= \log\left(\frac{D_b^{N_T}+1}{D_b^{N} + D_b^{N_T-N}}\right),
\end{equation*}
in which $N_T$ is the total number of sites in the disk region.
Therefore the Ising action contains two terms, an area law term and the bulk term which is a function of the volume of spin-down domain.
For simplicity, we can consider a fine-grained triangulation and approximate the area and volume by that in the continuum limit.
If we denote the average distance between two neighboring vertices as $l_g$, as in previous subsections, we obtain
\begin{equation*}
\mathcal{A}\left[M_\downarrow\right]=\log D \cdot l_g^{-1}\left|\partial M_\downarrow\right|+
\log\left(\frac{D_b^{V_T/l_g^2}+1}{D_b^{|M_\downarrow|/l_g^2} + D_b^{(V_T-|M_\downarrow|)/l_g^2}}\right).
\end{equation*}
Here $M_\downarrow$ is a spin-down region bounding a boundary region $A$, and $\partial M_\downarrow$ is the boundary of this region in the bulk (which does not include $A$). $V_T=N_Tl_g^2$ is the total volume of the disk region in the bulk.

Consider the Poincar\'{e} disk model of hyperbolic space, with the metric $ds^2 =4(dr^2+r^2d\theta^2)/(1-r^2)^2 $.
The boundary is placed at $r=1-\epsilon$ with $\epsilon>0$ a small cutoff parameter.
The disk region is defined by $r\leq b$.
Choose a boundary region $\theta\in[-\varphi,\varphi]$, with $\varphi\leq \pi/2$ so that the boundary region is smaller than half the system size.
(Boundary regions that exceed half the system size have the same entropy as their complement, since the whole system is in a pure state.)
If we assume the minimal surface $\partial M_\downarrow$ to be a curve described by $r=r(\theta)$ ({\it i.e.}, for each $\theta$ there is only one $r$ value), the volume and area of this curve can be written explicitly as
\begin{align}
\label{SBHQ}
 S_2(\varphi) &= \min_{r(\theta)}\left\{  \int_{-\varphi}^{\varphi} d\theta \frac{2l_g^{-1}\log D}{1-r^2(\theta)}\sqrt{\left(r^\prime(\theta)\right)^2+r^2(\theta)}
 + \log\frac{D_b^{V_T/l_g^2}+1}{D_b^{V_{r(\theta)}/l_g^2} + D_b^{\left(V_T-V_{r(\theta)}\right)/l_g^2}}\right\}\!\!, \\
\nonumber
V_{r(\theta)} &= \int_{-\varphi}^{\varphi} d\theta \int_{\min\{b,r(\theta)\}}^{b} dr\frac{4r}{(1-r^2)^2}, ~~~~~ V_T =  \int_{0}^{2\pi} d\theta \int_{0}^{b} dr\frac{4r}{(1-r^2)^2}= \frac{4\pi b^2}{1-b^2}.
\end{align}

\begin{figure}[!ht]
\centering
\includegraphics[width=0.95\textwidth]{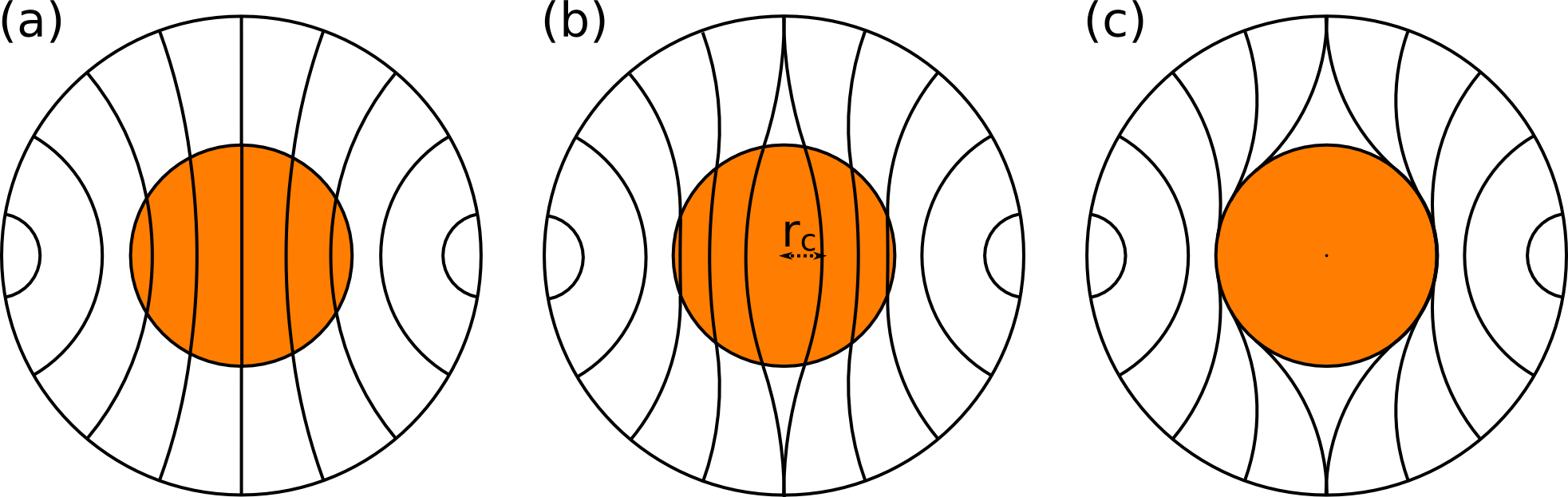}
\caption{Configuration of the minimal surfaces calculated
  numerically in the bulk for different boundary regions in the
  three phases. The random pure state is supported at the orange
  region. The parameters are set to $l_g^{-1}\log D=10$,
  $b=\tanh(1/2)$. Depending on the value of $l_g^{-2}\log D_b$,
  the phases of the system are given by (a) $l_g^{-2}\log D_b =1$,
  perturbed AdS phase; (b) $l_g^{-2}\log D_b=5$, small black hole
  phase; (c) $l_g^{-2}\log D_b= 15$, maximal black hole
  phase.} \label{SBHConf}
\end{figure}

For fixed $l_g^{-1} \log D$, when we gradually increase $l_g^{-2} \log D_b$, there are three distinct phases:
the perturbed AdS phase, the small black hole phase, and the maximal black hole phase.
The phase diagram can be obtained numerically, as shown in Fig.~\ref{SBHPro}\,(b).
In the calculation, we fix $b = \tanh(1/2)$, which means that the radius of the disk in proper distance is $1$ ({\it i.e.}, the AdS radius).
In the perturbed AdS phase, although the minimal surfaces are deformed due to the existence of the bulk random state, there is no topological change in the behavior of minimal surfaces.
As the size of the boundary region increases, the minimal surface swipes through the whole bulk
continuously (Fig.~\ref{SBHConf}\,(a)).
In the small black hole phase, the minimal surface experiences a discontinuous jump as the boundary region size increases.
There exists a region with radius $0<r_c<b$ that cannot be accessed by the minimal surfaces of any boundary regions (Fig.~\ref{SBHConf}\,(b)).
Qualitatively, the minimal surfaces therefore behave like those in a black hole geometry, which always stay outside the black hole horizon.
As $\log D_b$ increases, $r_c$ increases until it fills the whole disk ($r_c=b$).
Further increase of $\log D_b$ does not change the entanglement property of the boundary anymore, since the entropy in the bulk disk region has saturated at its maximum.
This is the maximal black hole phase (Fig.~\ref{SBHConf}\,(c)).

More quantitatively, the two phase boundaries in Fig.~\ref{SBHPro}\,(b) are fitted by $l_g^{-2}\log D_b = 0.937
\sqrt{l_g^{-1}\log D}$ (blue line) and $l_g^{-1} \log D_b/\log D =(1+b^2)/2b$ (red line), respectively.
The square root behavior of the blue line can be understood by taking the maximal boundary region of half the system size $\varphi=\frac{\pi}2$.
At the critical $l_g^{-2}\log D_b$, the diameter of the Poincar\'{e} disk goes from the minimal surface bounding the half system to a local maximum.
For more detailed discussion, see Appendix~\ref{SBHTran}.
The second transition at the red line is roughly where the entanglement entropy of the bulk region reaches its maximum. However, more work is required to obtain the correct coefficient $(1+b^2)/2b$, as we show in Appendix~\ref{SBHTran}.
In Fig.~\ref{SBHEntropy}\,(a), we present the evolution of the black hole size $r_c/b$ when $l_g^{-2}\log D_b$ increases and $l_g^{-1}\log D =10$ is fixed.

Fig.~\ref{SBHEntropy}\,(b) provides another diagnostic to differentiate the geometry with and without the black hole.
The entanglement entropy $S_2(\varphi)$ is plotted as a function of the boundary region size.
In the perturbed AdS phase (blue curve), $S_2(\varphi)$ is a smooth function of $\varphi$, just like in the pure AdS space.
In the small black hole phase (black curve) and the maximal black hole phase (red curve), there is a cusp in the function $S_2(\varphi)$ at $\varphi=\frac{\pi}2$, as a consequence of the discontinuity of the minimal surface.
For $\varphi \leq \frac{\pi}2$, $S_2(\varphi)$ shows a crossover from the AdS space behavior (which corresponds to the entanglement entropy of a CFT ground state) to a volume law.
Such behavior of $S_2(\varphi)$ is qualitatively consistent with the behavior of a thermal state (more precisely a pure state with finite energy density) on the boundary.


\begin{figure}[!ht]
\centering
\includegraphics[width=0.95\textwidth]{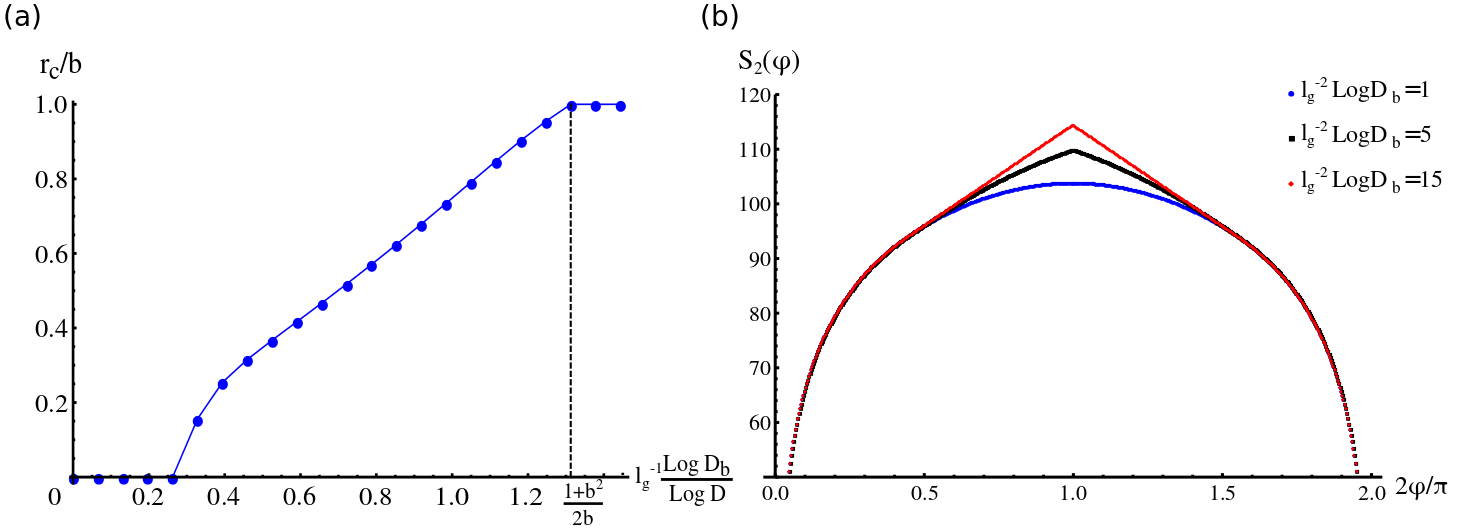}
\caption{(a) Evolution of the black hole size $r_c$ with respect to
  $l_g^{-2}\log D_b$, where $l_g^{-1}\log D$ is fixed to be
  10. (b)Entropy profile of the boundary system $r=1-\epsilon$,
  $\theta\in[-\varphi,\varphi]$ with respect to the different
  boundary region size $\varphi\leq \pi$. The blue data points,
  black data points and the red data points correspond to the
  boundary entropy profile in the perturbed AdS space, the small
  black hole phase and the maximal black hole case, respectively.
  The parameters are set as the same as the three phases in
  Fig.~\ref{SBHConf}.} \label{SBHEntropy}
\end{figure}

In summary, we see that a random state in the bulk region is mapped by the random tensor network to qualitatively different boundary states depending on the entropy density of the bulk.
This is a toy model of the transition between a thermal gas state in AdS space and a black hole.
In a more realistic model of the bulk thermal gas, the thermal entropy is mainly at the IR region (around the center of the Poincar\'{e} disk), but there is no hard cutoff.
Therefore there is no sharp transition between small black hole phase and maximal black hole phase.
The size of black hole will keep increase as a function of temperature.
In contrast, the lower phase transition between perturbed AdS phase and the small black hole phase remains a generic feature, since the minimal surface will eventually skip some region in the bulk when the volume law entanglement entropy of the bulk states is sufficiently high.
From this simple example we see how the bulk geometry defined by a random tensor network has nontrivial response to the variation of the bulk quantum state.
Finding a more systematic and quantitative relation between the bulk geometry and bulk entanglement properties will be postponed to future works.

\medskip

At last, we comment on the case of two-sided black holes.
As is well-known, an eternal black-hole in AdS space is the holographic dual of a thermofield double state~\cite{maldacena2003}, which is an entangled state between two copies of CFTs, such that the reduced density matrix of each copy is thermal.
As a toy model of the eternal black hole 
we consider a mixed bulk state with density matrix
\begin{equation*}
  \rho_b = \left(\bigotimes_{\lvert x\rvert>b} \rho_x^{\text{pure}}\right)\otimes \left(\bigotimes_{\lvert x \rvert<b} \rho_x^{\text{mix}} \right).
\end{equation*}
Here $\rho_x^{\text{pure}}$ is a pure state density matrix while $\rho_x^{\text{mix}}$ is a mixed state with finite entropy.
This density matrix described a bulk state in which all qudits in the disk region $|x|<b$ are entangled with some thermal bath.
The behavior of the geometry can be tuned by the entanglement entropy of $\rho_x^{\text{mix}}$ for each site, which plays a similar role as $\log D_b$ in the single-sided black hole case.
The analyis of minimal surfaces for a boundary region in this state can be done exactly in parallel with the single-sided case. 
Therefore, instead of repeating the similar analysis, we only comment on two major differences between the single-sided and two-sided case:
\begin{enumerate}
\item Because the bulk state is not a pure state, the entropy profile of the boundary system with respect to the different boundary region size is not symmetric at half the system size.
However, there is still a phase transition as a function of entropy density of the bulk, above which a cusp appears in the entropy profile.
This phase transition corresponds to the transition between thermal AdS geometry and AdS black hole geometry~\cite{witten1998b}. 
\item Similar to the single-sided case, there is a second phase transition where further increase of bulk entropy density does not change the boundary entanglement feature any more.
The transition point for two-sided case occurs at a slightly different value $l_g^{-2}S(\rho_x^{\text{mix}})=\frac{1}{b} l_g^{-1}\log D$.
When the bulk entropy exceeds this value, the boundary state is a mixed state with entropy $l_g^{-1}\log D\frac{4\pi b}{1-b^2}$, which is given by the boundary area of the disk region in the bulk.
The boundary of the disk plays the role of the black hole horizon.
\end{enumerate}

While the behavior observed here is consistent with black hole formation, it is important to stress that the conclusion is actually ambiguous. Geodesics can be excluded from regions of space even in the absence of a black hole.\footnote{We thank Aron Wall for bringing this point to our attention.}
The presence of a black hole is ultimately a feature of the causal structure, so resolving the ambiguity would require introducing time into our model.

\section{Random tensor networks as bidirectional holographic codes}
\label{sec:bhc}

In the previous section we discussed the entanglement properties of the boundary quantum state obtained from random tensor networks. In this section we will investigate the properties of random tensor networks interpreted as holographic mappings (or holographic codes).

In Ref.~\cite{yang2015}, the concept of a bidirectional holographic code (BHC) was introduced, which is a holographic mapping with two different kinds of isometry properties.
A BHC is a tensor network with boundary legs and bulk legs.
We denote the number of boundary legs as $L$ and the number of bulk legs ({\it i.e.}, the number of bulk vertices) as $V$, and denote the dimension of each boundary leg as $D$ and that of each bulk leg as $D_b$.
The first isometry is defined from the boundary Hilbert space with dimension $D^L$ to the bulk Hilbert space with dimension $D_b^V$.
The physical Hilbert space is identified with the image of this isometry from the boundary to the bulk, so that the full bulk Hilbert space is redundant in the sense that it contains many non-physical states.
The condition identifying these physical states can be formulated as a gauge symmetry.
The second isometry is defined from a subspace of the bulk Hilbert space to the boundary.
The physical interpretation of this subspace is as the low energy subspace of the bulk theory.
The bulk theory is intrinsically nonlocal in the space of all physical states, but locality emerges in the low energy subspace.
More precisely, the degrees of freedom at different locations of the low energy subspace are all independent, and a local operator acting in the low energy subspace can be recovered from certain boundary regions, satisfying the so-called ``error-correction property''~\cite{almheiri2014,pastawski2015}.
For this reason, the low energy subspace is also referred to as the code subspace.

In this section, we will investigate the properties of random tensor networks and show that they satisfy the BHC conditions in the large $D$ limit and moreover have properties that are even better than the BHC constructed using pluperfect tensors in Ref.~\cite{yang2015}.

\subsection{Code subspace}

We start from the holographic mapping in the low energy subspace, or ``code subspace'' in the language of quantum error correction~\cite{almheiri2014}.
Physically, the code subspace is a subspace of the Hilbert space which corresponds to small fluctuations around a classical geometry in the bulk.
More precisely, the criterion of ``small fluctuations'' states that these states are described well by a bulk quantum field theory with the given geometrical background.
In other words, in the code subspace the bulk fields (operators) at different spatial locations are independent and the Hilbert space seems to factorize with respect to the bulk position. The fact that one cannot take the code subspace to be the entire Hilbert space, {\it i.e.} that locality in the bulk fails if we consider the entire Hilbert space, is the essential feature of a theory of quantum gravity (defined as the holographic dual of a boundary theory), as compared to an ordinary quantum field theory in the bulk.

In general, the choice of code subspace is not unique. However, the random tensor network approach allows for a simple and explicit choice. We define the code subspace to be the tensor product of lower-dimensional subspaces at each vertex of the graph: $\mathbb{H}_{\rm code}=\bigotimes_{x}\mathbb{H}_{x}\left(D_b\right)$. Here, $\mathbb{H}_{x}\left(D_b\right)$ is a $D_b$-dimensional space at site $x$ in the bulk. The holographic mapping restricted to this subspace is simply a tensor network with a smaller bond dimension $D_b$ for each bulk leg. In the following, we investigate the condition for the bulk-to-boundary map to be an isometry, which thus determines the value of $D_b$ that makes such a subspace an eligible code subspace.

When we view the tensor network as a linear map $M$ from the bulk to the boundary, the isometry condition means $M^\dagger M = I$ is the identity operator.
To apply the results we obtained for the second R\'{e}nyi entropy, it is more convenient to view the tensor network as a pure state. Choose an orthonormal basis $\{\ket\alpha\}$ of the bulk and a basis $\{\ket a\}$ for the boundary.
The linear map $M$ with matrix element $M_{\alpha a}=\braket{\alpha | M | a}$ can then be identified with the pure quantum state
\begin{eqnarray}
    \ket{\Psi_M}=D_b^{-V/2}\sum_{\alpha,a}M_{\alpha a}\ket{\alpha} \otimes \ket{a}\label{PsiM}.
\end{eqnarray}
In terms of the state, the requirement that $M^\dagger M = I$ is equivalent to the statement that the bulk reduced density matrix $\rho_b = \tr _{\partial}\left(\ket{\Psi_M}\!\!\bra{\Psi_M}\right)=D_b^{-V} I$ is maximally mixed.
Therefore, the isometry condition can be verified by an entropy calculation.

For that purpose we calculate the second R\'{e}nyi entropy of the whole bulk. In the large $D$ limit, this is mapped to an Ising model partition function in the same way as in the RT formula discussion, except that there is now a pinning field everywhere in the bulk, in addition to the boundary:
\begin{eqnarray}
\mathcal{A}\left[\left\{s_x\right\}\right]=-\frac12\log D\left[\sum_{\langle xy\rangle}\left(s_xs_y-1\right)+
\sum_{x\in\partial}\left(h_xs_x-1\right)\right]-\frac12\log D_b\sum_{x}\left(b_xs_x-1\right). \label{IsingactionBulk}
\end{eqnarray}
For computation of the bulk-boundary entanglement entropy, we should take $b_x=-1$ for all $x$, and $h_x=+1$ for all boundary sites. (We have written Eq.~(\ref{IsingactionBulk}) in this general form because other configurations of $h_x,b_x$ will be used in our later discussion.)

In this action, the effect of the bulk pinning field $b_x$ competes with the boundary pinning field $h_x$.
The relative strength of these two pinning fields is determined by the ratio $\log D_b/\log D$.
If $\log D_b\ll \log D$, the lowest energy configuration will be the one with all spins pointing up. In the opposite limit $\log D_b\gg \log D$, all spins point down.
For the purpose of defining a code subspace with isometry to the boundary, we consider the limit $\log D_b\ll \log D$.
In that case all spins are pointing up, and the only energy cost in the Ising action~(\ref{IsingactionBulk}) comes from the last term, leading to the entropy
\begin{eqnarray}
S_{2,\text{bulk}}=V\log D_b,
\end{eqnarray}
which is the maximum possible for a state on the bulk Hilbert space since its dimension is $D_b^V$. In the limit $\log D_b\ll \log D,~D\rightarrow \infty$, the bulk is therefore in a maximally mixed state, so the corresponding holographic mapping from the bulk to the boundary is isometric. The isometry condition is equivalent to the condition that the lowest energy configuration of the Ising model has all spins pointing up.

Instead of requiring $\log D_b\ll \log D$, we can write down more precisely the isometry condition by requiring that the all-up configuration has the lowest energy. Consider a generic spin configuration with a spin-down domain $\Omega$. The energy of this configuration is $\mathcal{A}(\Omega)=\left(V-\left|\Omega\right|\right)\log D_b+\left|\partial \Omega\right|\log D$. Here $\left|\Omega\right|$ and $\left|\partial \Omega\right|$ are the volume and the surface area of $\Omega$, respectively. In order for the all-up configuration to be stable, we need $\mathcal{A}(\Omega)>V\log D_b$ for all nontrivial $\Omega$, which requires
\begin{eqnarray}
|\Omega|\log D_b<\left|\partial \Omega\right|\log D,~\text{for~all~regions~}\Omega.\label{Isometrycondition}
\end{eqnarray}
For example, if the bulk is a (triangulation of) hyperbolic space (with curvature radius $R=1$), a disk with boundary area $|\partial\Omega|=2\pi R/l_g$ has volume $|\Omega|=2\pi \left(\sqrt{R^2+1}-1\right)/l_g^2$. Here we have measured both area and volume by the triangulation scale $l_g$. Therefore the isometry condition requires
\begin{eqnarray}
\frac{\log D_b}{\log D}<l_g\frac{R}{\sqrt{R^2+1}-1},~\forall R \Rightarrow \frac{\log D_b}{\log D}\leq l_g.
\end{eqnarray}
There is a finite range of $D_b$ which satisfies the isometry condition, which is a consequence of the fact that the area/volume ratio is finite in hyperbolic space. For comparison, the same discussion for a disk in flat space with boundary area $2\pi R$ will require $\frac{\log D_b}{\log D}<\frac{2}{R}l_g$. Therefore the ratio $\log D_b/\log D$ must scale inversely with the size of the whole system $R_{\max}$.%
\footnote{In the pluperfect tensor work~\cite{yang2015}, the code subspace was defined by selecting some of the bulk sites, each having $D_b=D^2$. In contrast, the properties of random tensor networks considered in this work enable us to make a uniform choice of small $D_b$ at every site, which is more convenient.}

A useful remark is that the isometry condition~(\ref{Isometrycondition}) (or more precisely, a slightly weaker condition with $<$ replaced by $\leq$) is obviously necessary by a counting argument: In order for an operator defined in region $\Omega$ to be mapped to the boundary isometrically, it needs to be first mapped to the boundary of $\Omega$, so that the dimension of the Hilbert space at the boundary $D^{|\partial \Omega|}$ must be at least as large as the dimension of the bulk Hilbert space $D_b^{|\Omega|}$. With this observation, what we see from the Ising model representation is that the large-$D$ random tensor network is an {\it optimal} holographic code, in the sense that an isometry is defined as long as the counting argument does not exclude it. Of course one should keep in mind that this optimal property is only true asymptotically in the large $D$ limit.

\subsection{Entanglement wedges and error correction properties}
\label{subsec:errcorr}

Having shown that the holographic mapping $M$ defines an isometry from the bulk to the boundary degrees of freedom for suitable ratios $\log D_b / \log D$, it is natural to ask whether this isometry has the error correction properties proposed in Ref.~\cite{almheiri2014}, {\it i.e.}, whether operators in the bulk can be recovered from parts of the boundary instead of from the whole boundary.
Specifically, consider an operator $\phi_C$ in the bulk which only acts nontrivially in a region $C$.
Denote the complement of $C$ in the bulk by $\overline C$. We say that $\phi_C$ can be recovered from a boundary region $A$ if there exists a boundary operator $O_A$ such that~\cite{pastawski2015}
\begin{equation}
\label{errorcorrectioncond}
O_A M = M \phi_C.
\end{equation}
We note that condition~\eqref{errorcorrectioncond} is composable: For example, if $\phi_C$ and $\phi'_{C'}$ can be recovered from $A$ and $A'$, respectively, then $O_A O'_{A'} M = O_A M \phi'_{C'} = M \phi_C \phi'_{C'}$ for the corresponding boundary obervables $O_A$ and $O'_{A'}$. It follows that
\[ \braket{\phi_C \phi'_{C'}}_{\rho_b}
= \tr \rho_b \phi_C \phi'_{C'}
= \tr \rho_b M^\dagger M \phi_C \phi'_{C'} = \tr \rho_b M^\dagger O_A O'_{A'} M = \braket{O_A O'_{A'}}_{\rho}, \]
for any bulk state $\rho_b$ and the corresponding boundary state $\rho = M \rho_b M^\dagger$.
In the same way, an arbitrary $n$-point function in the bulk can be obtained from a corresponding correlation function on the boundary.

In the language of quantum error correction, Eq.~\eqref{errorcorrectioncond} states that the logical operator $\phi_C$ acting on the degrees of freedom in $C$ can be realized by an equivalent physical operator acting on the degrees of freedom in $A$ only.
We are now interested in understanding when \emph{all} operators $\phi_C$ in the region $C$ can be recovered from $A$.
That is, we would like the quantum information stored in subsystem $C$ to be protected against erasure of the degrees of freedom in $B$, the complement of $A$ on the boundary.
This amounts to another entropic condition, namely, that in the pure state $\ket{\Psi_M}$ defined in Eq.~\eqref{PsiM} there is no mutual information between $C$ and the region $B\overline C$~\cite{nielsen2007algebraic}, which ensures that the mutual information between $A$ and $C$ is maximal:
\begin{equation}
\label{ACisometry}
    S(C) + S(B\overline C) = S(BC\overline C).
\end{equation}
For the reader's convenience, we recount a short proof of this fact in Appendix~\ref{app:qec}.

In general it is important that Eq.~\eqref{ACisometry} is evaluated in terms of von Neumann entropies rather than R\'{e}nyi entropies.
In the limit of large $D$, however, both entropies are closely approximated by the Ryu-Takayanagi formula as long as the minimal surfaces are unique (see Section~\ref{sec:finiteD}).
What is more, we may even arrange for the Ryu-Takayanagi formula to be satisfied exactly, without any assumption on the uniqueness of minimal surfaces, by using ensembles of random stabilizer states instead of Haar random states (see Section~\ref{sec:stabs}).
In the following we shall therefore evaluate the quantum error correction condition~\eqref{ACisometry} in terms of second R\'{e}nyi entropies and assume (for simplicity) that the RT formula holds exactly.

\begin{figure}[!ht]
  \centering
\includegraphics[width=0.95\textwidth]{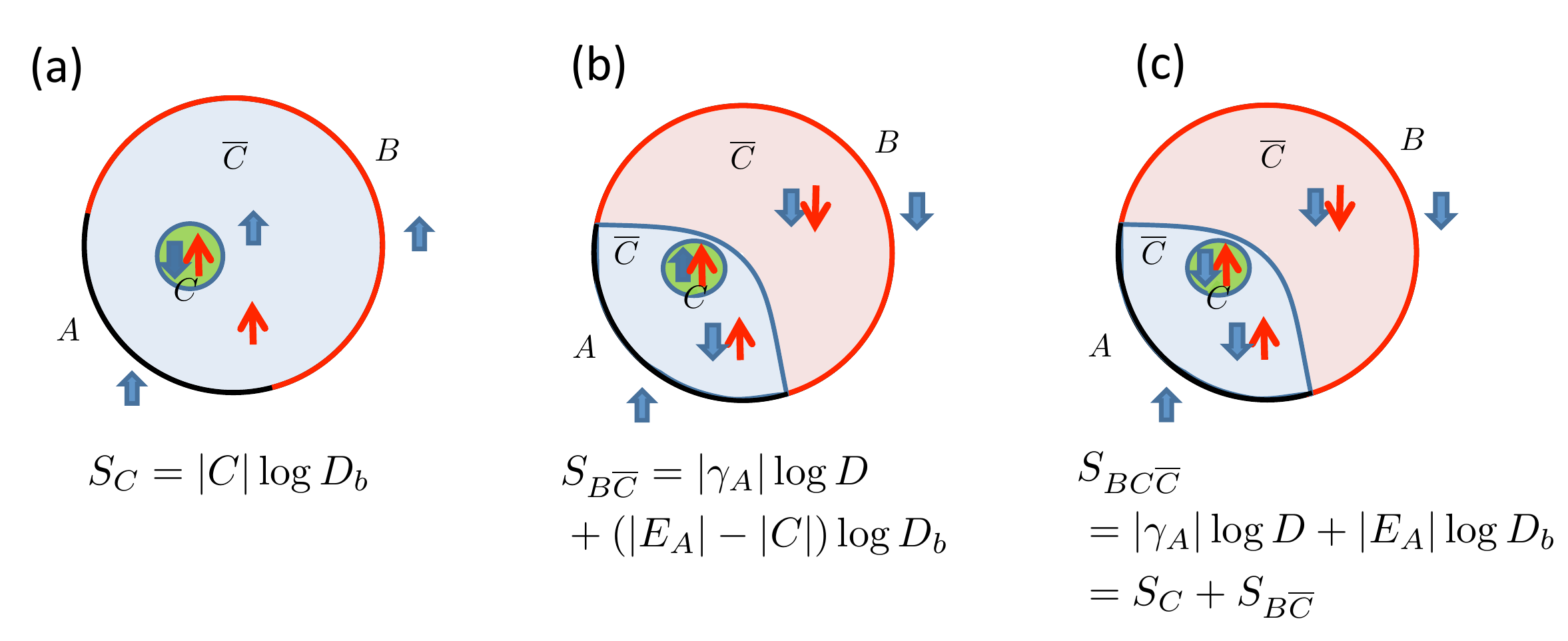}
\caption{The bulk-to-boundary isometry from a code subspace with small $D_b$ satisfying condition~(\ref{Isometrycondition}). The three panels show the Ising spin configurations for the calculation of (a) $S_2(C)$, (b) $S_2(B\overline C)$ and (c) $S_2(BC\overline C)$. $\overline C$ is the complement of $C$ in the bulk, and $B$ is the complement of $A$ on the boundary. The blue arrows are pinning fields ($h_x$ on the boundary and $b_x$ in the bulk), and the red arrows are the direction of Ising spins. The entropy is given by the energy of the configuration, which is contributed by the region with Ising spins anti-parallel with the pinning field. The blue (orange) regions are Ising spin-up (down) domains, respectively.} \label{fig:causalwedge}
\end{figure}

To understand when the error correction condition holds, we consider the configuration shown in Fig.~\ref{fig:causalwedge}. The calculation of $S_2(C)$ is straightforward. Given the isometry condition~\eqref{Isometrycondition}, the whole bulk is in a maximally mixed state after tracing over the boundary, so that $S_2(C)$ also takes the maximal value $|C|\log D_b$. In the calculation of $S_2(B\overline C)$, the pinning field is set to $b_x=-1$ for $x\in B$ and $h_x=-1$ for $x\in \overline C$. The boundary spin-down field in $B$ will pin a spin-down domain (orange region in Fig.~\ref{fig:causalwedge}). We consider the case when $C$ is in the spin-up (blue) domain, in which case the energy cost gives the entropy $S_2(B\overline C)=\left|\gamma_A\right|\log D+\left(\left|E_A\right|-\left|C\right|\right)\log D_b$. Here $\gamma_A$ is the domain wall bounding region $A$, and $E_A$ is the spin-up domain, which is the entanglement wedge of $A$.
The first term is the area law energy cost of the domain wall, and the second term is the volume law energy cost. $S_2(BC\overline C)$ can be computed similarly by flipping the pinning field in $C$ to downwards. Due to the isometry condition~(\ref{Isometrycondition}), flipping the field in $C$ does not create new spin-down domains, so that the only difference between $S_2(BC\overline C)$ and $S_2(B\overline C)$ is an additional energy cost in the $C$ region that is exactly $S_2(C)$. Therefore condition~(\ref{ACisometry}) holds, and the operators in $C$ can be recovered from $A$. As a final note, observe that the domain wall $\gamma_A$ is generally not the minimal surface, due to the presence of the bulk pinning field, but our conclusion holds as long as $C$ is in the spin-up domain and is disconnected from $\gamma_A$.

For comparison, we can consider the same configuration in Fig.~\ref{fig:causalwedge} and ask whether operators in $C$ can be recovered from $B$. This requires the calculation of $S_2(C) + S_2(A\overline C) - S_2(AC\overline C)$. Following an analysis similar to the previous paragraph, one can obtain $S_2(A\overline C)=\left|\gamma_A\right|\log D+\left(\left|E_B\right|+\left|C\right|\right)\log D_b$, and $S_2(AC\overline C)=\left|\gamma_A\right|\log D+\left|E_B\right|\log D_b$. Here $E_B$ is the complement of $E_A$ in the bulk, which is the entanglement wedge of $B$. Therefore the mutual information $I_2(C:A\overline C) = 2 \, S_2(C)>0$, so that $C$ cannot be recovered from $B$.

From the two cases studied above, we can see that operators in a bulk region $C$ can be recovered from a boundary region $A$ if and only if $C$ is included in the entanglement wedge $E_A$ of $A$. It should be noted that this statement only applies to small bulk $D_b$, or for sufficiently small regions $C$ if $D_b$ is larger, when the entanglement wedge $E_A$ (spin-down domain in the Ising model) is independent of the direction of the pinning field in $C$.

\subsection{Gauge invariance and absence of local operators}

In the two subsections above, we showed how a large $D$ and small $D_b$ random tensor network defines bulk-to-boundary isometries with error correction properties.  In this subsection we would like to investigate the other direction of the BHC, {\it i.e.}, the boundary-to-bulk isometry. To define this isometry, we need to require that the boundary-bulk entanglement entropy be equal to $\lvert A \rvert \log D$,
which is the maximum possible entropy for the boundary. This requires the opposite condition from Eq.~(\ref{Isometrycondition}):
\begin{eqnarray}
|\Omega|\log D_b>\left|\partial \Omega\right|\log D,~\text{for~all~regions~}\Omega. \label{Isometrycondition2}
\end{eqnarray}

To satisfy this condition, we can take $\Omega=\left\{x\right\}$ as a single site in the bulk, for which the condition is reduced to $D_b>D^{n_x}$, with $n_x$ the number of links connected to $x$. If this condition is satisfied for each site, Eq.~(\ref{Isometrycondition2}) also applies to other regions, since $|\partial B|\leq \sum_{x\in B}n_x$ always holds. Therefore the condition ensuring a boundary-to-bulk isometry is
\begin{eqnarray}
D_b>D^{n_x},~\forall x. \label{Isometrycondition3}
\end{eqnarray}
This is similar to the condition proposed in Ref.~\cite{yang2015}, with the difference that Ref.~\cite{yang2015} has $D_b=D^{n_x}$ because each tensor is required to be rigorously a unitary mapping from the in-plane legs to the bulk leg.

When this isometry condition is satisfied, the boundary-to-bulk isometry maps each boundary state isometrically to a bulk state in a larger Hilbert space with dimension $D_b^V$. It should be clarified that the physical Hilbert space is always that of the boundary, and that the $D_b^V$-dimensional Hilbert space, which is factorizable into a direct product of each bulk site, is just an auxiliary tool. The situation is very similar to a gauge theory, in which one can embed gauge invariant states into a larger auxiliary Hilbert space by treating the gauge vector potential as a physical field. In fact, it was shown in Ref.~\cite{yang2015} that the physical Hilbert space -- the image of the boundary Hilbert space under the holographic mapping -- can be defined by a gauge invariance condition. The discussion also applies to the random tensor network satisfying condition~\eqref{Isometrycondition3}.

\begin{figure}[!ht]
  \centering
\includegraphics[width=0.95\textwidth]{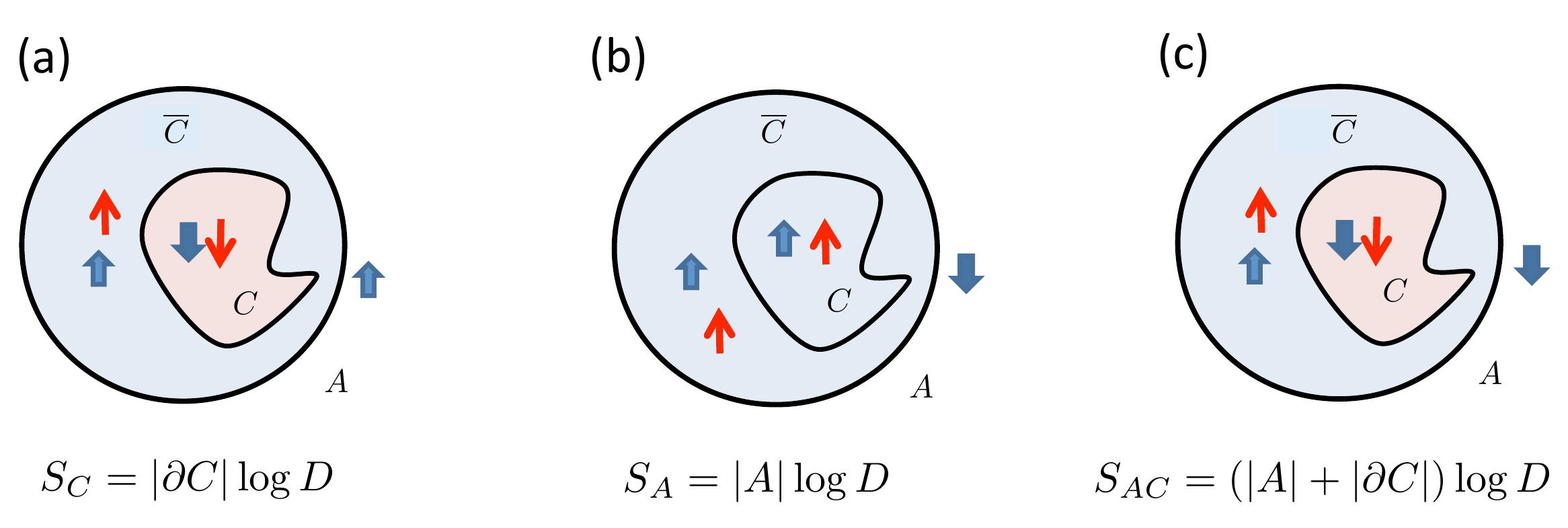}
\caption{Boundary-to-bulk isometry with a large $D_b$ satisfying condition~(\ref{Isometrycondition3}). The three panels show the Ising spin configurations for the calculation of (a) $S_2(C)$, (b) $S_2(A)$ and (c) $S_2(AC)$. Here $A$ represents the whole boundary. The blue arrows are pinning fields, and the red arrows are the direction of Ising spins. The blue (orange) regions are Ising spin-up (down) domains, respectively.} \label{fig:gaugesymmetry}
\end{figure}

The main property of the boundary-to-bulk isometry is that the bulk theory is intrinsically nonlocal. To be more precise, consider an arbitrary region $C$ that disconnected from the boundary, as shown in Fig.~\ref{fig:gaugesymmetry}. We would like to show that any operator $\phi_C$ supported in $C$ is mapped to the boundary trivially, {\it i.e.},
\begin{equation*}
M \phi_C M^\dagger = c I_{A}.
\end{equation*}
Here, we have denoted the whole boundary as region $A$, while $I_A$ is the identity operator on the boundary, and $c$ is a constant. 
This statement is equivalent to the statement $I(A:C)=S(A) + S(C) - S(AC) = 0$, which means there is no mutual information between $C$ and the whole boundary. Following an argument similar to that of the previous subsection, and using condition~(\ref{Isometrycondition2}) one can easily conclude that
\begin{equation*}
S_2(C) = \left|\partial C\right|\log D,~S_2(A) = |A|\log D,~S_2(AC) = S_2(A) + S_2(C),
\end{equation*}
as is illustrated in Fig.~\ref{fig:gaugesymmetry}.
Therefore all purely bulk operators are trivial, and only those in regions adjacent to the boundary contain nontrivial information about boundary physical operators.
As was discussed in Ref.~\cite{yang2015}, this property is a consequence of the gauge symmetry of the tensor network.
For all tensor networks, there is a gauge symmetry induced by acting unitarily on each internal leg while preserving the physical state after contraction. However, for tensor networks with the boundary-to-bulk isometry property, this gauge symmetry is isometrically mapped to constraints on the bulk legs.

In summary, we have shown that a BHC can be built from a large $D$ random tensor network with bulk leg dimension $D_b$ satisfying condition~(\ref{Isometrycondition3}).
The boundary theory is mapped isometrically to a nonlocal theory in the bulk, with the physical (boundary) Hilbert space defined by gauge constraints.
A code subspace is defined by a local projection at every bulk site to a smaller subspace with dimension $D_b'$ which satisfies condition~(\ref{Isometrycondition}).
A bulk-to-boundary isometry is defined in the code subspace, and a bulk local operator in the code subspace can be recovered from a boundary region as long as the entanglement wedge of this region encloses the support of this bulk operator.
In this way, random tensor networks can be used to define a bulk theory with intrinsic nonlocality and emergent locality in a subspace, as is desired for a theory of quantum gravity.

\section{Higher R\'{e}nyi entropies}\label{sec:nthRenyi}

In this section, we will generalize the second R\'{e}nyi entropy calculation to higher R\'{e}nyi entropies, and show that the higher R\'{e}nyi entropies of a random tensor network are also mapped to partition functions of classical spin models, with the spin now living in a different target space, the permutation group $\Sym_n$ of $\{1,\dots,n\}$. For $n=2$, the permutation group $\Sym_2 = \mathbb Z_2$ reduces to the target space of the Ising model.

The derivation is in exact parallel with that for the second R\'{e}nyi entropy in Section~\ref{sec:secondRenyi}
For the random tensor network state given by Eq.~(\ref{peps3}), the $n$-th R\'{e}nyi entropy is:
\begin{equation*}
S_n(A)=\frac1{1-n}\log \frac{\tr \rho^n_A}{\left(\tr \rho\right)^n}.
\end{equation*}
Again we use the natural logarithm to define higher R\'{e}nyi entropies. We now define:
\begin{align}
Z_1^{(n)}&= \tr \rho_A^n = \tr \left[\rho^{\otimes n} \, \mathcal C_A^{(n)} \right], \label{eq:Z1ndefn} \\
Z_0^{(n)}&= \left(\tr \rho\right)^n = \tr \left(\rho^{\otimes n}\right). \nonumber
\end{align}
Here $\rho^{\otimes n}$ denotes the direct product of $n$-copies of $\rho$, and $\mathcal C_A^{(n)}$ is the permutation operator that permutes the $n$ copies cyclically in $A$ region. For a basis $\ket{m_A}$ of region $A$, a basis of the direct product space is given by $\ket{{m_1}_{A}}\otimes \ket{{m_2}_{A}}\otimes \dots \otimes \ket{{m_n}_{A}}$, and the action of $\mathcal C_A^{(n)}$ is given by $\mathcal C_A^{(n)} (\ket{{m_1}_{A}}\otimes \ket{{m_2}_{A}}\otimes \dots \otimes \ket{{m_n}_{A}}) = \ket{{m_2}_{A}}\otimes \ket{{m_3}_{A}}\otimes \dots \otimes \ket{{m_n}_{A}}\otimes \ket{{m_1}_A}$.

As in Section~\ref{sec:secondRenyi}, we approximate the typical R\'{e}nyi entropy by
\begin{equation*}
S_n(A) = \frac1{1-n} \log\frac{Z_1^{(n)}}{Z_0^{(n)}} \simeq \frac1{1-n}\log \frac{\overline{Z_1^{(n)}}}{\overline{Z_0^{(n)}}}
\end{equation*}
and compute $\overline{Z_1^{(n)}}$ and $\overline{Z_0^{(n)}}$.
By inserting the definition of $\rho$ in Eq.~\eqref{peps3} into Eq.~\eqref{eq:Z1ndefn}, the first average can be written as
\begin{eqnarray}
\overline{Z_1^{(n)}}=\tr \left[\rho_P^{\otimes n} \, \mathcal C_A^{(n)} \prod_x \overline{\ket{V_x}\!\!\bra{V_x}^{\otimes n}}\right].\label{Z1n}
\end{eqnarray}
The average of $\ket{V_x}\!\!\bra{V_x}^{\otimes n}$ results in a projector onto the symmetric subspace of the $n$-fold tensor power Hilbert space (e.g.,~\cite{harrow2013}):
\begin{eqnarray}
\overline{\ket{V_x}\!\!\bra{V_x}^{\otimes n}}=\frac1{C_{n,x}}\sum_{g_x\in \Sym_n}g_x.\label{nStateaverage}
\end{eqnarray}
Here $g_x$ runs over all permutation group elements and we identify $g_x$ with its action on the $n$-copy single site Hilbert space.
This action is defined by permuting the $n$ copies of systems, similar to the definition of $\mathcal C_A^{(n)}$.
The normalization constant is $C_{n,x}=\sum_{g\in \Sym_n}\tr g=\sum_{k=1}^n c(n,k)D_x^{k} = (D_x+n-1)!/(D_x-1)!$, with $c(n,k)$ the Stirling number of the first kind.

Using this result in Eq.~(\ref{Z1n}), we obtain a sum over permutation elements $\left\{g_x\right\}$ on each vertex, and thus $\overline{Z_1^{(n)}}$ becomes a partition function of classical spin model:
\begin{align}
    \overline{Z_1^{(n)}} &= \sum_{\left\{g_x\right\}}e^{-\mathcal A^{(n)}\left[\left\{g_x\right\}\right]} , \label{eq:SnAverage} \\
    e^{-\mathcal A^{(n)}\left[\left\{g_x\right\}\right]} &= \frac1{\prod_xC_{n,x}}\tr \left[\rho_P^{\otimes n} \, \mathcal C_A^{(n)} \bigotimes_x g_x\right]. \label{eq:Snweight}
\end{align}
The statistical weight of a configuration $\left\{g_x\right\}$ is determined by the expectation value of this permutation (multiplied by the additional cyclic permutation $\mathcal C_A^{(n)}$ on the boundary) in the state $\rho_P^{\otimes n}$.
In general, this expectation value is not related to the R\'{e}nyi entropy of any bulk region, which is a key difference from the case of second R\'{e}nyi entropy.
One can view such expectation values of permutation operators as generalized multi-partite entanglement measures that contains more information than R\'{e}nyi entropies.%
\footnote{This is because they are invariant under local unitary transformations that only act on a domain with the same permutation value $g_x=g$. Such quantities are known as LU invariants in the quantum information literature (see, e.g., Ref.~\cite{leifer2004measuring} and references therein).}

In our case, $\rho_P = \rho_b \otimes \prod_{\langle xy\rangle}\ket{xy}\!\!\bra{xy}$ by Eq.~(\ref{rhoP}).
Thus the action $\mathcal A^{(n)}\left[\left\{g_x\right\}\right]$ becomes a sum of bond contributions and contributions of the bulk state $\rho_b$, similar to \eqref{Isingaction2} in the case of the second R\'{e}nyi entropy:
\begin{equation}
\label{SnAction}
\begin{split}
\mathcal A^{(n)}\left[\left\{g_x\right\}\right] &=-\sum_{\langle xy\rangle}\log D_{xy}\left(\chi(g_x^{-1} g_y) - n\right)-\sum_{x\in \partial}\log D_{x\partial}\chi(g_x^{-1} h_x) \\
&-\log\left[\tr \left(\rho_b^{\otimes n} \bigotimes_x g_x\right)\right]+\sum_x\log C_{n,x}.
\end{split}
\end{equation}
Here, $\chi(g)$ denotes the number of cycles in a permutation $g$ (including cycles of length one).
The boundary pinning field $h_x$ takes the value
\begin{equation*}
h_x = \begin{cases}
    \mathcal C_x^{(n)}, & x\in A \\
    I_x, & x\in \overline{A}
\end{cases}
\end{equation*}
with $C_x^{(n)}$ the cyclic permutation acting on site $x$.
For $n=2$, it is straightforward to check that \eqref{SnAction} reduces to the Ising action~\eqref{Isingaction2}.
The EPR pairs on the internal legs contribute a two-spin interaction energy
\begin{equation}
\label{eq:epr energy}
  -\log D_{xy} \left(\chi(g_x^{-1} g_y)-n\right),
\end{equation}
which vanishes only if $g_x = g_y$.
In this sense, the interaction is ``ferromagnetic'', which prefers all $g_x$ to align.

\begin{figure}[!ht]
  \centering
\includegraphics[width=0.95\textwidth]{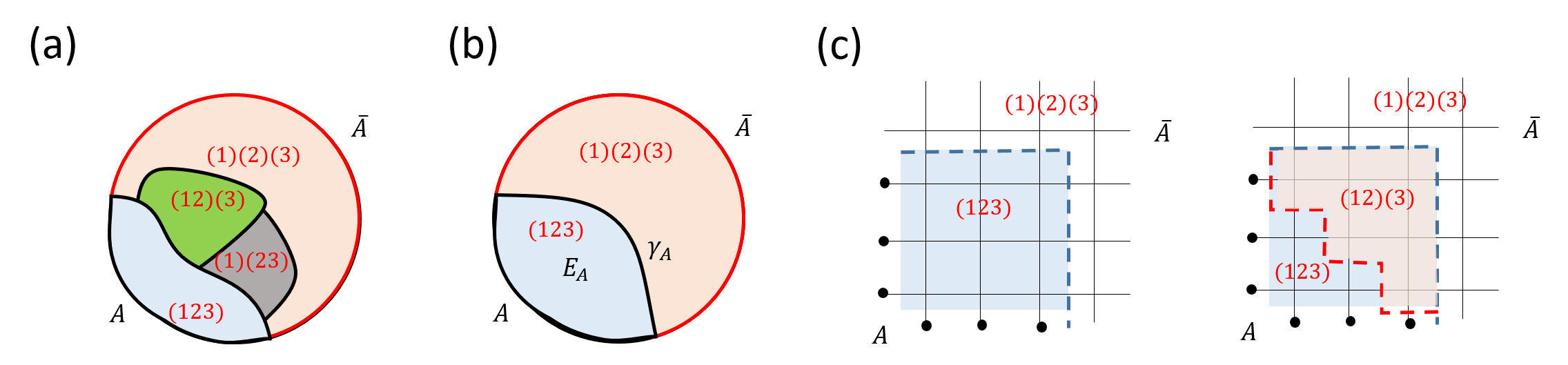}
\caption{Examples of different spin configurations $\left\{g_x\right\}$ for $n=3$. The permutation elements $g_x\in \Sym_3$ are denoted by their cycle structure. For example $(1)(2)(3)$ is identity and $(123)$ is the cyclic permutation. The boundary pinning field is $h_x=(123)$ in the $A$ region and $h=(1)(2)(3)$ in the complement. When the bulk is a pure direct-product state, the minimal energy configuration is given in panel (b), with the two domains separated by the minimal surface $\gamma_A$. (c) is an example illustrating that there can be multiple configurations with the same contribution to the partition function $\overline{Z_1^{(n)}}$ if the minimal surface is degenerate.} \label{fig:nthRenyi}
\end{figure}

We first consider the case when the bulk is a pure direct-product state, so that the contribution of $\rho_b$ to Eq.~(\ref{SnAction}) vanishes.
We also take $D_{xy}=D_{x\partial}=D$ for simplicity, as in previous sections. The action~(\ref{SnAction}) describes a $\Sym_n$-spin model with ferromagnetic interaction and a boundary pinning field, at inverse temperature $\beta=\log D$. Some examples of $\left\{g_x\right\}$ configurations are shown in Fig.~\ref{fig:nthRenyi}. Each domain wall between two different values of $g_x$ has an energy cost which is proportional to the area of the domain wall and $\chi(g_x^{-1} g_y)$.
Up to this point, the derivation applies to arbitrary values of $D$. In the large $D$ limit, the partition function is dominated by the lowest energy contribution.
If the entanglement wedge $E_A$ is unique (i.e., if the Ising model used to evaluate the second R\'enyi entropy has a unique minimal energy configuration) then the spin model with action~\eqref{SnAction} likewise has a unique minimal energy configuration.
It is given by setting $g_x$ equal to the cyclic permutation throughout region $E_A$ and to the identity elsewhere (see Fig.~\ref{fig:nthRenyi}\,(b) for an illustration in the case $n=3$).
We give a detailed proof of this fact in Appendix~\ref{app:unique}.
Since the boundary of $E_A$ in the bulk is the geodesic surface $\gamma_A$, 
we obtain
\begin{equation*}
\left.\overline{Z_1^{(n)}}\right|_{D\rightarrow \infty} \simeq \text{const.} \times e^{(1-n)\log D\left|\gamma_A\right|},
\end{equation*}
where the constant prefactor is independent from the choice of region $A$ and will be canceled by the same factor in the denominator $\overline{Z_0^{(n)}}$. The factor $(1-n)$ comes from the fact that the cyclic permutation contains one loop.
Therefore we conclude that the typical $n$-th R\'{e}nyi entropy of the random tensor network state is given by
\begin{equation*}
  S_n(A) \simeq \log D\left|\gamma_A\right|.
\end{equation*}

If the bulk is in an entangled state $\rho_b$ then we need to consider the corresponding contribution to the action~\eqref{SnAction}, which is given by
\begin{equation}
\label{eq:Sn bulk action piece}
    -\log\left[\tr \left(\rho_b^{\otimes n} \bigotimes_x g_x\right)\right].
\end{equation}
As long as the bulk dimension is not too large, we may think of \eqref{eq:Sn bulk action piece} as a perturbation to the statistical model for the direct product case.
In the minimal energy configuration of the unperturbed model, the contribution~\eqref{eq:Sn bulk action piece} is precisely equal to $(n-1) S_n(E_A; \rho_b)$, {\it i.e.}, $(n-1)$ times the $n$-th R\'{e}nyi entropy of the reduced density matrix of the entanglement wedge in the bulk state.
For a general configuration $\{g_x\}$, however, \eqref{eq:Sn bulk action piece} cannot be interpreted as an entropy.
In fact, the corresponding statistical weight $\tr \left(\rho_b^{\otimes n} \bigotimes_x g_x\right)$ can even be a complex number, so that the interpretation of the action~\eqref{SnAction} requires suitable care.
However, we note that the partition function~\eqref{eq:SnAverage} is by definition an average of the positive quantities~\eqref{eq:Z1ndefn} and therefore always positive.
The choice of branch for the logarithm in \eqref{eq:Sn bulk action piece} is also irrelevant for the resulting statistical weight and so does not concern us further.
The key observation now is that $\lvert \tr \left(\rho_b^{\otimes n} \bigotimes_x g_x\right) \rvert \leq 1$ by the Cauchy-Schwarz inequality.
Thus the real part of \eqref{eq:Sn bulk action piece} is always non-negative:
The bulk correction only ever increases the real part of the energy levels.
In particular, the only way that the (real part of the) energy gap can decrease in the perturbed model is due to the bulk corrections in the minimal energy configuration of the unperturbed model.
Since $S_n(E_A; \rho_b) \leq \log D_b \lvert E_A \rvert$, the energy gap can therefore be lower bounded by $\log D - (n-1) \log D_b \lvert E_A \rvert$.
As long as this gap diverges for large $D$, the minimal energy configuration remains unchanged and dominates the partition sum, so that
\begin{equation*}
    \left.\overline{Z_1^{(n)}}\right|_{D\rightarrow \infty} \simeq \text{const.} \times e^{(1-n) \log D \lvert \gamma_A \rvert + (1-n) S_n(E_A; \rho_b)}.
\end{equation*}
We conclude that, for an entangled bulk state of sufficiently low dimension and large $D$, the typical $n$-th R\'{e}nyi entropy of the random tensor network state is given by
\begin{equation}
\label{eq:nth renyi with bulk}
    S_n(A) \simeq \log D \lvert \gamma_A \rvert + S_n(E_A; \rho_b).
\end{equation}
It should be noted that these conclusions only hold when there is a unique minimal geodesic surface, as discussed above. When there are multiple degenerate minimal surfaces, the entropy is reduced by a factor $\log N$ with $N$ the number of minimal energy configurations. An example of degenerate minimal energy configurations are shown in Fig.~\ref{fig:nthRenyi}\,(c) for a square lattice in flat space.

The fact that in leading order all R\'{e}nyi entropies approach the same value in the large $D$ limit tells us an important difference between a large $D$ random tensor network state and a large central charge CFT ground state.%
\footnote{We would like to thank Juan Maldacena and Aron Wall for reminding us of this point.}
When $A$ is a length-$l$ interval, the R\'{e}nyi entropy of $A$ in a $(1+1)$-dimensional CFT ground state is given by~\cite{holzhey1994,calabrese2004}
\begin{eqnarray}
S_n(A)=\left(1+\frac 1n\right) \frac{c}6\log l,\label{CFTRenyi}
\end{eqnarray}
which shows that in the large central charge limit, the $n$ dependence remains nontrivial. In term of the eigenvalue spectrum of the reduced density matrix $\rho_A$, this difference tells us that in a random tensor network state, the eigenvalues of $\rho_A$ are more strongly concentrated than that in a CFT ground state, although the density of states is also highly peaked in the latter case.

From the point of view of
  the dual gravity theory, the nontrivial $n$-dependence is enforced
  by the requirement that the dual geometry of $\tr[\rho^n]$ ought to satisfy the equations of motion.
  To be more specific, if we naively constructed the dual
  geometry of $\tr[\rho^n]$ by gluing $n$ copies of the original bulk geometry around the minimal surfaces (in view of Eq.~\eqref{eq:Z1ndefn}, this is quite literally what the calculation of the R\'enyi entropy of the tensor network state amounts to) then there
  will be no $n$-dependence of the R\'enyi entropy~\cite{fursaev2006proof}. The problem here is that this naively replicated geometry does not satisfy the equations of motion. In other words, the geometry does not backreact and converge to the saddle point of some gravitational action.

If we are interested in modifying the random tensor network to realize the same R\'{e}nyi entropy behavior (\ref{CFTRenyi}) as a CFT ground state, the simplest way is by replacing the maximally entangled EPR pair state $\ket{xy}$ at each internal leg by a more generic state. For a more generic link state $\ket{L_{xy}}$, the calculation of $\overline{Z_1^{(n)}}$ still applies, with the link terms~\eqref{eq:epr energy} in the action~\eqref{SnAction} replaced by
\begin{eqnarray}
-\log \tr \left[\ket{L_{xy}}\!\!\bra{L_{xy}}^{\otimes n}g_x\otimes g_y\right].
\end{eqnarray}
Since such terms are always non-negative, and only vanish if $g_x = g_y$, the qualitative behavior of the $\Sym_n$-spin model remains ferromagnetic, and the lowest energy configuration in large $D$ limit still only contains a single domain wall with minimal area bounding $A$, as shown in Fig.~\ref{fig:nthRenyi}\,(b). In this case one obtains the following formula for the R\'{e}nyi entropies,
\begin{eqnarray}
S_n(A)&\simeq &S_n\left(\ket{L_{xy}}\right)\left|\gamma_A\right|\approx \frac{2}{l_g}S_n\left(\ket{L_{xy}}\right)\log l,
\end{eqnarray}
with $S_n\left(\ket{L_{xy}}\right)$ the R\'{e}nyi entropy of the bond
state $\ket{L_{xy}}$ with the partition between $x$ and $y$. As a
reminder, $l_g$ is the area occupied by each leg of the tensor, in
units of the AdS radius $R$. Therefore the R\'{e}nyi entropy behavior
is identical to that of a CFT ground state if the tensor network is a
triangulation of the hyperbolic space, and the bond state
$\ket{L_{xy}}$ satisfies
$S_n\left(\ket{L_{xy}}\right)=l_g\frac{c}{12}\left(1+\frac
  1n\right)$. For example, one can take $l_g=\frac2c$, and define the
state $\ket{L_{xy}}$ as a thermofield double state of the free boson
CFT. In this case, the reduced density matrix
$\rho_x=\tr _y\ket{L_{xy}}\!\!\bra{L_{xy}}$ is a thermal density
matrix of the free boson CFT on the torus with aspect ratio
$\frac{\beta}L$ of order one. The aspect ratio can be chosen to fit
the entropy $S_n=\frac 16\left(1+\frac 1n\right)$.  The random tensors
on each site impose random projections acting on these simple free CFT
states, each of which is defined on a small circle.  This random
projection defines a state on the boundary which for single intervals
has the R\'{e}nyi entropy behavior of a strongly correlated CFT in $1+1$ dimensions.
  However, for more complicated subsystems, such as a
disjoint union of two distant
intervals~\cite{faulkner2013entanglement}, the R\'{e}nyi entropy $S_n$
exhibits a dependence on $n$ that cannot be accommodated by a suitable
choice of link state (which only affects the R\'{e}nyi entropy per
bond but not the minimal surface or the replica geometry itself).%
\footnote{We would like to thank Xi Dong for explaining this to us.}
Besides, it is not quite clear whether the above
  modification will reproduce the correct R\'enyi entropies
  even for a single region if we go to the higher dimensions. More
systematic investigation of the comparison between random tensor
networks and CFT ground states will be reserved for future works.

The mapping we derived from the calculation of $n$-th R\'{e}nyi entropy to classical $\Sym_n$-spin models applies to more general situations. For example, studying the second-order correction terms in the calculation of average second R\'{e}nyi entropy, Eq.~(\ref{S2approximation}), involves the computation of $\overline{\delta Z_1^2}$ and $\overline{\delta Z_0^2}$. These quantities are quartic in $\prod_x\ket{V_x}\!\!\bra{V_x}$, so that one can apply formula~(\ref{nStateaverage}) and translate $\overline{\delta Z_1^2}$ into a partition function of $\Sym_4$-spin model. The only difference between $\overline{\delta Z_1^2}$ and $\overline{Z_1^{(4)}}$ in the $4$-th R\'{e}nyi entropy calculation is the value of boundary field. In the calculation of $\overline{\delta Z_1^2}$ the value of boundary field should be chosen as permutation $(12)(34)$ in region $A$, and identity elsewhere. We will use this strategy in Section~\ref{sec:finiteD} to bound the fluctuations of the R\'{e}nyi entropies around their semiclassical value \eqref{eq:nth renyi with bulk}. Another important application of this mapping with $n>2$ is the calculation of boundary two-point functions, which will be discussed in next section.

\section{Boundary two-point correlation functions}\label{sec:correlationspec}

As we discussed in Section~\ref{sec:RTformula}, the mutual information between two distant regions (Fig.~\ref{fig:Isingdomain}\,(b)) does not grow with $\log D$ (if the bulk state $\rho_b$ remains $D$-independent), although the entropy of each region is proportional to $\log D$. This observation indicates that two-point correlation functions between $A$ and $B$ are suppressed, as a consequence of strong multi-partite correlation in the random tensor network state. In this section we will investigate the behavior of two-point correlation functions more systematically, making use of the state averaging techniques.


We consider two small regions $A$ and $B$, whose entanglement wedges are disconnected, as was shown in Fig.~\ref{fig:Isingdomain}\,(b). Here the bulk has a given state $\rho_b$ and the entanglement wedges are defined with respect to the Ising action~(\ref{Isingaction2}) corresponding to this bulk state. For two operators $O_A$ and $O_B$ supported in $A$ and $B$ respectively, the correlation function is given by $\left\langle O_AO_B\right\rangle=\tr \left[\rho O_AO_B\right]/\tr \rho$. It is not appropriate to directly consider the state average of this quantity for fixed $O_A$ and $O_B$, since local unitary transformations acting on vertices in $A$ or $B$ will transform $O_A$ and $O_B$ and thus may average over two-point functions with very different behaviors. To define a more refined measure of two-point correlations, we introduce a complete basis of Hermitian operators in $A$ labeled by $O_A^\alpha$, and similarly a basis in $B$ labeled by $O_B^\beta$. When the Hilbert space dimension of $A$ region is $D_A$, the index $\alpha$ runs from $1,2,\dots,D_A^2$. Choose these operators to satisfy the orthonormality conditions
\begin{equation}
\label{orthonormal}
\tr \left[O_A^{\alpha}O_A^{\beta}\right]=\delta^{\alpha\beta},~\sum_{\alpha}\left[O_A^{\alpha}\right]_{ab}\left[O_A^{\alpha}\right]_{cd}=\delta_{ad}\delta_{bc}.
\end{equation}
For example, if $A$ consists of $N$ qubits, a choice of $O_A^\alpha$ are the $4^N$ direct products of Pauli matrices or identity operator acting on each site, with proper normalization.

Given a choice of basis operators, we define the correlation matrix
\begin{eqnarray}
M^{\alpha\beta}=\frac{\tr \left[\rho \, O_A^{\alpha}O_B^\beta\right]}{\tr \rho}. \label{correlationmatrix}
\end{eqnarray}
Two arbitrary operators $F_A,~F_B$ in $A$ and $B$ can be expanded in this basis as $F_A=\sum_\alpha f_{A\alpha}O_A^\alpha,~F_B=\sum_{\beta}f_{B\beta}O_B^\beta$, such that $\langle F_AF_B\rangle=\sum_{\alpha\beta}f_{A\alpha}M^{\alpha\beta}f_{B\beta}$ is determined by the correlation matrix. Therefore $M_{\alpha\beta}$ contains complete information about correlation functions between $A$ and $B$.

To define a basis-independent measure of correlation, one natural choice is the singular value spectrum of $M_{\alpha\beta}$. Denote the singular value decomposition of $M$ as $M_{\alpha\beta}=\sum_sU_{\alpha s}\lambda_sV_{\beta s}$, with $\lambda_s\geq 0$ real singular values, and $U,V$ unitary matrices.
This decomposition tells us that there is a particular set of operators
\begin{eqnarray}
K_{As}&=&\sum_\alpha U_{\alpha s}^*O_{A}^\alpha,~K_{Bs}=\sum_\beta V_{\beta s}^*O_{B}^\beta,
\end{eqnarray}
which satisfies
\begin{eqnarray}
\langle K_{As}K_{Bt}\rangle=\delta_{st}\lambda_s.
\end{eqnarray}
This set of operators can be considered as the analogs of the quasi-primary fields in a conformal field theory, and the singular values $\lambda_{s}$ are basis-independent measures of two-point correlations between $A$ and $B$.

\begin{figure}[t]
  \centering
\includegraphics[width=0.75\textwidth]{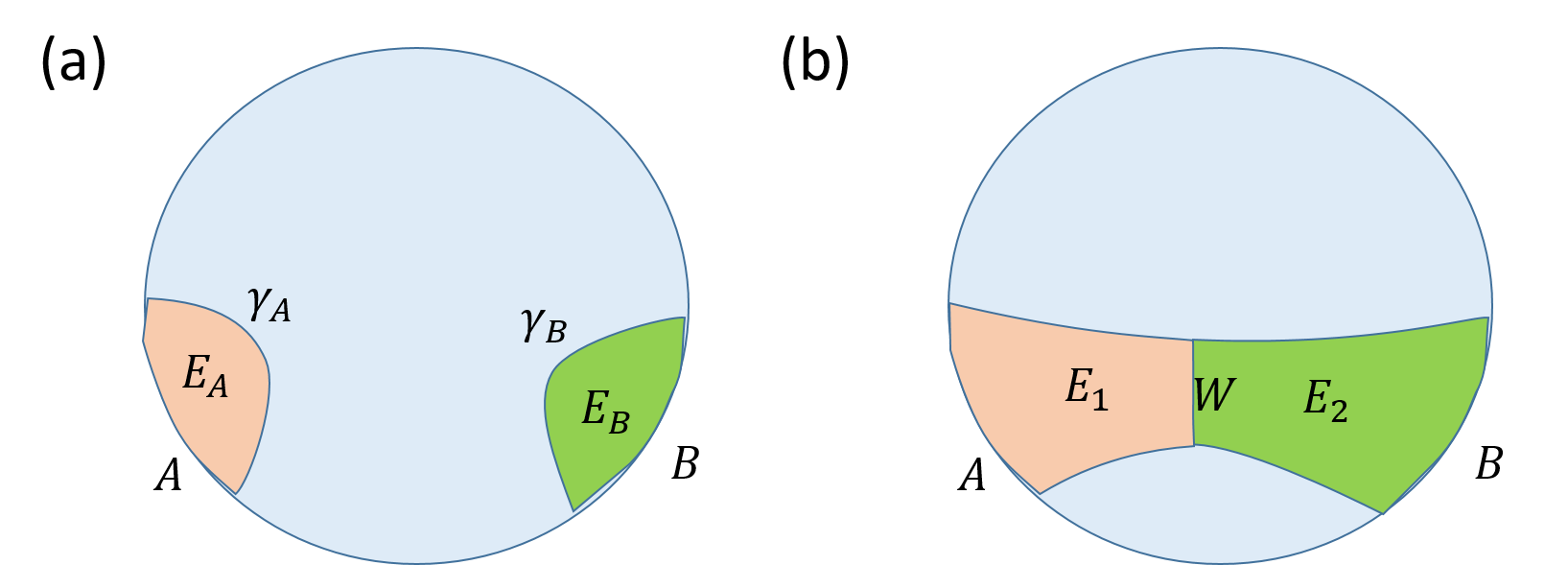}
\caption{(a) The minimal energy spin configuration in the calculation of $C_{2n}$ in Eq.~(\ref{C2n}). The red and green regions are domains with permutation $\mathcal X_A$ and $\mathcal Y_B$ defined in Eq.~(\ref{permutationsforM}), respectively. The blue region is the identity domain. (b) The lowest energy spin configuration that contributes nontrivially to the connected correlation between $A$ and $B$. The regions are defined in the same way as in (a). $W$ is the narrowest throat separating the two geodesic surfaces connecting $A$ and $B$.} \label{fig:correlationmatrix}
\end{figure}

Instead of directly carrying out singular value decomposition of $M$ and studying $\lambda_s$, it is more convenient to consider the following quantity:
\begin{eqnarray}
C_{2n}=\tr \left[\left(M^\dagger M\right)^n\right]\equiv \sum_s\lambda_s^{2n}.\label{C2ndef}
\end{eqnarray}
Knowing $C_{2n}$ for all integers $n$ determines the singular values $\lambda_s$, in the same way that the eigenvalue spectrum of a density matrix is determined by all R\'{e}nyi entropies.
On the other hand, using the orthonormality condition~(\ref{orthonormal}), $C_{2n}$ can be reexpressed into the following form:
\begin{eqnarray}
C_{2n}=\frac{\tr \left[\rho^{\otimes 2n} (\mathcal X_A \otimes \mathcal Y_B) \right]}{\left(\tr \rho\right)^{2n}}.\label{C2n}
\end{eqnarray}
Here $\mathcal X_A$ and $\mathcal Y_B$ are two permutation operators
\begin{eqnarray}
    \mathcal X_A=(1~2)(3~4) \dots (2n\!-\!1~2n), ~
    \mathcal Y_B=(2~3)(4~5) \dots (2n-2~2n-1)(2n~1),\label{permutationsforM}
\end{eqnarray}
which means $\mathcal X_A$ permutes each copy with an odd label $2k-1$ with copy $2k$, and $\mathcal Y_B$ permutes each copy $2k+1$ with copy $2k$.
The details of this derivation are presented in Appendix~\ref{app:correlation}. In this way, we have expressed $C_{2n}$ in a form similar to the $2n$-th R\'{e}nyi entropy, with a different permutation operator. Once we obtain Eq.~(\ref{C2n}) it is straightforward to perform the state average, which maps $C_{2n}$ in large $D$ limit to the same classical $\Sym_{2n}$-spin model as in the $2n$-th R\'{e}nyi entropy calculation:
\begin{eqnarray}
\overline{C_{2n}}\simeq \frac{\overline{Z_1^{(2n)}\left[h_x\right]}}{\overline{Z_0^{(2n)}}},
\end{eqnarray}
in which $\overline{Z_1^{(n)}\left[h_x\right]}$ is the same partition function defined in Eq.~(\ref{SnAction}), with a different boundary field $h_x$. $h_x$ takes the value of the two permutations in Eq.~(\ref{permutationsforM}) for $x$ in $A$ and $B$ respectively, and identity elsewhere. The denominator is the same as that of the $2n$-th R\'{e}nyi entropy. In the large $D$ limit, the minimal energy configuration that dominates $\overline{Z_1^{(2n)}\left[h_x\right]}$ is shown in Fig.~\ref{fig:correlationmatrix}. The minimal energy domain walls are the same as in the $2n$-th R\'{e}nyi entropy, which are minimal sufaces bounding $A$ and $B$. However, the prefactor of the area law term is different, since $\tr \mathcal X_A = \tr \mathcal Y_B =D^ n$.
Therefore we obtain
\begin{eqnarray}
\left.\overline{C_{2n}}\right|_{D\rightarrow \infty}\simeq D^{-n\left(\left|\gamma_A\right|+\left|\gamma_B\right|\right)}\tr \left[\rho_b^{\otimes 2n} (\mathcal X_{E_A}\otimes \mathcal Y_{E_B}) \right].\label{C2nLargeD}
\end{eqnarray}
Here $\mathcal X_{E_A}$ and $\mathcal Y_{E_B}$ are the same permutations as $\mathcal X_A$ and $\mathcal Y_B$ in Eq.~(\ref{permutationsforM}), respectively, but acting in the bulk regions $E_A$ and $E_B$. Interestingly, the bulk state contribution $\tr[\rho_b^{\otimes 2n} (\mathcal X_{E_A} \otimes \mathcal Y_{E_B})]$ is exactly the same expression as $C_{2n}$ in Eq.~(\ref{C2n}), but for the bulk state $\rho_b$. In other words, we can define an orthonormal basis $\phi_{E_A}^\alpha$ and $\phi_{E_B}^\beta$ in the bulk regions $E_A$ and $E_B$, and define the bulk correlation matrix
\begin{eqnarray}
M^{\alpha\beta}_b=\tr \left[\rho_b \, \phi_{E_A}^\alpha \phi_{E_B}^\beta\right].
\end{eqnarray}
Following the same derivation as Eq.~(\ref{C2ndef}) and Eq.~(\ref{C2n}) we obtain the bulk correlation moments
\begin{eqnarray}
C_{2n}^{\text{bulk}}={\rm tr}\left[\left(M_b^\dagger M_b\right)^n\right]=\sum_{s}\lambda_{s,\text{bulk}}^{2n}=\tr \left[\rho_b^{\otimes 2n} (\mathcal X_{E_A}\otimes \mathcal Y_{E_B}) \right],
\end{eqnarray}
where $\lambda_{s,bulk}$ are the singular values of the bulk correlation matrix $M_b$. (Note that
$\tr \left(\rho_b\right)=1$ so that the denominator for
$C_{2n}^{\text{bulk}}$ is trivial.)  Therefore Eq.~(\ref{C2nLargeD})
can be interpreted as the following relation between the boundary
correlation matrix and the boundary one:
\begin{eqnarray}
\overline{\sum_s\lambda_s^{2n}}&\simeq &D^{-n\left(\left|\gamma_A\right|+\left|\gamma_B\right|\right)}\sum_s\lambda_{s,\text{bulk}}^{2n}\nonumber\\
\Rightarrow \lambda_s&\simeq &D^{-\frac12\left(\left|\gamma_A\right|+\left|\gamma_B\right|\right)}\lambda_{s,\text{bulk}}.\label{bulkboundaryrelation}
\end{eqnarray}
Therefore in large $D$ limit, the singular values of the boundary correlation matrix $M^{\alpha\beta}$ are given by those of the bulk correlation matrix $M^{\alpha\beta}_b$ between the two entanglement wedges $E_A$ and $E_B$, multiplied by a constant factor that is independent of the distance between the two regions (as long as their joint entanglement wedge $E_{AB}$ stays disconnected).

Eq.~(\ref{bulkboundaryrelation}) has several important consequences. Firstly, it tells us that in a proper basis choice, there is a one-to-one correspondence between bulk two-point correlators and boundary ones. When we take the limit that both $A$ and $B$ are small (compared with the extrinsic curvature radius of the boundary), the entanglement wedges $E_A$ and $E_B$ become narrow regions near the boundary. In this limit the two-point correlation functions between $E_A$ and $E_B$ can be viewed as the boundary limit of bulk two-point functions. Therefore Eq.~(\ref{bulkboundaryrelation}) shows that the boundary two-point functions between local operators are, up to a constant prefactor, equal to the bulk two-point functions with both points approaching the boundary. In other words, the holographic mapping defined by a random tensor network gives a bulk-boundary correspondence consistent with the usual ``dictionary'' of holographic duality. Secondly, if $\rho_b$ is taken to be independent from $D$, the bulk correlation spectrum $\left\{\lambda_{s,\text{bulk}}\right\}$ is $D$-independent. Therefore the boundary correlation spectrum is also $D$-independent (except for the prefactor), although the total number of operators in $A$ and $B$ are both increasing with $D$.

To understand the consequences of this observation more explicitly, we consider the special case that the bulk is the Poincar\'{e} patch of hyperbolic space, and the state $\rho_b$ is also invariant with respect to the isometry group of the bulk geometry. If we take the limit of small $A$ and $B$ (much smaller than the AdS radius), the bulk entanglement wedges approach the boundary, and the bulk two-point functions all decay as a power law of the boundary distance due to scale invariance. Therefore, in this limit
\begin{equation}
\lambda_{s,\text{bulk}}=\frac{C_s}{|x-y|^{\Delta_s}},
\end{equation}
with $\left\{\Delta_s\right\}$ defining the spectrum of scaling dimensions. According to Eq.~(\ref{bulkboundaryrelation}), the boundary two-point functions also decay as a power law, with the same set of scaling dimensions. Compared with the situation in the AdS/CFT corresondence, we see that the boundary operators with scaling dimension $\Delta_s$ are analogs of low-dimensional operators with scaling dimensions independent of $N$. The number of low-dimensional operators is determined by the bulk theory.

It is natural to ask whether there are also high-dimensional operators in the random tensor network state, which are the analog of ``stringy'' operators in AdS/CFT with scaling dimensions growing with $N$. To address this question, one needs to consider the finite $D$ fluctuations. In the following we will provide some arguments about finite $D$ corrections to the correlation spectrum which are not rigorous but may be helpful for physical understanding. At finite $D$ the partition function $\overline{Z_1^{(2n)}}$ receives a contribution from other spin configurations with higher energy. Many low energy spin configurations are separate deformations of the domain walls bounding $A$ and that bounding $B$. Although such fluctuations will renormalize the correlation functions, they do not change the correlation length since there is no distance dependence. The lowest energy configuration which contributes nontrivially to the distance dependence of correlation function is the one shown in Fig.~\ref{fig:correlationmatrix}\,(b). This configuration contains two domains $E_1$ and $E_2$ with permutations the same as $\mathcal X_A$ and $\mathcal Y_B$ in Eq.~(\ref{permutationsforM}). The boundary of $E_1\cup E_2$ consists of the connected geodesics bounding $A$ and $B$, and the interface between $E_1$ and $E_2$ is chosen as the narrowest ``throat'' between the two geodesics. There are many configurations with similar energy, so that it is difficult to give a quantitative estimate of the finite $D$ correction to $C_{2n}$. However as a rough estimate if we only consider the contribution of this configuration, we obtain
\begin{eqnarray}
\overline{C_{2n}}\sim D^{-n\left(\left|\gamma_A\right|+\left|\gamma_B\right|\right)}C_{2n}^{\text{bulk}}+
{\text{const.}}\times D^{-2n \, d_{AB}}.
\end{eqnarray}
The constant term is of order $1$, given by $D^{-(2n-2)|W|}\tr [\rho_b^{\otimes 2n} (\mathcal X_{E_1}\otimes \mathcal Y_{E_2})]$, with $|W|$ the width of the throat. Since $E_1$ and $E_2$ are adjacent to each other, the bulk correlation term $\tr[\rho_b^{\otimes 2n} (\mathcal X_{E_1}\otimes \mathcal Y_{E_2})]$ will be dominated by short-range correlations, and thus does not decay with the distance $d_{AB}$. For hyperbolic space, at long distance $d_{AB}\propto \frac1{l_g}\log |x-y|$, with $l_g$ the discretization scale. Therefore the finite $D$ correction due to this domain configuration contributes new power laws with the scaling dimension $\Delta=\frac1{l_g}\log D$. With this new contribution to $C_{2n}$, the scaling dimension spectrum of the boundary state now contains $\left\{\Delta_s,~\frac1{l_g}\log D\right\}$, which consists of the low-lying scaling dimensions $\Delta_s$ that are $D$-independent, and the high scaling dimension that grows linearly with $\log D$. Such a separation in scaling dimensions is consistent with the requirement in AdS/CFT for CFT's with a gravitational dual, known as the scaling dimension gap~\cite{heemskerk2009,elshowk2012,benjamin2015}. Although the analysis here is clearly incomplete, it is reasonable to believe that the separation of two types of operators remains valid in a more detailed analysis, since there are two different origins of power law correlations, those from the bulk state and those from the spin fluctuations in the classical statistical model.


\section{Fluctuations and corrections for finite bond dimension}\label{sec:finiteD}

In preceding sections, we have shown that the unnormalized state averages $\overline{Z_1^{(n)}}$ and $\overline{Z_0^{(n)}}$ are mapped to Ising partition functions with inverse temperature $\beta=\log D$ and different boundary conditions.
In the large $D$ limit, these Ising partition function are dominated by the contribution of the lowest energy spin configuration, which gave rise to the Ryu-Takayanagi formula for the R\'{e}nyi entropies assuming that $S_n(A) = \log Z_1^{(n)} / Z_0^{(n)} \simeq \log \overline{Z_1^{(n)}} / \overline{{Z_0^{(n)}}}$.
In this section we will make this step precise 
and quantify how well the R\'{e}nyi entropies $S_n(A)$ are approximated by the Ryu-Takayanagi formula. Before going into the details, we first present our conclusion:

\begin{enumerate}
\item
For a system with volume ({\it i.e.}, number of bulk vertices) $V$, for an arbitrary small deviation $\delta>0$, one can define a critical bond dimension
\begin{equation*}
D_c=\alpha\delta^{-2}e^{c_{2n}V},
\end{equation*}
with $\alpha$ and $c_{2n}$ constants independent from the volume.
The meaning of the exponent $c_{2n}$ will be explained below.
In the limit $D\gg D_c$ the deviation satisfies
\begin{equation}
\label{eq:what we show}
\lvert S_n(A) - S_n^{RT}(A) \rvert < \delta,~\text{with a high probability}~P(\delta)=1-\frac{D_c}{D},
\end{equation}
where
\begin{equation}
\label{eq:rt nice}
    S_n^{RT}(A) \equiv \log D \, \lvert \gamma_A \rvert + S_n(E_A; \rho_b)
\end{equation}
is the RT formula for the $n$-th R\'{e}nyi entropy, including the bulk correction.
We will always assume that the bulk dimension $D_b$ is finite, so that in the large $D$ limit the minimal surface $\gamma_A$ is determined by minimizing the area.
\item
We subsequently show that under a plausible physical assumption on the free energy of the statistical models, the bound given in Eq.~(\ref{eq:what we show}) can be improved by reducing the critical bond dimension to
\begin{equation*}
D_c=\alpha'\delta^{-2}V^{2/\Delta_{2n}},
\end{equation*}
with $\alpha'$ a non-universal constant.
The meaning of the exponent $\Delta_{2n}$ will be explained below.
\end{enumerate}

\subsection{The general bound on fluctuations}

To start, we denote the Ising action \eqref{SnAction} of the minimal energy spin configuration with a boundary field $h_x$ by $\mathcal A^{(n)}_{\min}\left[\left\{h_x\right\}\right]$.
We shall assume throughout this section that the minimal energy configuration is unique (otherwise see Section~\ref{sec:stabs} and Appendix~\ref{app:second-moments}).
In the large $D$ limit, $\overline{Z_{0,1}^{(n)}}$ approaches $Z^{(n),\infty}_{0,1} \equiv e^{-\mathcal A^{(n)}_{\min}[h_{0,1}]}$, with $h_{0,1}$ denoting the boundary field configuration for the calculation of $\overline{Z_0^{(n)}}$ and $\overline{Z_1^{(n)}}$, respectively. We note that
$\mathcal A^{(n)}_{\min}[h_1] = (n-1) \log D \lvert \gamma_A \rvert + (n-1) S_n(E_A;\rho_b) + \log C_{n,x}$
and
$\mathcal A^{(n)}_{\min}[h_0] = \log C_{n,x}$, with $C_{n,x}$ defined in Eq.~(\ref{nStateaverage}) and the text below it.
Thus the RT formula~\eqref{eq:rt nice} can also be written as
\begin{equation}
\label{eq:rt in terms of min energies}
  S_n^{RT}(A)
  = -\frac 1 {n-1} \log \frac {Z_1^{(n),\infty}} {Z_0^{(n),\infty}}
  = \frac 1 {n-1} \Big( \mathcal A^{(n)}_{\min}[h_1] - \mathcal A^{(n)}_{\min}[h_0] \Big).
\end{equation}
To bound the fluctuations of $Z_1^{(n)}$ away from $Z_1^{(n),\infty}$, we consider
\begin{equation}
\label{eq:fluct first}
      \overline{\left( \frac {Z_1^{(n)}} {Z^{(n),\infty}_1} - 1 \right)^2}
    = \left[ \frac {\overline{(Z_1^{(n)})^2}} {(Z^{(n),\infty}_1)^2} - 1 \right]
    - 2 \left[ \frac {\overline{Z_1^{(n)}}} {Z^{(n),\infty}_1} - 1 \right]
    \leq \frac {\overline{(Z_1^{(n)})^2}} {(Z^{(n),\infty}_1)^2} - 1,
\end{equation}
where we have used that $\overline{Z_1^{(n)}} \geq Z_1^{(n),\infty}$ since at finite temperature the partition function receives contributions from all spin configurations, not just the minimal energy configuration.

The key insight now is that the second moment of $Z_1^{(n)}$, $\overline{( Z_1^{(n)})^2}$, can be interpreted as the partition function of an $\Sym_{2n}$-spin model with boundary field $h_x=(1\dots{}n)(n\!+\!1\dots{}2n)$ for $x\in A$ and $h_x = I$ elsewhere, as was discussed at the end of Section~\ref{sec:nthRenyi}.
In the large $D$ limit, the lowest energy spin configuration is given by the same minimal energy surface as that in the Ising model for the $\overline{Z_1^{(n)}}$ calculation, with corresponding energy
\begin{equation}
    \mathcal A^{(2n)}_{\min}[\{h_x\}] = 2 \mathcal A^{(n)}_{\min}[h_1] + \sum_x \log \frac {C_{2n,x}} {C_{n,x}^2}. \label{eq:min action comparison}
\end{equation}
The last term comes from the different normalization factors in the average $\overline{Z_1^{(n)}}$ and $\overline{\left(Z_1^{(n)}\right)^2}$.
Thus the ground state energy of this $\Sym_{2n}$-spin model is essentially two times that of the $\Sym_n$-Ising model.
More precisely, it follows from \eqref{eq:min action comparison} that
\begin{equation}
\label{eq:fluct second}
    \frac {\overline{(Z_1^{(n)})^2}} {(Z^{(n),\infty}_1)^2} - 1
    = \frac {\overline{(Z_1^{(n)})^2}} {e^{-2\mathcal A^{(n)}_{\min}[h_1]}}
      - 1
    = \frac {\overline{(Z_1^{(n)})^2}} {e^{-\mathcal A^{(2n)}_{\min}[\{h_x\}]}}
      \left( \prod_x \frac {C_{n,x}^2} {C_{2n,x}} \right)
      - 1
    \leq \frac {\overline{(Z_1^{(n)})^2}} {e^{-\mathcal A^{(2n)}_{\min}[\{h_x\}]}}
      - 1.
\end{equation}
To bound the right-hand side term, we use that by assumption the minimal energy configuration is unique; all other configurations incur an additional energy cost of at least $\log D - (2n-1) \log D_b \, V$ (cf.\ the discussion before Eq.~\eqref{eq:nth renyi with bulk}).
Since there are $(2n)!$ configurations at each bulk site this leads to the conservative upper bound
\begin{equation}
\label{eq:conservative ising}
    \frac {\overline{(Z_1^{(n)})^2}} {e^{-\mathcal A^{(2n)}_{\min}[\{h_x\}]}} - 1 \leq ((2n)!)^V \frac {D_b^{(2n-1)V}} D = \frac {e^{c_{2n} V}} D,
\end{equation}
where $c_{2n} \equiv \log\,(2n)! + (2n-1) \log D_b$.
By combining Eqs.~\eqref{eq:fluct first}, \eqref{eq:fluct second} and \eqref{eq:conservative ising} we obtain that
\begin{equation}
\label{eq:conservative fluct bound}
    \overline{\left( \frac {Z_1^{(n)}} {Z^{(n),\infty}_1} - 1 \right)^2} \leq \frac {e^{c_{2n} V}} D.
\end{equation}
The same conclusion holds for $Z_0^{(n)} = (\tr \rho)^n$ (corresponding to a boundary field with $h_x = I$ everywhere).
By Markov's inequality, it follows that
\begin{equation}
\label{eq:markov}
  \Prob\left( \left\lvert \frac {Z_1^{(n)}} {Z^{(n),\infty}_1} - 1 \right\rvert \geq \frac \delta 4 \right) \leq \frac {\overline{\left( \frac {Z_1^{(n)}} {Z^{(n),\infty}_1} - 1 \right)^2}} {\left( \frac \delta 4 \right)^2} = \frac {16} {\delta^2} \frac {e^{c_{2n} V}} D,
\end{equation}
and likewise for $Z_0^{(n)}$.
The union bound thus implies that both $\lvert Z_1^{(n)} / Z^{(n),\infty}_1 - 1 \rvert < \delta/4$ and $\lvert Z_0^{(n)} / Z^{(n),\infty}_0 - 1 \rvert < \delta/4$ with probability at least $1 - 32 \, e^{c_{2n} V} / D \delta^2$.
In this case we can bound the deviation of the $n$-th R\'{e}nyi entropy from the Ryu-Takayanagi formula \eqref{eq:rt in terms of min energies} by
\begin{align*}
    \lvert S_n(A) - S_n^{RT}(A) \rvert
    = &\frac 1 {n-1} \left\lvert \log \frac {Z_1^{(n)}} {Z_0^{(n)}} - \log \frac {Z_1^{(n),\infty}} {Z_0^{(n),\infty}} \right\rvert
    = \frac 1 {n-1} \left\lvert \log \frac {Z_1^{(n)}} {Z_1^{(n),\infty}} - \log \frac {Z_0^{(n)}} {Z_0^{(n),\infty}} \right\rvert \\
    \leq &\frac 1 {n-1} \left( \left\lvert \log \frac {Z_1^{(n)}} {Z_1^{(n),\infty}} \right\rvert + \left\lvert \log \frac {Z_0^{(n)}} {Z_0^{(n),\infty}} \right\rvert \right)
    \leq \frac 1 {n-1} \left( \frac \delta 2 + \frac \delta 2 \right) \leq \delta,
\end{align*}
where we have used $\delta \leq 2$ such that $\log(1\pm\delta/4) \leq \delta/2$.
We have thus proved that the desired bound \eqref{eq:what we show} holds with probability at least $1-\frac{D_c}{D}$, where $D_c=32\delta^{-2}e^{c_{2n}V}$.

Interestingly, the above results for the R\'{e}nyi entropies can be used to show corresponding statements for the von Neumann entropy.
For a bulk direct product state, this is easy to see: Here, the Ryu-Takayanagi formula amounts to $S_n^{RT}(A) \equiv \log D \, \lvert \gamma_A \rvert$.
Since $S(A) \geq S_n(A)$ for any quantum state and $S(A) \leq \log \rank \rho_A \leq \log D \lvert \gamma_A \rvert$ in any tensor network state, we have essentially matching upper and lower bounds for the von Neumann entropy, and hence $S(A) \simeq \log D \lvert \gamma_A \rvert$ with high probability.
This result can be established more generally even in the presence of an entangled bulk state as long as $D_b \ll D$ by adapting the techniques of~\cite{dutil2010one} (cf.~Section~\ref{sec:split-transfer}).

\subsection{Improvement of the bound under a physical assumption}

The above results establish rigorously that the entropies approximate the Ryu-Takayanagi formula in the limit of large $D$.
However, the technique lead to a rather conservative estimate of the finite $D$ correction, since it only proves that the entropy is close to the RT value for exponentially large bond dimension $D \gg e^{c_{2n} V}$.
In this subsection we would like to argue based on a plausible physical assumption that actually the RT formula applies to a much larger range of $D$, as long as $D$ is bigger than some power law function of $V$.

To start, let us reinvestigate Eq.~(\ref{eq:conservative ising}), which was the basis of the general bound~(\ref{eq:what we show}). In obtaining Eq.~(\ref{eq:conservative ising}) we replaced the energy of all higher energy spin configurations by their minimum $\log \frac{D}{D_b^V}$. This leads to a very conservative bound since most configurations certainly have an energy much higher than that. Since the statistical model has a local action, the number of excitations with lowest energy is actually proportional to $V$ rather than exponential of $V$. Although the number of slightly higher energy excitations are super-extensive, it is still true that the free energy of the spin model is extensive at finite temperature. Furthermore, the free energy approaches the ground state energy in the lo temperature (large $D$) limit exponentially, since the probability of lowest energy excitation with energy $E_g$ is suppressed by the Boltzman weight $e^{-\beta E_g}=D^{-E_g/2}$. Using these plausible physical observations we can write the asymptotic form of the free energy $F=-\log\overline{\left(Z_1^{(n)}\right)^2}\simeq \mathcal A^{(2n)}_{\min}[\{h_x\}]-C(\log D)^aD^{-E_g/2}V$, with $C$ a constant. Note that there is generically a power law term $(\log D)^a$ multiplying the exponential factor in the free energy density. However, this power law correction is not important for our bound, since we can choose an energy $\Delta_{2n}$ slightly smaller than $E_g$, such that
\begin{equation}
\label{eq:physical ising}
  -\log\overline{\left(Z_1^{(n)}\right)^2}\geq \mathcal A^{(2n)}_{\min}[\{h_x\}]-CD^{-\Delta_{2n}/2}V.
\end{equation}
In the limit $D \gg V^{1/\Delta_{2n}}$ this implies that
\begin{equation*}
\frac {\overline{(Z_1^{(n)})^2}} {e^{-\mathcal A^{(2n)}_{\min}[\{h_x\}]}} - 1 \leq \frac {C'V} {D^{\Delta_{2n}/2}}
\end{equation*}
by choosing a constant $C'$ slightly larger than $C$. If we substitute this estimate for the conservative bound \eqref{eq:conservative ising} then Eq.~\eqref{eq:conservative fluct bound} becomes
\begin{equation*}
    \overline{\left( \frac {Z_1^{(n)}} {Z^{(n),\infty}_1} - 1 \right)^2} \leq \frac {C'V} {D^{\Delta_{2n}/2}},
\end{equation*}
and likewise for $Z^{(n)}_0$.
We may now proceed as above and conclude that, under the assumption \eqref{eq:physical ising} on the Ising models, the R\'{e}nyi entropies satisfy the RT formula to arbitrary precision and with arbitrarily high probability if $D \gg V^{2/\Delta_{2n}}$.
This improves the dependency of the bond dimension on the system size from an exponential function of $V$ to a power law.

\medskip

To illustrate the behavior of the free energy~\eqref{eq:physical ising} in an explicit example, we consider an Ising model on the square lattice.
(It should be noted that the Ising model case does not directly apply to the discussion above since ${\rm Sym}_{2n}$-spin models are used there.
However the behavior of free energy is generic for gapped spin models.)
Specifically, we shall consider a cylindrical geometry given by an $M\times N$ square lattice with periodic boundary conditions along the first direction and open boundary conditions along the second one.
In this setup, the minimal surface bounding a boundary region is unique.
As the boundary region we choose a single interval of length $L < M$.
${\overline{Z_1}} / {Z_1^\infty}$ and ${\overline{Z_0}} / {Z_0^\infty}$ can be computed exactly using Onsager's solution~\cite{Onsager1944,McCoy1967}.
The asymptotic behavior for large $D$ is given by
\begin{align*}
  \frac{\overline{Z_0}}{Z_0^\infty} &\simeq 1+D^{-4}MN+o(D^{-4}), \\
  \frac{\overline{Z_1}}{Z_1^\infty} &\simeq 1+2LD^{-1}+o(D^{-1}).
\end{align*}
Therefore Eq.~\eqref{eq:physical ising} holds with exponent $\Delta_{2n}=2$.
More details about this calculation are presented in Appendix~\ref{app:Ising}.

\subsection{Possible effects of even smaller bond dimension}

When $D$ does not satisfy the condition $D\gg (CV)^{1/\Delta}$, the deviation of the entropy from the RT value can be large.
An interesting question is whether the correction to the RT formula is simply a renormalization of the coefficient of the area law, or if there is a qualitative change.
For the second R\'{e}nyi entropy, the quantity $-\log {\overline{Z_1}}/{\overline{Z_0}}$ is the free energy cost induced by the boundary pinning field $h_x$ in the Ising model.
The behavior of this free energy cost depends on the strength of the fluctuations of the domain wall configuration.
If the domain wall only fluctuates mildly around the minimal energy configuration, one can naturally expect the energy cost of the domain wall is still proportional to its area, although the coefficient may be renormalized to be different from the bare value given by the lowest energy configuration.
In contrast, if the domain wall is strongly fluctuating, the energy of the domain wall may have a qualitatively different dependence in the minimal area $\left|\gamma_A\right|$.
Interestingly, the behavior of the domain wall in the Ising model was studied a long time ago.
If the bulk spatial dimension $d\geq 3$, it was found that there is a finite critical temperature $T_r$ (which is lower than the phase transition temperature $T_c$ of the Ising model), below which the fluctuations of a domain wall configuration have a finite range.
The transition at $T_r$ is known as the roughening transition~\cite{burton1949,burton1951}.\footnote{We would like to thank Steven Kivelson for teaching us this result.}
It is natural to expect that the RT formula for second R\'{e}nyi entropy applies for any $\log D>T_r^{-1}$, which remains finite even if the system size $V$ goes to infinity.
However, it is not clear how to bound the deviation of $\overline{S_2(A)}$ from $-\log\frac{\overline{Z_1}}{\overline{Z_0}}$, given by the fluctuation terms in Eq.~(\ref{S2approximation}).
When the bulk spatial dimension is $d=2$, the domain wall is one-dimensional, and thus the fluctuation of its position is always strong.
Consequently, the RT formula does not apply to any finite $D$ if we take $V\rightarrow \infty$ first.

\section{Relation to random measurements and the entanglement of assistance}
\label{sec:split-transfer}

\begin{figure}
\begin{center}
\includegraphics[width=0.92\textwidth]{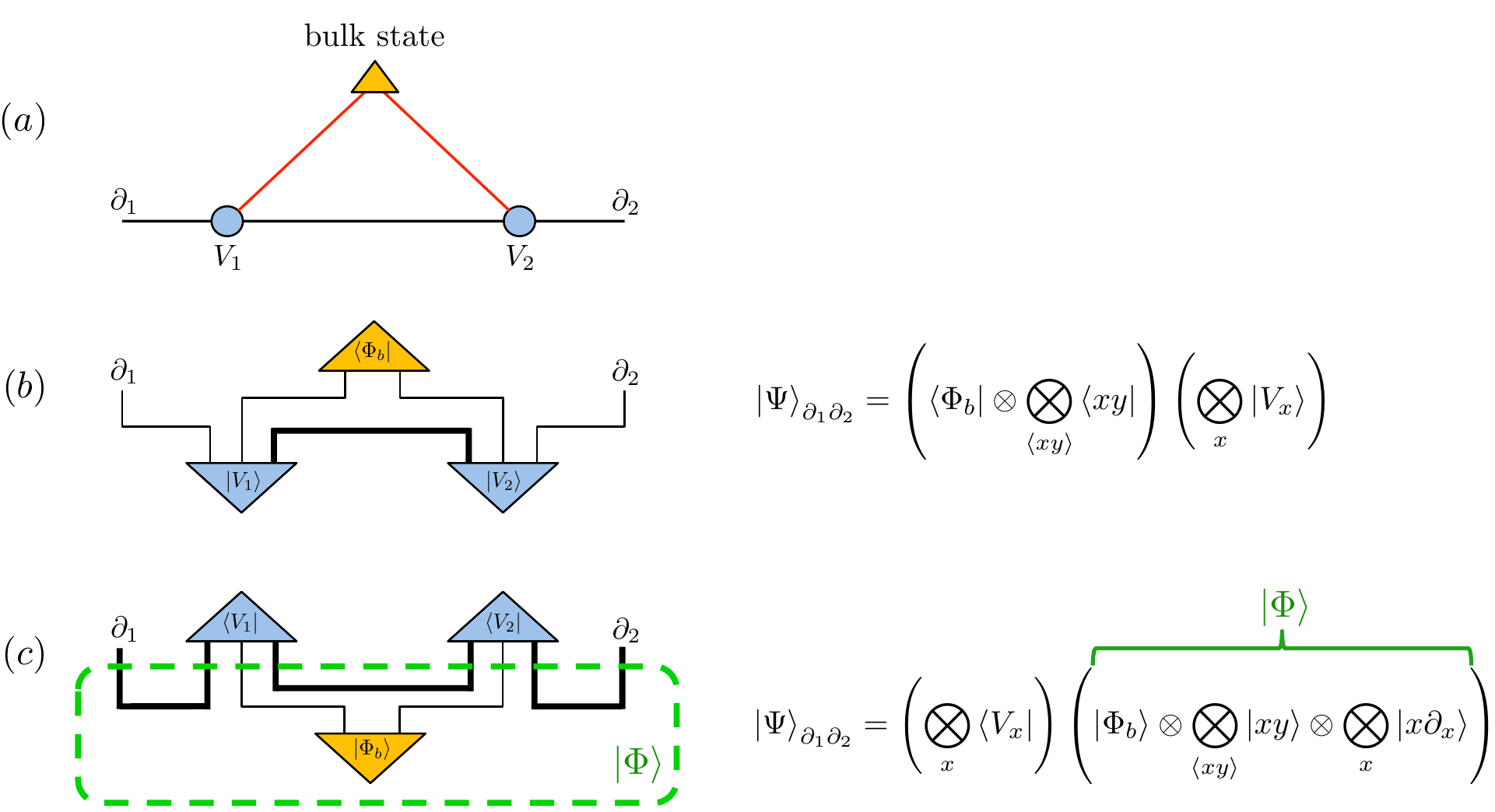}
\caption{{(a) A simple tensor network consisting of two vertices $V_1$ and $V_2$ of degree 3, a pure bulk state and two boundary dangling legs $\partial_1$ and $\partial_2$. (b) The construction of the Section~\ref{sec:setup}. The state of the random tensors $\bigotimes_x \ket{V_x}$ is contracted with Bell pairs and the bulk state to obtain the boundary state $\ket \Psi _{\partial_1\partial_2}$ on the Hilbert spaces of $\partial_1 \partial_2$. Bell pairs are shown by thick lines. (c) The construction of Section~\ref{sec:split-transfer}. Here we have a large background state $\ket \Phi$ and contract with random vertex states to obtain the state $\ket \Psi_{\partial_1 \partial_2}$ on the boundary. Note that we need to add extra boundary Bell pairs $ \bigotimes_x \left |  x \partial_x \right \rangle $.}}\label{fig:twopres}
\end{center}
\end{figure}

The average over random tensors that has played a central role in this work has appeared previously in the quantum information literature, but with a very different motivation. The definition of the boundary state $\ket{\Psi}$ in Eq.~(\ref{eqn:boundary-state}) involves contracting the random vertex states $\bigotimes_x \ket{V_x}$ at the bulk vertices with a bulk state $\ket{\Phi_b}$ as well as a collection of Bell pairs $\bigotimes_{\langle xy\rangle} \ket{xy}$ for the internal edges and $\otimes_{x} \ket{x\partial_x}$ connecting boundary vertices to their boundary connecting points $\partial_x$.
To obtain a new physical interpretation for the state $\ket \Psi$, one can start with the state
\begin{equation} \label{eqn:bulk-boundary-state}
\ket{\Phi} = \ket{\Phi_b} \otimes  \bigotimes_{\langle xy\rangle} \ket{xy} \otimes \bigotimes_{x} \ket{x\partial_x},
\end{equation}
and perform a random measurement at every bulk vertex $x$. The post-measurement state on the unmeasured boundary vertices will then have the same distribution as $\ket{\Psi}$.
Note that the state $\ket \Phi$ in Eq.~(\ref{eqn:bulk-boundary-state}) is supplemented by new bulk-boundary Bell pairs $\bigotimes_{x} \ket{x\partial_x}$ as compared to Eq.~(\ref{eqn:boundary-state}). The reason lies in the change of perspective; in Section~\ref{sec:setup} the random vertex states were being projected to Bell pairs and the bulk state, but here the Bell pairs and the bulk state are being projected to the random states, and therefore we need a larger Hilbert space to get a non-empty Hilbert space after projection. See Fig.~(\ref{fig:twopres}).
These two perspectives are mathematically equivalent in our examples.

The post-measurement state on the unmeasured boundary vertices will then have the same distribution as $\ket{\Psi}$. From this point of view, boundary entanglement is being induced by performing a suitable measurement on a joint bulk-boundary state.

One of the basic problems of quantum information theory is how to establish as much high-quality entanglement as possible between spatially separated parties. One scenario that had been considered was to start with a pure state $\ket{\Phi}_{ABC}$ of three systems and to ask how much entanglement could be induced on average between $A$ and $B$ upon measuring $C$, optimized over all possible $C$ measurements. Because the party in possession of $C$ is helping $A$ and $B$ establish entanglement, this quantity is known as the \emph{entanglement of assistance} $E_A(A;B)_\Phi$~\cite{divincenzo1999entanglement}. Concavity of the entropy implies a trivial upper bound: $E_A(A;B)_\Phi \leq S(\Phi_A)$, and likewise for $Y$. In a remarkable paper, Smolin \emph{et al.}\ showed that this upper bound was asymptotically achievable~\cite{smolin2005entanglement}:
\begin{equation}
\lim_{k\rightarrow\infty} \frac{1}{k} E_A(A^k;B^k)_{\Phi^{\otimes k}} = \min[ S(\Phi_A), S(\Phi_B) ].
\end{equation}

Going further, one can imagine partitioning $C$ into subsystems $C_1, C_2, \ldots, C_m$ and allowing only local measurements of each $C_j$ instead of joint measurements of the entire $C$ system. From an engineering perspective, such a scenario could arise naturally if $A$ and $B$ are distant and the $C_j$ represent intermediate ``repeater'' stations in a network~\cite{dutil2011assisted}. The additional locality restrictions will reduce the amount of entanglement that can be induced between $A$ and $B$. While the concavity upper bound still applies, it can be applied here with a bit more finesse. If we choose any subset $S \subseteq \{C_1, \ldots, C_m\}$, then the bound implies that this multipartite version of the entanglement of assistance, $E^{\text{multi}}_A(A;B)_\Phi$, will be bounded above by $S(\Phi_{AS^c})$ since the total entanglement generated between $A$ and $B$ will be no more than the entanglement between $AS^c$ and $B$ after measuring $S$ but prior to measuring $S^c$. Likewise, $E^{\text{multi}}_A(A;B)_\Phi \leq S(\Phi_{BS^c})$. Therefore
\begin{equation}
E^{\text{multi}}_A(A;B)_\Phi \leq  \min_{S \subseteq \{C_1,\ldots, C_m\}} \min[ S(\Phi_{AS}), S(\Phi_{BS}) ]
	= \min_{S \subseteq \{C_1,\ldots, C_m\}} S(\Phi_{AS}),
\end{equation}
where the equation follows from the fact that the entropies of two complementary subsystems of a pure state are always the same.
The Smolin \emph{et al.}\ result applied inductively gives in turn~\cite{smolin2005entanglement}
\begin{equation} \label{eqn:smolin-multi}
\lim_{k\rightarrow\infty} \frac{1}{k} E^{\text{multi}}_A(A^k;B^k)_{\Phi^{\otimes k}}
	= \min_{S \subseteq \{C_1,\ldots, C_m\}} S(\Phi_{AS}).
\end{equation}

Consider now the special case in which $\ket{\Phi}$ has the form of Eq.~(\ref{eqn:bulk-boundary-state}) used in this paper. Set $A$ to be any boundary region, $B$ the complement $A^c$ of $A$ in the boundary and identify the different subsystems $C_j$ with the bulk vertices $x$. The righthand side of Eq.~(\ref{eqn:smolin-multi}) is then nothing other than the Ryu-Takayanagi formula with corrections due to the bulk state $\ket{\Phi_b}$, since minimizing over subsets $S$ amounts to minimizing over cuts in the tensor network:
\begin{equation} \label{eqn:multi-correct}
\min_{S \subseteq \{C_1,\ldots, C_m\}} S(\Phi_{AS}) = |\gamma_A | \log D + S(E_A; \Phi_b),
\end{equation}
where $E_A$ is the bulk region corresponding to the minimizing set $S$ and $|\gamma_A|$ is the size of the cut separating $E_A$ from its complement.
This matches Eq.~(\ref{eq:second renyi}) up to the substitution of the von Neumann entropy for the second R\'enyi entropy. (The reason for taking the $k \rightarrow \infty$ limit in~(\ref{eqn:smolin-multi}) is essentially to make all R\'enyi entropies equal after suitable small perturbations to the state. For reasonable physical choices of $\ket{\Phi_b}$ such as quantum field theory ground states, it should be sufficient to take $k=1$ and include a small correction on the righthand side of $(\ref{eqn:smolin-multi})$. This has been shown, for example, in the case that $A$ is an interval in a 1+1 dimensional CFT~\cite{czech2014}.)

While the original proof of the multipartite entanglement of assistance formula used classically-inspired random coding information theory techniques, subsequent proofs proceeded by performing appropriate isotropic measurements of the $C$ subsystems~\cite{horodecki2007quantum,dutil2010one}. Because of the equivalence between contracting random tensors and performing random measurements, the analyses in the quantum information theory literature are mathematically very similar to the calculations in this article. The analog of the calculations justifying reconstruction of a bulk operator contained in the entanglement wedge of a boundary region $A$ has even appeared, again with a different motivation, as the ``split-transfer protocol''~\cite{dutil2010one,dutil2011multiparty}.

One could go as far as to rename the one-shot multipartite entanglement of assistance formula of~\cite{dutil2011multiparty} the ``fully-quantum Ryu-Takayanagi'' formula, in that it captures the essence of Ryu-Takayanagi without making any prior assumptions about the geometrical interpretation of the bulk state.
Aside from connecting to pre-existing literature, one virtue of this change of perspective is that it suggests a possible physical justification for the random tensor networks in our model. One could imagine taking the state in a quantum theory of gravity and measuring the Planckian degrees of freedom of a large ``bulk'' subset, leaving some ``boundary'' degrees of freedom and bulk fields unmeasured. If the Hilbert spaces are large and the measurements generic, then the measurements should reveal almost no information about the bulk, inducing a nontrivial mapping between non-Planckian bulk degrees of freedom and the boundary. In this way, fixing the bulk Planckian degrees of freedom in the bulk-boundary state through measurement generically produces a holographic correspondence. Expanding around a particular background geometry in this picture amounts to choosing a bulk-boundary state with the correct area law entropy and randomly fixing the Planckian degrees of freedom through projection.

\section{Random tensor networks from 2-designs}\label{sec:stabs}

In the construction of our random tensor network state~\eqref{peps}, the tensors $\ket{V_x}$ were chosen to be Haar-random, {\it i.e.}, drawn from to the unitarily invariant ensemble of pure states.
However, our calculations for the second R\'{e}nyi entropy in Section~\ref{sec:secondRenyi} made use only of the second moments of the Haar measure.
This calculation led to the emergence of a classical Ising model and thereby to the Ryu-Takayanagi formula.
It is therefore natural to consider other ensembles of pure states whose first two moments agree with those of the Haar measure, known collectively as complex projective \emph{2-designs}~\cite{renes2003symmetric}.

It follows from the discussion in Section~\ref{sec:finiteD} that for a tensor network state with Haar-random tensors and bulk direct product state $\rho_b$, the Ryu-Takayanagi formula $S(A) \simeq \log D \lvert \gamma_A \rvert$ will be satisfied with high probability in the limit of large $D$ if the minimal geodesic is unique.
This conclusion was obtained from considering higher moments of the Haar measure and therefore does not apply for a general 2-design.
Another complication arises from the fact that the tensor network state can be zero (i.e., $\rho=0$) with nonzero probability, in which case its entropies are not well-defined.
In Appendix~\ref{app:second-moments} we show that for any 2-design the boundary state is nonzero with high probability and that, moreover,
\begin{equation}
\label{eq:two design}
  S_{RT}(A) - \log k - o(1) \leq \nav {S_2(A)}\leq S(A) \leq S_{RT}(A),
\end{equation}
where $k$ denotes the number of minimal geodesics, where $S_{RT}(A) = \log D \lvert \gamma_A \rvert$ since we consider the case of a direct product bulk state, and where we write $\nav{S_2(A)}$ for the average second R\'enyi entropy conditioned on the boundary state being nonzero.
We note that, since the lower bound in \eqref{eq:two design} matches the \emph{deterministic} upper bound up to a constant, it follows that $S(A)$ is at most constantly away from $S_{RT}(A)$ with high probability.

One random ensemble of particular interest is given by stabilizer states.
Stabilizer states, defined as common eigenvectors of generalized Pauli operators, are quantum states that can be highly entangled, but whose particular algebraic structure allows for efficient simulation and effective reasoning~\cite{gottesman97}.
It has been shown in~\cite{chau2005unconditionally} that pure stabilizer states in prime power dimension $D=p^n$ form a 2-design when drawn uniformly at random.
Thus \eqref{eq:two design} applies to the entropies of the corresponding tensor network state~\eqref{peps} constructed from random stabilizer states.
Such a state is again a stabilizer state, as we argue in Appendix~\ref{app:stabcontract}.
The particular algebraic structure of stabilizer states implies that their reduced density matrices not only have flat spectrum (so that all R\'{e}nyi entropies agree with the von Neumann entropy) but in fact that all their entropies are quantized in units of $\log p$.
It follows that for large $p$ the Ryu-Takayanagi formula
\begin{equation}
\label{eq:exact rt}
  S(A) = S_{RT}(A) = \log D \lvert \gamma_A \rvert
\end{equation}
will hold \emph{exactly} with high probability (even in the presence of multiple minimal geodesics).

In particular, we may use this construction to obtain random holographic codes and evaluate their error correcting properties by using \eqref{ACisometry} and the exact Ryu-Takayanagi formula \eqref{eq:exact rt} purely from the structure of the tensor network.
In~\cite{pastawski2015}, holographic codes were constructed from perfect tensors, {\it i.e.}, states that are maximally entangled across any bipartition, and it was shown that under certain circumstances this already implies the Ryu-Takayanagi formula (such as for single intervals in nonpositively curved space).
Random stabilizer states are perfect tensors with high probability,\footnote{This follows as a special case of our result for a tensor network with a single vertex. We thank Fernando Pastawski for explaining an alternative proof of this fact to us.}
and so the analysis and results of~\cite{pastawski2015} can likewise be applied to our random tensor networks constructed from stabilizer states with high bond dimension.
However, our tensors are not only perfect or pluperfect~\cite{yang2015} but also generically so and therefore can achieve the Ryu-Takayanagi formula for arbitrary subsystems.

Another consequence of \eqref{eq:exact rt} is that any entropy inequality that is valid for arbitrary quantum states, or even just for stabilizer states~\cite{gross2013stabilizer,linden2013quantum}, is also valid for the Ryu-Takayanagi entropy formula, thereby establishing a conjecture from~\cite{bao2015cone}.
This can be understood as consistency check of the Ryu-Takayanagi formula, generalizing~\cite{headrick_takayanagi_2007}, where the validity of strong subadditivity was verified for the Ryu-Takayanagi formula.
We refer to~\cite{nezami2016multipartite} for a detailed analysis of the entanglement properties of tensor networks built from random stabilizer states.

\section{Conclusion and discussion}\label{sec:conclusion}

In this work we have studied the quantum information theoretic properties of random tensor networks with large bond dimension.
In the following we will revisit our method from a more general perspective and summarize our findings.
Viewing each tensor as a quantum state $\left|V_x\right\rangle$, the tensor network state $\rho = \rho\left(\left|V_x\rangle\langle V_x\right|\right)$ obtained by contracting these tensors is a linear function of each tensor.
Denote by $f_n(\rho)$ an arbitrary function that is a monomial function of the state $\rho$ with degree $n$.
Then the state average of $f_n$ over all possible choices of $\left|V_x\right\rangle$ is exactly mapped to the partition function of an classical spin model, with degrees of freedom in the permutation group $\Sym_n$, with the spins defined on the vertices of the same graph that underlies the tensor network.
Different physical quantities can be translated to different functions $f_n(\rho)$.
When the tensor network is used as a quantum state of the boundary, one can consider $\tr\left(\left[\tr_{\overline{A}}\rho\right]^n\right)$ for an arbitrary region $A$, which corresponds to the $n$-th Renyi entropy of $A$.
When the tensor network is used as a linear map, it can be viewed as a ``holographic mapping'' between two parts of the degrees of freedom (boundary and bulk, respectively).
In this case, in addition to the Renyi entropies one can study the entanglement entropy of a given region while another region is projected to a certain quantum state.
For example, one can project the bulk into a given quantum state and study the entanglement properties of the resulting boundary state.
We can also define basis-independent measures of correlation functions and relate that to a calculation of monomial functions, which allows us to study the behavior of two-point functions in the boundary state.

The mapping between the random state average and the spin-model partition function has rich consequences.
For a random tensor network state, in the large $D$ limit the Ryu-Takayanagi formula can be proven for all Renyi entropies, where the minimal surface area condition comes naturally from minimizing the energy of the spin model with given boundary conditions.
The Ryu-Takayanagi formula also generalizes naturally to include bulk state corrections when there is nontrivial quantum entanglement in the bulk.
As a particular example, we study the behavior of minimal surfaces in the presence of a bulk random state, and show how the minimal surface behavior can change topologically upon increase of the bulk entanglement entropy, in a way that is qualitatively consistent with black hole formation.
In addition to entanglement entropy, we also studied the behavior of two-point correlation functions.
The boundary correlation functions between two regions are directly determined by bulk correlation functions between two corresponding regions known as the entanglement wedges of the boundary regions.
In the special case of hyperbolic space, our results on correlation functions imply that the boundary theory has power law correlations with a large scaling dimension gap.
In the large $D$ limit there are two types of scaling dimensions, those which does not scale with $D$ coming from the bulk quantum state, and those which scale with $D$ coming from the tensor network contribution.
Such behavior of the scaling dimension gap is consistent with those of CFT ground states with a gravity dual, although the condition is necessary but not sufficient.

Random tensor networks provide a new framework for understanding holographic duality.
Besides the properties studied in this paper, many other physical properties can be evaluated by the mapping to classical spin models. Compared to other tensor network models, properties of the random tensor networks can be studied much more systematically. The large dimension $D$ limit is an analog of the large $N$ limit in gauge theories. The fact that a random tensor network with large dimension automatically satisfies many desired properties for holographic duality further supports the point of view that semi-classical gravity is deeply related to scrambling and chaos.

There are a several open questions that shall be studied in future works.
One question is whether it is possible to use a random tensor network to describe the ground state of a conformal field theory.
The underlying graph of random tensor networks on hyperbolic space is invariant under a subgroup of discrete isometries of the bulk which do not involve transformation in time.
Therefore we expect the distribution of tensor network states on the boundary to remain invariant under the subgroup of boundary conformal transformations that correspond to the bulk discrete isometries, modulo complications arising from the cut-off.
It is an open question whether we can modify the tensor network state to preserve the whole conformal symmetry.
Related to the discussion of R\'enyi entropies, this may require modification of the state on links between vertices.
It would also be interesting to consider random tensor network models where the same tensor is placed at each vertex.%
\footnote{After the first version of this manuscript had appeared, Matthew Hastings showed that for large $D$ the entanglement spectra of reduced density matrices have the same limiting behavior in both models~\cite{hastings2016asymptotics}. Therefore typical R\'enyi entropies in the model with identical tensors are also given by the Ryu-Takayanagi formula if $D$ is sufficiently large.}
Another question is how to generalize this formalism to include dynamics. What Hamiltonians of the boundary theory can be mapped to local Hamiltonians in the bulk ``low energy'' subspace? How to see that conserved currents on the boundary correspond to massless fields in the bulk? The answers to these questions will also be essential for understanding how the bulk gravity equation emerges.

\section*{Acknowledgments}

We would like to thank Xi Dong, David Gross, Daniel Harlow, Steve Kivelson, Juan Maldacena, Fernando Pastawski, Brian Swingle and Aron Wall for their helpful insights.
PH and MW gratefully acknowledge support from CIFAR, FQXI and the Simons Foundation.
SN is supported by a Stanford Graduate Fellowship.
XLQ and ZY are supported by the National Science Foundation through grant No.~DMR-1151786, and by the David and Lucile Packard Foundation.

\appendix

\section{Analytic study of the three phases for a random bulk state}\label{SBHTran}

In this appendix, we will provide an analytical explanation of when the transition happens between the perturbed AdS phase and the small black hole phase, and between the small black hole phase and the maximal black hole phase.
In particular, we will show
a) why at the transition between the perturbed AdS phase and the small black hole phase, $l_g^{-2}\log D_b$ scales as the square root of $l_g^{-1}\log D$,
b) why in the large $D$ limit, the transition between the small black hole phase to the maximal black hole phase happens at $l_g^{-2}\log D_b =  l_g^{-1}\log D (1+b^2)/(2b)$.

In fact, the problem we are going to solve has already been set up in Eq.~(\ref{SBHQ}).
The transition between the perturbed AdS phase to the small black hole phase is decided by the stability of the solution that covers half of the boundary system and goes through the center of the Poincar\'{e} disk.
Such a solution is the extremal solution of Eq.~(\ref{SBHQ}), since it minimizes the area contribution from the domain wall and maximizes the volume contribution from the bulk random state.
However, when this solution becomes a local maximum instead of a minimum, it means that the minimal surfaces of all the boundary regions would avoid the center of the Poincar\'{e} disk.
In other words, there exists a region in the bulk inaccessible to any measurements from the boundary smaller than half system size.

For convenience, we use $(x,y)$ coordinates instead of $(r,\theta)$ in this problem. Thus what we care about is
\begin{eqnarray*}
  \left.\frac{\delta^2 S_2(\pi/2)}{\delta y(x_1) \delta y(x_2)}\right|_{y=0}
  &=&l_g^{-1}\log D \left[ \frac{4}{(1-x_1^2)(1-x_2^2)} -\left.\frac{d}{dx}\left(\frac{2}{1-x^2} \delta^\prime(x-u)\right)\right|_{x=v} \right] \\
  &&- \left(l_g^{-2}\log D_b\right)^2\frac{2}{\left(1-x_1^2\right)^2}\frac{2}{\left(1-x_2^2\right)^2}\Theta(b-|x_1|)\Theta(b-|x_2|)
\end{eqnarray*}
where $\Theta(x) =1$ when $x>0$ and $0$ otherwise.
It is obvious that, if we treat the above expression as a matrix, the first term is always a positive definite matrix after integrating by parts of the derivative term, and the second term is a negative definite matrix, which corresponds to the fact that $y=0$ minimizes the area contribution from the domain wall and maximizes the volume contribution from the bulk random pure state.
Although it is hard to analytically diagonalize $\left.\frac{\delta^2 S_2(\pi/2)}{\delta y(x_1) \delta y(x_2)}\right|_{y=0}$, it is straightforward to observe that the instability happens at $l_g^{-1}\log D
\sim \left(l_g^{-2}\log D_b\right)^2$.

Now we turn to the second question, the transition between the small black hole phase and the maximal black hole phase.
In order to understand the formation of the maximal black hole, we need a more detailed investigation of Eq.~(\ref{SBHQ}). We first focus on the random pure state region $r\leq b$, and assume the minimal surface enters this region at angle $\varphi$ and $-\varphi$.
The minimization problem in Eq.~(\ref{SBHQ}) can be solved by asking
\begin{eqnarray*}
 0 &=&  l_g^{-1}\log D \int_{-\varphi}^{\varphi} d\theta ~\delta\left(\frac{2}{1-r^2(\theta)}\sqrt{\left(r^\prime(\theta)\right)^2+r^2(\theta)}\right) \\
&+&  l_g^{-2}\log D_b \frac{ D_b^{V_T/l_g^{2}}- D_b^{2V_{r(\theta)}/l_g^{2}}}{ D_b^{V_T/l_g^{2}}+D_b^{2V_{r(\theta)}/l_g^{2}}} \int_{-\varphi}^{\varphi} d\theta \frac{4r(\theta) \delta r(\theta)}{(1-r^2(\theta))^2}.
\end{eqnarray*}
The above variational equation contains both the derivative and the integration (contained in $V_{r(\theta)}$) of $r(\theta)$.
But in the large $D$ limit, which indicates that the transition happens when $D_b$ is also big, as long as $2 V_{r(\theta)} < V_T$, $D_b^{2V_{r(\theta)}} << D_b^{V_T}$ near transition point.
Thus in this limit, the above equation can be simplified with only $r(\theta)$ and its derivatives left.
\begin{eqnarray*}
   l_g^{-1}\log D \int_{-\varphi}^{\varphi} d\theta ~\delta\left(\frac{2}{1-r^2(\theta)}\sqrt{\left(r^\prime(\theta)\right)^2+r^2(\theta)}\right) +  l_g^{-2} \log D_b \int_{-\varphi}^{\varphi} d\theta \frac{4r(\theta) \delta r(\theta)}{(1-r^2(\theta))^2} = 0.
\end{eqnarray*}
The trick we use to solve this equation is to transform it back to a minimization problem, $I[r(\theta)]$ is the objective function to be minimized with respect to $r(\theta)$.
\begin{eqnarray*}
  &&I[r(\theta)] =  \int_{-\varphi}^{\varphi} d\theta \frac{2 l_g^{-1}\log D}{1-r^2(\theta)}\sqrt{\left(r^\prime(\theta)\right)^2+r^2(\theta)} + l_g^{-2}\log D_b\int_{-\varphi}^{\varphi} d\theta \left(\frac{2}{1-b^2} - \frac{2}{1-r^2(\theta)}\right)
\end{eqnarray*}
Because $I[r(\theta)]$ does not explicitly contain $\theta$, thus using a Legendre transformation, we only need to solve a first order differential equation.
\begin{eqnarray*}
  && r^\prime(\theta)\frac{\partial I[r(\theta)]}{\partial r^\prime(\theta)} - I[r(\theta)] = \frac{l_g^{-2}\log D_b}{1-r^2(\theta)} - \frac{r^2(\theta)l_g^{-1}\log D}{(1-r^2(\theta))\sqrt{\left(r^\prime(\theta)\right)^2+r^2(\theta)}} = C,
\end{eqnarray*}
whose analytic solution is
\begin{eqnarray}
  \nonumber&&  r^2(\theta)= \frac{1}{2 C^2} \Bigg[2 C  \left(C-l_g^{-2}\log D_b\right)\cos(2\theta) + \left(l_g^{-1}\log D\right)^2 \cos^2\theta\\
  \nonumber &&-\cos \theta  \sqrt{4 C \left(C-l_g^{-2}\log D_b\right)+\left(l_g^{-1}\log D\right)^2}
  \sqrt{4 C \left(l_g^{-2}\log D_b -C\right)\sin^2\theta +\left(l_g^{-1}\log D\right)^2 \cos^2\theta }\Bigg]\\
  \nonumber&&C = \frac{l_g^{-2}\log D_b}{(b^2-1)^2+4 b^2 \sin^2\varphi } \Bigg( 1 -b^2 \cos (2 \varphi ) \\
  \nonumber &&- \frac{b\cos\varphi}{l_g^{-2}\log D_b}\sqrt{  (b^2-1)^2 \left(l_g^{-1}\log D\right)^2 +4 b^2 \sin^2\varphi \left(\left(l_g^{-1}\log D\right)^2-\left(l_g^{-2}\log D_b\right)^2\right) }\Bigg) \label{solutbh}
\end{eqnarray}
where $C$ is fixed by asking $r(\pm \varphi) = b$.
In order for $r^2(\theta)$ not to be an extraneous root, we ask
\begin{eqnarray*}
  b^2= r^2(\varphi) \leq \frac{1}{2 C^2} \Bigg[2 C  \left(C-l_g^{-2}\log D_b\right)\cos(2\varphi) + \left(l_g^{-1}\log D\right)^2 \cos^2\varphi\Bigg]
\end{eqnarray*}
which can be satisfied if
\begin{equation*}
   l_g^{-2}\log D_b \leq \frac{1+b^2}{2b} l_g^{-1}\log D.
\end{equation*}
What is interesting is that this condition is independent of $\varphi$, the angle at which the minimal surfaces enters the random pure state region.
In other words, when $l_g^{-2}\log D_b \leq l_g^{-1}\log D (1+b^2)/2b $, the above solution $r(\theta)$ always exists for all angle $\varphi$, which means the minimal surfaces will enter the random pure state region.
However, when $l_g^{-2}\log D_b = l_g^{-1}\log D (1+b^2)/2b$ then for all $\varphi$ the minimal surfaces are repelled to the boundary of the random pure region, indicating that the formation of the single sided black hole is complete.
Thus we have proved in the large $D$ limit, the transition between the small black hole phase and the maximal black hole phase happens at $l_g^{-2}\log D_b = l_g^{-1}\log D (1+b^2)/2b $.

\section{Derivation of the error correction condition}
\label{app:qec}

In this appendix we give a short proof that the vanishing of the mutual information $I(C:B\overline{C})$, Eq.~\eqref{ACisometry}, implies that any operator $O_C$ in bulk region $C$ can be recovered from the boundary region $A$.
We do so for the reader's convenience as the proof will describe the construction of the boundary operator rather explicitly, but note that the result can be readily extracted from the literature~\cite{schumacher1996quantum,schumacher2002approximate,nielsen2007algebraic}.

\begin{figure}
\centering
\includegraphics[width=0.8\textwidth]{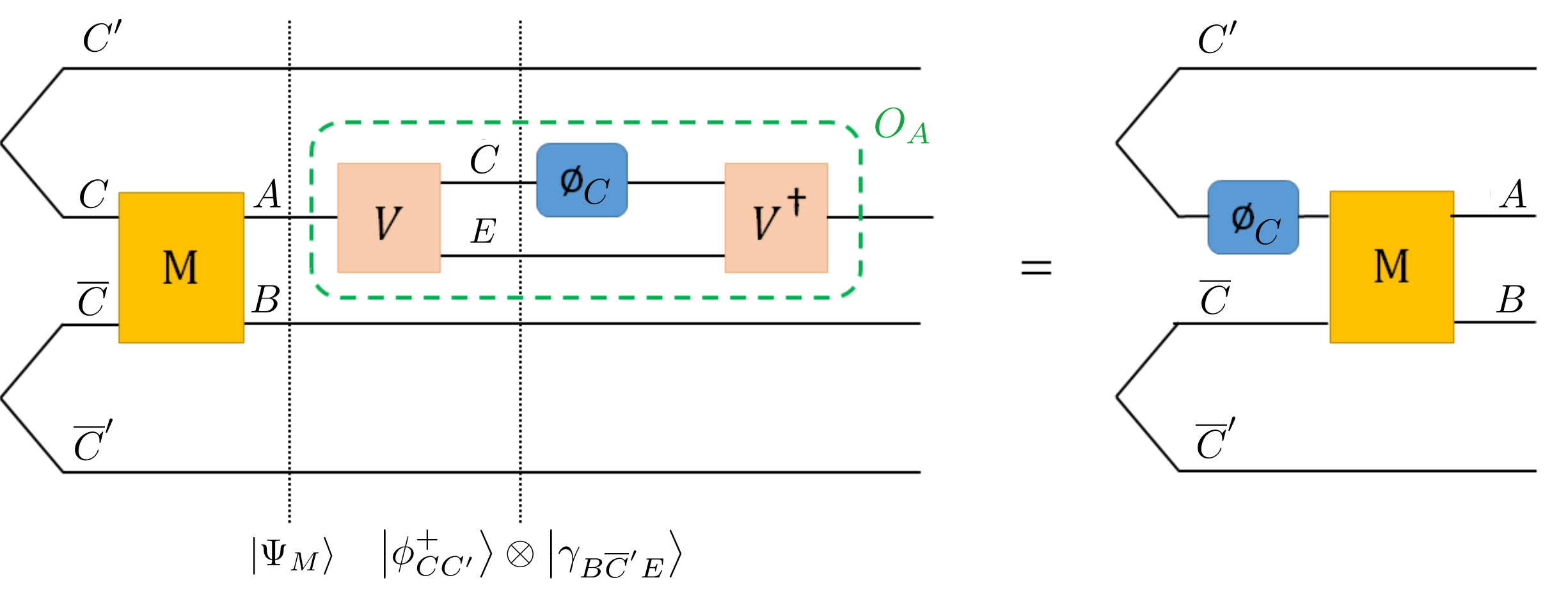}
\caption{Construction of the boundary operator $O_A$ corresponding to a bulk operator $\phi_C$ and illustration of the recovery equation~\eqref{errorcorrectioncond}.}
\label{fig:qec}
\end{figure}

In the following it will be crucial to distinguish the input systems $C$ and $\overline C$ of the bulk-to-boundary isometry $M$ from the corresponding subsystems of the pure state $\ket{\Psi_M}$ defined in \eqref{PsiM}.
We will thus denote the latter by $C'$ and $\overline C'$, so that
\[ \ket{\Psi_M} = M (\ket{\phi^+_{CC'}} \otimes \ket{\phi^+_{\overline C\, \overline C'}}), \]
where $\ket{\phi^+_{CC'}}$ and $\ket{\phi^+_{\overline C\, \overline C'}}$ denote maximally entangled states between $C$ and $C'$ and between $\overline C$ and $\overline C'$, respectively.
Eq.~\eqref{ACisometry}, which becomes $I(C':B\overline C') = 0$, implies at once that
\begin{equation}
\label{eq:qec reduced}
    \tr_A(\Psi_M) = \tau_{C'} \otimes \tr_{AC'}(\Psi_M),
\end{equation}
where $\tau_{C'} = \tr_{AB\overline C'}(\Psi_M)$ is a maximally mixed state (since $M$ is an isometry).
By definition, $\ket{\Psi_M}$ is a purification of \eqref{eq:qec reduced}, but we can also find a purification that respects the product structure $\ket{\phi^+_{CC'}} \otimes \ket{\gamma_{B\overline C'E}}$, obtained by purifying $\tau_{C'}$ to a maximally entangled state and $\tr_{AC'}(\Psi_M)$ to an arbitrary pure state $\ket{\gamma_{B\overline C'E}}$. If we choose the dimension of $E$ to be sufficiently large then the two purifications can be related by an isometry $V$ from $A$ to $CE$:
\begin{equation}
\label{eq:uhlmann}
    V \ket{\Psi_M} = \ket{\phi^+_{CC'}} \otimes \ket{\gamma_{B\overline C'E}}
\end{equation}
It can now be readily verified that any bulk operator $\phi_C$ can be recovered from $A$ by using the boundary operator $O_A = V^\dagger \phi_C V$.
Indeed, Eq.~\eqref{errorcorrectioncond}, which states that $O_A M = M \phi_C$, is a direct consequence of the following calculation:
\begin{equation*}
\begin{split}
    O_A \ket{\Psi_M} &= V^\dagger \phi_C V \ket{\Psi_M}
    = V^\dagger \phi_C (\ket{\phi^+_{CC'}} \otimes \ket{\gamma_{B\overline C'E}})
    = (\phi^T_{C'} \otimes V^\dagger) (\ket{\phi^+_{CC'}} \otimes \ket{\gamma_{B\overline C'E}}) \\
    &= \phi^T_{C'} \ket{\Psi_M}
    = (\phi^T_{C'} \otimes M) (\ket{\phi^+_{CC'}} \otimes \ket{\phi^+_{\overline C \,\overline C'}})
    = M \phi_C (\ket{\phi^+_{CC'}} \otimes \ket{\phi^+_{\overline C\, \overline C'}}),
\end{split}
\end{equation*}
where we have used \eqref{eq:uhlmann} and that $\phi_C \ket{\phi^+_{CC'}} = \phi^T_{C'} \ket{\phi^+_{CC'}}$ (twice). We refer to Fig.~\ref{fig:qec} for an illustration.

\section{Uniqueness of minimal energy configuration for higher R\'enyi models}
\label{app:unique}

In this appendix we give a formal proof of the assertion made in Section~\ref{sec:nthRenyi} that the spin model with action~\eqref{SnAction} has a unique minimal energy configuration, given by setting the entanglement wedge $E_A$ to the cyclic permutation $\mathcal C^{(n)}$ and its complement to the identity, provided the entanglement wedge $E_A$ is unique.
For simplicity, we assume that $D_{xy} = D_{x\partial} = D$ (but it is easy to see that the same conclusions hold true if all bond dimensions are powers of a fixed integer), and we consider the equivalent spin model with energy
\[ E[\{g_x\}] = \sum_{x,y} \bigl( n - \chi(g_x^{-1} g_y) \bigr), \]
where the $g_x$ are variables in $\Sym_n$, with $x$ and $y$ ranging over both bulk and boundary vertices, subject to the boundary conditions $g_x = \mathcal C^{(n)}$ in $A$ and $g_x = 1$ in $\bar A$ (cf.\ Section~\ref{sec:split-transfer}).

The first observation is that $n-\chi(g)$ is equal to the minimal number of transpositions (i.e., permutations that exchange only two indices) required to write a permutation $g$. This implies that
\[ d(g_x, g_y) := n - \chi(g_x^{-1} g_y) \]
defines a metric. In particular, it satisfies the triangle inequality.
The second ingredient is that, by the integral flow theorem, we can decompose a maximal flow between $A$ and $\bar{A}$ into edge-disjoint paths.
Each path starts in $A$, ends in $\bar{A}$, and by the max-flow/min-cut theorem there are $\lvert\gamma_A\rvert$ many such paths $P_1,\dots,P_{\lvert\gamma_A\rvert}$.

Now consider an arbitrary configuration $\{g_x\}$ that satisfies the boundary conditions.
We can bound its energy by looking only at those edges that occur in one of the paths, resulting in the lower bound
\begin{equation}
\label{eq:path bound}
	E[\{g_x\}] \geq \sum_{k=1}^{\lvert\gamma_A\rvert} \sum_{\langle xy\rangle \in P_k} d(g_x, g_y).
\end{equation}
Along each path $P_k$, the first spin is assigned the cyclic permutation $\mathcal C^{(n)}$ and the last spin the identity permutation 1.
Therefore, the triangle inequality (invoked once for each path) implies that
\begin{equation}
\label{eq:pathwise triangles}
	\sum_{k=1}^{\lvert\gamma_A\rvert} \sum_{\langle xy\rangle \in P_k} d(g_x, g_y) \geq \sum_{k=1}^{\lvert\gamma_A\rvert} d(\mathcal C^{(n)}, 1) = (n-1) \lvert\gamma_A\rvert.
\end{equation}
Note that the right-hand side is just the energy cost of the configuration where we assign $\mathcal C^{(n)}$ to the spins in $E_A$ and 1 to all other spins.
We claim that this is the unique minimal energy configuration.
To see this, suppose that $\{g_x\}$ is an arbitrary configuration that achieves this energy cost.

{\em Case 1:} The only permutations that appear in $\{g_x\}$ are $\mathcal C^{(n)}$ and $1$.
Then the domain where $g_x = \mathcal C^{(n)}$ is a minimal cut between $A$ and $\bar A$, i.e., an entanglement wedge for $A$.
Since we have assumed that the entanglement wedge is unique, it must be equal to $E_A$.
Thus $\{g_x\}$ is the configuration described above.

{\em Case 2:} The configuration $\{g_x\}$ contains some other permutations.
Since it is a minimal energy configuration, both inequalities~\eqref{eq:path bound} and \eqref{eq:pathwise triangles} above must be tight.
The fact that the first inequality is tight means that if an edge is not contained in any of the paths $P_k$ then the configuration $\{g_x\}$ necessarily assigns the same permutation in $\Sym_n$ to its endpoints.
It follows that the first inequality remains tight if we modify the configuration $\{g_x\}$ by changing an entire domain from one permutation to another.
For the second inequality, we can use the triangle inequality to see that the sequence of permutations in any path $P_k$ must always be of the form $\mathcal C^{(n)},\dots,\mathcal C^{(n)},***,1,\dots,1$, where $***$ denotes a sequence of permutations that are neither $\mathcal C^{(n)}$ nor $1$.
Indeed, if this were not the case then the energy cost of the corresponding path would be higher than $(n-1)$.
But this implies that by either changing all other permutations to $\mathcal C^{(n)}$, or by changing all of them to $1$, we obtain two \emph{distinct} minimal energy configurations that only contain $\mathcal C^{(n)}$ and $1$.
By case 1, this is a contradiction.

\section{\texorpdfstring{Calculation of $C_{2n}$ in Section~\ref{sec:correlationspec}}{Calculation of C\_2n in Section~\ref{sec:correlationspec}}}
\label{app:correlation}

\begin{figure}
  \centering
\includegraphics[width=0.85\textwidth]{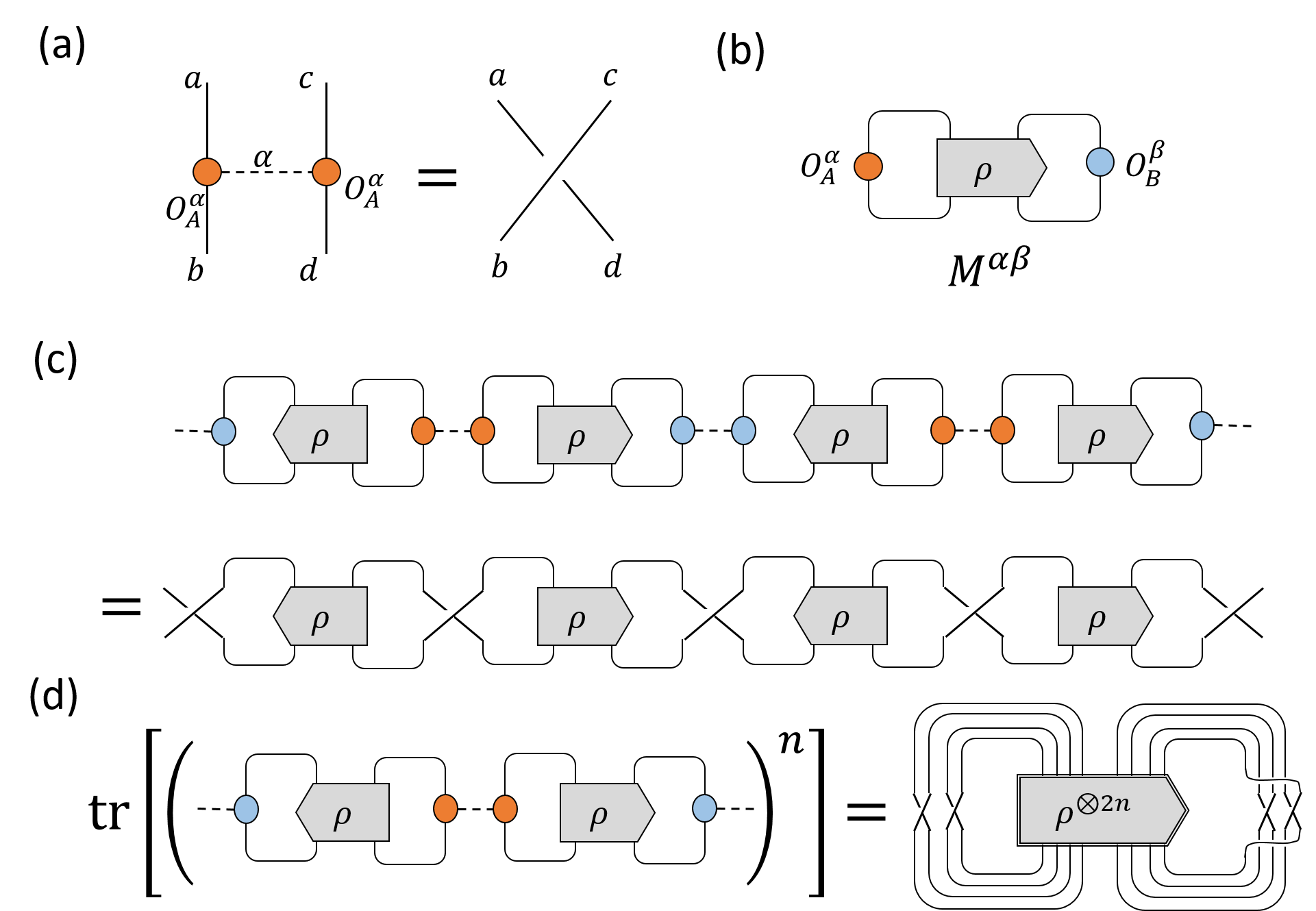}
\caption{(a) The graph representation of the orthonormality condition~(\ref{orthonormal}) of $O_A^\alpha$. A similar condition holds for $O_B^\alpha$. (b) The graph representing the matrix $M^{\alpha \beta}={\rm tr}\left[O_A^\alpha O_B^\beta \rho\right]$. We use red and blue dots to represent the basis operators $O_A^\alpha$ and $O_B^\beta$, respectively. We have drawn $\rho$ in a slightly assymetric shape to keep track of the difference between $A$ and $B$ regions. (c) Using the orthonormality condition in subfigure (a), the quantity $(M^\dagger M)^n$ is tranformed to a contraction of $2n$ copies of $\rho$. (d) A compact way of drawing ${\rm tr}\left[(M^\dagger M)^n\right]$, which corresponds to Eq.~(\ref{C2n}). } \label{fig:appC2n}
\end{figure}

In this appendix, we will present the derivation from Eq.~(\ref{C2ndef}) to Eq.~(\ref{C2n}) in Section~\ref{sec:correlationspec}. We first calculate $M^\dagger M$ using the orthonormality condition~(\ref{orthonormal}).
\begin{eqnarray}
\left[M^\dagger M\right]^{\alpha\beta}&=&{M^{\gamma \alpha}}^*M^{\gamma\beta}=\frac{{\rm tr}\left[O_B^\alpha O_A^\gamma \rho\right]{\rm tr}\left[O_A^\gamma O_B^\beta\rho\right]}{\left({\rm tr}\rho\right)^2}\nonumber\\
&=&\frac1{\left({\rm tr}\rho\right)^2}{\rm tr}_B\left[\rho O_B^\alpha\right]_{ab}{\rm tr}_B\left[O_B^\beta \rho\right]_{ba}
\end{eqnarray}
Similarly we can apply the orthonormality condition in the $B$ region when we multiply $M^\dagger M$. For example,
\begin{eqnarray}
{\rm tr}\left[\left(M^\dagger M\right)^2\right]=\frac1{({\rm tr}\rho)^4}\rho_{am,bn}\rho_{bk,al}\rho_{cl,dk}\rho_{dn,cm}
\end{eqnarray}
in which $a,b,c,d$ are indices in the Hilbert space of $A$, and $m,n,k,l$ are those in $B$. The best way of visualizing this calculation is by introducing a diagrammatic representation, as is shown in Fig.~\ref{fig:appC2n}. In the trace of $(M^\dagger M)^n$, there are $2n$ copies of the density matrix $\rho$. The $n$ contractions of $A$ indices lead to pairwise permutations between pairs of density matrices $1\leftrightarrow 2$, $3\leftrightarrow 4$,\dots,$2n-1\leftrightarrow 2n$. Similarly, the contractions of $B$ indices lead to pairwise permutations between $2\leftrightarrow 3$, $4\leftrightarrow 5$, \dots, $2n\leftrightarrow 1$. This concludes the proof of Eq.~(\ref{C2n}).

\section{Partition function of Ising model on the square lattice}
\label{app:Ising}

In this appendix, we calculate the partition function of the Ising model in the large $D$ (low temperature) limit on a 2D rectangular lattice of size $M\times N$, with periodic boundary conditions along the first direction and open boundary conditions along the second one.
We will use several results that can be found in~\cite{McCoy1967}.
As in the main text, we let $2 \beta = \log D$.
We denote the partition function of the Ising model at temperature $1/\beta$, with boundary pinning field pointing down everywhere, by $Z_0(\beta)$, and its zero-temperature limit by $Z_0^\infty = Z_0(\beta\rightarrow\infty)$.
When system size is large, $M,N \gg 1$,
\begin{eqnarray}
    \frac{\overline{Z_0}}{Z_0^\infty}
    &=& \left(\frac{2}{e^{2\beta}} \exp\left[ \frac{1}{2(2\pi)^2}\int_0^{2\pi} d\theta_1 d\theta_2 \log\left[(\cosh 2\beta)^2 -\sinh 2\beta (\cos\theta_1 + \cos\theta_2) \right] \right]\right)^{MN}\nonumber \\
    &\cdot&\left( \frac{e^\beta}{\cosh\beta}
     \exp \left[\frac{1}{2\pi} \int_{-\pi}^{\pi} d\theta \log\left(\frac{1+ \bar{\phi}(\theta)}{2} \right) \right]\right)^N  \left( \frac{\cosh\beta}{e^\beta}
     \exp \left[\frac{1}{4\pi} \int_{-\pi}^{\pi} d\theta \log\left(1 - W(\theta) \right) \right]\right)^{2N}\nonumber \\
    &=& \left(1+ D^{-4} +O(D^{-6})\right)^{MN}\left(1+ \frac{1}{5} D^{-5} +O(D^{-4})\right)^N, \nonumber
\end{eqnarray}
where
\begin{align*}
  \bar{\phi}(\theta) &= -\sqrt{\frac{(1-\alpha_1(\beta) e^{i \theta})(\alpha_2(\beta)- e^{-i \theta})}{(1-\alpha_1(\beta) e^{-i \theta})(\alpha_2(\beta)- e^{i \theta})}}, \\
  W(\theta) &= \frac{\tanh(\beta)^2\left|1+e^{i\theta}\right|^2}{\tanh(\beta) \left|1+\tanh(\beta) e^{i\theta}\right|^2-(1-\tanh(\beta)^2\alpha(\theta)))}, \\
  \alpha_1(\beta) &= \tanh(\beta)\frac{1-\tanh(\beta)}{1+\tanh(\beta)}, ~~~~ \alpha_2(\beta)=\frac{1}{\tanh(\beta)}\frac{1-\tanh(\beta)}{1+\tanh(\beta)}, \\
  \alpha(\theta) &= \frac{(1+\tanh(\beta)^2)^2}{2\tanh(\beta)(1-\tanh(\beta)^2)}-\cos(\theta) +\frac{\left|\left(1-\alpha_1(\beta) e^{i \theta}\right) \left(1-\alpha_2^{-1}(\beta) e^{i \theta}\right)\right|}{2\tanh(\beta)}.
\end{align*}
Thus the leading order correction in the large $D$ limit is $M N D^{-4}$.

Now we turn to $\overline{Z_1}(\beta)$, the partition function of the Ising model at temperature $1/\beta$, with boundary pinning field down everywhere except for in a single interval of length $L$.
Similarly as above, we denote the corresponding zero-temperature limit by $Z_1^\infty = Z_1(\beta\rightarrow\infty)$.
Using the duality of Ising model, we know that
$\overline{Z_1}(\beta)/\overline{Z_0}(\beta) = \langle S_{0,0} S_{0,L} \rangle(\beta^\prime)$, where $ e^{-2\beta^\prime}=\tanh \beta $.
Here, $\langle S_{0,0} S_{0,L} \rangle(\beta^\prime)$ denotes the two-point correlation function on the boundary of the dual lattice at temperature $1/\beta^\prime$, whose analytical form is also provided in~\cite{McCoy1967}.
When $L \gg 1$, we can expand $\overline{Z_1}(\beta)/ Z_1^\infty$ to leading order in $D$ and $L$,
\begin{eqnarray}
 \frac{\overline{Z_1}(\beta)}{ Z_1^\infty}&=& \frac{\langle S_{0,0} S_{0,L}\rangle(\beta^\prime)}{\langle S_{0,0} S_{0,L}\rangle(\beta^\prime\rightarrow0)} \frac{\overline{Z_0}(\beta)}{ Z_0^\infty} \nonumber\\
&=&\left(1 + D^{-4} +O(D^{-6})\right)^{MN}\left(1+ \frac{1}{5} D^{-5} +O(D^{-4})\right)^N\left(1+2 D^{-1} + O(D^{-2})\right)^L. \nonumber
\end{eqnarray}
Thus the leading order correction is $2L D^{-1}$.

\section{Average second R\'{e}nyi entropy for 2-designs}
\label{app:second-moments}

In the following, we will show that \eqref{eq:two design} holds for an arbitrary 2-design in the limit of large bond dimension.
We recall from Section~\ref{sec:finiteD} that the inequality $S_2(A) \leq S(A)$ holds for arbitrary quantum states, while $S(A) \leq \log \rank \rho_A \leq \log D \lvert \gamma_A \rvert$ in any tensor network state. Therefore it remains to prove the lower bound on the average of the second R\'{e}nyi entropy.

The first moments of the Haar measure are given by $\overline{\ket{V_x}\!\!\bra{V_x}} = I/D_x$, and so $\overline{T} = 1/\prod_x D_x$.
Together with our calculation in Sections~\ref{sec:setup} and \ref{sec:RTformula} it follows that
\begin{equation}
\label{eq:first piece is rt}
  -\log \frac {\overline{Z_1}} {\overline{T}^2} \rightarrow S_{RT}(A) - \log k
\end{equation}
in the large $D$ limit, 
where have introduced $T = \tr \rho$ and recall that $Z_1 = \tr\rho_A^2$.
We now bound the fluctuations in the trace $T$.
Noting that $T^2 = Z_0$, we can bound the variance of $T / \overline{T}$ as follows:
\[ \overline{\left( \frac T {\overline T} - 1 \right)^2}
 = \frac {\overline{Z_0}} {\overline{T}^2} - 1
 = \sum_{\{s_x\}} e^{-\mathcal A_0[\{s_x\}] + \sum_x \log D_x^2} - 1
 \leq \sum_{\{s_x\}} e^{-\mathcal A[\{s_x\}]} - 1
 = \sum_{\{s_x\} \not\equiv +1} e^{-\mathcal A[\{s_x\}]}
\]
where $\mathcal A_0$ refers to the Ising action in its original form~\eqref{Isingaction2} and $\mathcal A$ to the simplified form~\eqref{Isingaction3} with constants removed.
But any nontrivial spin configuration incurs an energy cost of at least $\log D$, so that we obtain the upper bound
\[ \overline{\left( \frac T {\overline T} - 1 \right)^2} \leq 2^V / D. \]
By Chebyshev's inequality, it follows that, for any $\varepsilon > 0$,
\begin{equation}
\label{eq:p good}
	p_{\text{good}} \equiv \Prob\left( \frac T {\overline T} \geq 1 - \varepsilon \right) \geq 1 - O\left(\frac 1 {D\varepsilon^2}\right).
\end{equation}
We now condition on the event that $T/\overline{T} \geq 1-\varepsilon$.
Writing $\gav X$ for corresponding averages, we obtain the following bound using concavity of the logarithm,
\begin{align}
  \nonumber&\gav {S_2(A)}
  = \gav{-\log \frac {Z_1} {T^2}}
  = -\log \frac {\gav{Z_1}} {\overline{T}^2} + \gav{2 \log \frac {T} {~\overline{T}~}} - \gav{\log \frac {Z_1} {\gav{Z_1}}}\\
  \nonumber&\geq -\log \frac {\gav{Z_1}} {\overline{T}^2} + \gav{2 \log \frac {T} {~\overline{T}~}}
  \geq-\log \frac {\overline{Z_1}} {\overline{T}^2} + \log {p_{\text{good}}}+ \gav{2 \log \frac {T} {\overline{T}}} \\
  \label{eq:cond avg}&\geq-\log \frac {\overline{Z_1}} {\overline{T}^2} + \log {p_{\text{good}}}+ 2\log(1-\varepsilon),
\end{align}
since $\gav{Z_1} \leq \overline{Z_1} / p_{\text{good}}$.
On the other hand, $T/\overline{T} \geq 1-\varepsilon$ implies that $\rho\neq0$.
Thus,
\begin{equation}
\label{eq:uncond to cond}
    \nav{S_2(A)}
    \geq \frac{p_\text{good}}{p_{\neq 0}} \, \gav{S_2(A)}
    \geq p_\text{good} \, \gav{S_2(A)}
    \geq \gav{S_2(A)} - O\Big(\frac {\log D} {D \varepsilon^2}\Big),
\end{equation}
where we have used~\eqref{eq:p good} and $S_2(A) \leq \lvert\gamma_A\rvert \log D = O(\log D)$, as follows from the deterministic upper bound in~\eqref{eq:two design}, which holds for an arbitrary tensor network state.
The desired lower bound,
\begin{align*}
  \nav{S_2(A)}
	&\geq S_{RT}(A) - \log k - o(1),
\end{align*}
now follows by combining \eqref{eq:uncond to cond}, \eqref{eq:cond avg} and \eqref{eq:first piece is rt} and choosing, e.g., $\varepsilon = D^{-1/4}$.

\section{Contractions of stabilizer states}
\label{app:stabcontract}

In this appendix we will show that a tensor network state built by contracting stabilizer states is again a stabilizer state.
More generally, let $\ket\phi_A$ and $\ket\psi_{AB}$ denote two stabilizer states, where $A = (\mathbb C^p)^{\otimes a}$ and $B = (\mathbb C^p)^{\otimes b}$, with stabilizer groups $G$ and $H$, respectively, such that $\ket{\psi'_B} \equiv \braket{\phi_A | \psi_{AB}} \neq 0$.
We will show that in this case $\ket{\psi'_B}$ is a stabilizer state, a fact that is certainly well-known to experts.
To see this, we start by writing the contracted state as
\begin{equation*}
\begin{split}
  \ket{\psi'_B}\!\!\bra{\psi'_B}
  &= \braket{\phi_A | \psi_{AB}}\!\!\braket{\psi_{AB} | \phi_A}
   = \tr_A \big[ \ket{\psi_{AB}}\!\!\bra{\psi_{AB}} \ket{\phi_A}\!\!\bra{\phi_A} \big] \\
  &= \frac 1 {\lvert G \rvert} \frac 1 {\lvert H \rvert} \sum_{g_A  \in G} \sum_{h_{AB} \in H} \tr_A(g_A h_{AB}) \\
  &= \frac 1 {\lvert H \rvert} \sum_{g_A \in G} \sum_{h_{AB} \in H} \varphi(g_A, h_{AB}),
\end{split}
\end{equation*}
where we have introduced the function $\varphi(g_A,h_{AB}) \equiv \frac 1 {\lvert G \rvert} \tr_A(g_A h_{AB})$.
We claim that
\[ K = \{ (g_A,h_{AB}) \in G \times H : \varphi(g_A,h_{AB}) \neq 0 \} \]
is a subgroup of $G \times H$ and that the restriction of $\varphi$ to $K$ is a group homomorphism.
To see this, note that any $h_{AB} \in H$ can be written as $h_{AB} = h_A h_B$, where $h_A$ and $h_B$ are elements of the generalized Pauli groups of $A$ and $B$, respectively.
Thus $\varphi(g_A,h_{AB}) = (\tr g_A h_A) h_B / \lvert G \rvert$, which is either zero or equal to some Pauli operator.
In the latter case, $h_A = \lambda g_A^{-1}$ for some overall phase $\lambda$; in particular, $h_A$ commutes with $G$.
If also $\varphi(g'_A,h'_{AB}) \neq 0$ then likewise $h'_A = \lambda' (g'_A)^{-1}$ for some phase $\lambda'$, and it is now easy to verify that
\[ \varphi(g_A g'_A, h_{AB} h'_{AB}) = \lambda \lambda' h_B h'_B = \varphi(g_A, h_{AB}) \varphi(g'_A, h'_{AB}). \]
This implies both that $K$ is a subgroup of $G \times H$ and that $\varphi\big|_K$ is a group homomorphism.
Thus $L \equiv \varphi(K)$ is a (commutative) subgroup of the Pauli group; it follows that
\begin{equation*}
  \ket{\psi'_B}\!\!\bra{\psi'_B}
  = \frac 1 {\lvert H \rvert} \sum_{g_A \in G} \sum_{h_{AB} \in H} \varphi(g_A, h_{AB})
  = \frac {\lvert \ker \varphi \rvert} {\lvert H \rvert} \sum_{g_B \in L} g_B
  = \frac {\lvert K \rvert} {\lvert H \rvert} \frac 1 {\lvert L \rvert} \sum_{g_B \in L} g_B.
\end{equation*}
Thus $\ket{\psi'_B}$ is indeed a subnormalized stabilizer state, as we set out to show.


Since maximally entangled states are stabilizer states, it follows at once that a tensor network state~\eqref{peps} constructed by contracting stabilizer states $\ket{V_x}$ is again a stabilizer state.

\bibliographystyle{JHEP}
\bibliography{holography}

\end{document}